\documentclass[fleqn,usenatbib]{mnras}

\usepackage{newtxtext,newtxmath}

\usepackage[T1]{fontenc}
\usepackage{ae,aecompl}

\usepackage{graphicx}	
\usepackage{amsmath}	
\usepackage{amssymb}	
\usepackage{multirow}

\newcommand{\hdtwenty}{HD~209458~b}
\newcommand{\hdeighteen}{HD~189733~b}
\newcommand{\Teq}{T_{\textrm{eq}}}
\newcommand{\Teff}{T_{\textrm{eff}}}
\newcommand{\Tint}{T_{\textrm{int}}}
\newcommand{\Porb}{P_{\textrm{orb}}}
\newcommand{\Prot}{P_{\textrm{rot}}}
\newcommand*\chem[1]{\ensuremath{\mathrm{#1}}}
\newcommand{\kzz}{K_{zz}}
\newcommand{\timescale}[1]{\tau_{\textrm{#1}}}
\newcommand{\degrees}{^\circ}

\title[Grid of Pseudo-2D Chemistry Models]{Grid of Pseudo-2D Chemistry Models for Tidally-Locked Exoplanets.
I. The Role of Vertical and Horizontal Mixing}

\author[R. Baeyens et al.]{%
Robin Baeyens,$^{1}$\thanks{E-mail: robin.baeyens@kuleuven.be}
Leen Decin,$^{1}$
Ludmila Carone,$^{2}$
Olivia Venot,$^{3}$
Marcelino Ag\'undez$^{4}$ \newauthor
and Paul Molli\`ere$^{2}$
\\
$^{1}$Institute of Astronomy, KU Leuven, Celestijnenlaan 200D, 3001, Leuven, Belgium\\
$^{2}$Max-Planck-Institut f\"ur Astronomie, K\"onigstuhl 17, 69117 Heidelberg, Germany\\
$^{3}$Laboratoire Interuniversitaire des Syst\`{e}mes Atmosph\'{e}riques (LISA), UMR CNRS 7583, Universit\'{e} Paris-Est Cr\'{e}teil, Universit\'{e} de Paris, Institut Pierre
\\ Simon Laplace, Cr\'{e}teil, France\\
$^{4}$Instituto de F\'isica Fundamental, CSIC, C/ Serrano 123, 28006 Madrid, Spain
}

\date{Accepted XXX. Received YYY; in original form ZZZ}

\pubyear{2020}

\begin{document}
\label{firstpage}
\pagerange{\pageref{firstpage}--\pageref{lastpage}}
\maketitle

\begin{abstract}
The atmospheres of synchronously rotating exoplanets are intrinsically three-dimensional, and fast vertical and horizontal winds are expected to mix the atmosphere, driving the chemical composition out of equilibrium. 
Due to the longer computation times associated with multi-dimensional forward models, horizontal mixing has only been investigated for a few case studies.
In this paper, we aim to generalize the impact of horizontal and vertical mixing on the chemistry of exoplanet atmospheres over a large parameter space. We do this by applying a sequence of post-processed forward models for a large grid of synchronously rotating gaseous exoplanets, where we vary the effective temperature (between 400~K and 2600~K), surface gravity, and rotation rate. We find that there is a dichotomy in 
the horizontal homogeneity of the chemical abundances. Planets with effective temperatures below 1400~K tend to have horizontally homogeneous, vertically quenched chemical compositions, while planets hotter than 1400~K exhibit large compositional day-night differences for molecules such as \chem{CH_4}. Furthermore, we find that the planet's rotation rate impacts the planetary climate, and thus also the molecular abundances and transmission spectrum.
By employing a hierarchical modelling approach, we assess the relative importance of disequilibrium chemistry on the exoplanet transmission spectrum, and conclude that the temperature has the most profound impact. Temperature differences are also the main cause of limb asymmetries, which we estimate could be observable with the \textit{James Webb Space Telescope}. 
This work highlights the value of applying a consistent modelling setup to a broad parameter space in exploratory theoretical research.
\end{abstract}





\section{Introduction}

The atmospheres of exoplanets hold great potential for answering questions about a planet's history and habitability. Indeed, accurately determining their bulk chemical composition can help constrain the conditions of planet formation, and provide insight into the potential migration history \citep{Oberg2011, Madhusudhan2016, Notsu2020}. In addition, determining the atmospheric composition is a crucial step in achieving one of the ultimate goals of exoplanet astronomy: detecting biosignatures on extrasolar planets \citep{Schwieterman2018}. We are entering the era of comparative exoplanetary science, as developments of new observational facilities, such as the \textit{James Webb Space Telescope} \citep{Greene2016} and \textit{Ariel} \citep{Tinetti2018}, will result in progressively more accessible atmospheric detections. Therefore, understanding the complex physical and chemical processes occurring in the atmospheres of exoplanets will be essential to properly interpret new observational results.

The atmospheres of tidally locked hot Jupiters, i.e.~giant close-in planets that are the most suitable for atmospheric observations, are highly dynamic environments due to the strong one-sided irradiation from their host star. Acting as a planetary heat engine, this irradiation is the driver for atmospheric circulation with very fast ($\sim$km/s) wind flows, which redistribute the heat from the day side toward the night side of the planet \citep{Koll2018}. These high wind speeds have been detected with high-resolution transmission spectroscopy in the hot Jupiters \hdtwenty{} \citep{Snellen2010} and \hdeighteen{} \citep{Brogi2016, Flowers2019}. The dynamic climate is tightly coupled to the atmospheric chemistry through the temperature and the wind field that causes mixing. This interplay potentially yields very different chemical environments between the day and night sides \citep{Ehrenreich2020}. All these aspects make exoplanet atmospheres intrinsically three-dimensional and motivate the need for sophisticated modelling tools. 

General circulation models (GCMs), i.e.~three-dimensional hydrodynamics codes used to simulate the large-scale climate of planets, have been frequently applied to tidally locked gaseous exoplanets \citep[e.g.][]{Showman2009, Rauscher2012, Charnay2015, Amundsen2016, Mendonca2018_phasecurves, Carone2020}. A prominent result of many hot Jupiter GCMs is the consistent prediction of a fast equatorial jet stream, advecting air, and thus heat, eastward. The mechanism for this jet stream was identified to be the up-gradient pumping of angular momentum through planetary waves \citep{Showman2011, Tsai2014}, and its existence has been observationally inferred from the offsets in the peak flux emission of several hot Jupiter phase curves \citep[e.g.][and references therein]{Knutson2007, Komacek2017, Zhang2018_phasecurves}. The 3D structure of the atmosphere and the heat advection through fast winds have been demonstrated to introduce temperature gradients and chemical heterogeneities between the day-night terminators, having an important effect on transit observations \citep{Caldas2019, Pluriel2020}. 

Studies with chemical kinetics models -- initially one-dimensional -- have demonstrated that disequilibrium chemistry significantly changes the atmospheric mixing ratios compared to local equilibrium chemistry \citep[e.g.][]{Moses2011, Venot2012, Tsai2017}. Vertical mixing, if efficient enough, can cause quenching in the molecular abundances of prevalent species like \chem{CO} and \chem{CH_4}. If this happens, chemical reactions are too slow to achieve chemical equilibrium, and the gas gets replenished dynamically, potentially with dramatic changes in the atmospheric composition at pressures lower than a critical level. Furthermore, photochemical reactions in the upper atmosphere day side of these highly irradiated planets can cause dissociation of molecules like \chem{CH_4} and \chem{NH_3}, especially in cooler planets, where the reaction rates are comparatively slow and these photochemically active molecules tend to be favoured \citep{Moses2014}. As a next step, pseudo-2D chemical codes employ the same chemical kinetics, but in addition introduce a longitude-dependent background temperature and uniform horizontal advection \citep{Agundez2012, Agundez2014}. It has been demonstrated that species can also be efficiently mixed horizontally, in addition to vertically, as they are advected by the equatorial jet stream \citep{Agundez2014, Venot2020_wasp43b, Moses2021}. This can result in contamination of the night side by photochemically produced species, such as HCN, and can cause a horizontal homogenization of the chemical composition in general. 
As observational efforts are increasingly directed towards cooler objects, accurately taking into account disequilibrium chemistry will become more important when deriving bulk elemental abundances \citep{Line2013, Baudino2017, Venot2020_wasp43b}.

Due to the high computational cost associated with both GCMs and chemical kinetics models, previous efforts of coupling dynamics and chemistry in 1D, 2D or 3D models have often relied on simplifying assumptions. An ubiquitous, parametrized approach of implementing dynamic disequilibrium chemistry is to approximate vertical mixing as a one-dimensional diffusion process, tuned by the eddy diffusion coefficient $\kzz$. Previous studies have either considered $\kzz$ to be a free parameter \citep[e.g.][]{Miguel2014, Drummond2016, Tsai2017}, or derived a numerical value for $\kzz$ from the Eulerian mean overturning circulation \citep{Heng2011_frierson_Phillipps_II_kzz} or from mixing length theory \citep[e.g.][]{Cooper2006, Moses2011, Agundez2012, Venot2012, Lee2015, Rimmer2016, Kitzmann2018, Molaverdikhani2020_hatp7b, Shulyak2020}. In the latter case, $\kzz = w(z) L(z)$ is a commonly used assumption, where $w(z)$ is the root mean square of the horizontally averaged vertical velocity, and $L(z)$ is a characteristic mixing length scale, often taken to be a (fraction of) the atmospheric pressure scale height. 
It has, however, been argued that the mixing length approximation is only applicable in convective regimes, and is thus too crude to describe the complex motions in the radiative dynamic atmospheres of hot Jupiters. \cite{Parmentier2013} have equipped their GCM with passive tracers, which are advected by the flow, in order to better constrain the amount of (vertical) mixing in the canonical hot Jupiter \hdtwenty{}. Likewise, \cite{Charnay2015} have used the method of passive tracers to investigate the mixing in GJ~1214~b. Both studies found $\kzz$ values up to two orders of magnitude lower than those derived from mixing length theory, and have derived 1D parametrized, pressure-dependent expressions for the vertical mixing strength in these planets.
Some theoretical studies have aimed at refining and generalizing the theory of vertical mixing, by deriving analytic expressions for the quench level in brown dwarfs \citep{Bordwell2018}, and for the $\kzz$ of both fast-rotating 
planets \citep{Zhang2018_and_showman_I_fastrotating}, and tidally-locked, irradiated planets \citep{Zhang2018_and_showman_II_tidallylocked, Komacek2019}. These studies have shown that $\kzz$ generally depends on properties of the chemical species, such as the mean chemical lifetime $\tau_{\rm chem}$, in addition to the (local) velocity field. Furthermore, \cite{Komacek2019} have demonstrated a positive trend between $\kzz$ and equilibrium temperature, as well as different regimes of vertical transport, depending on whether the atmospheric climate is governed by a fast equatorial jet stream or day-to-night-type winds. Recently, statistical evidence has been found for variable mixing efficiencies across the range of hot Jupiter atmospheres \citep{Baxter2021}.

Another, more direct method of coupling chemistry to the dynamic atmosphere has been to implement the interconversion of \chem{CO}, \chem{CH_4} and \chem{H_2O}, dominant opacity sources in hot Jupiter atmospheres, in GCMs by prescribing a fixed chemical relaxation time-scale towards equilibrium \citep{Cooper2006, Drummond2018_HD209458b, Drummond2018_HD189733b}. This approach was expanded upon by \cite{Mendonca2018_disequilibrium} by using different and more accurate chemical relaxation times for each molecule, as determined by the rate-limiting reaction \citep{Tsai2018}. Finally, the most recent development has been the full coupling of a GCM with a reduced chemical kinetics scheme \citep{Drummond2020}. These studies demonstrate the potential of disequilibrium chemistry to homogenize the atmospheric composition both vertically and horizontally. 
A caveat to chemistry-coupled GCMs is that, due to their limited resolution, only the large-scale advection of chemical species is taken into account. This could underestimate the mixing efficiency if microphysical or sub-grid scale processes, such as turbulence, are important \citep{Menou2019}.

As forward models of exoplanet atmospheres are becoming increasingly more sophisticated and complex, many studies have focused on a small number of exoplanets to apply these models to. Most prominently, these include \hdtwenty{} \citep{Showman2008, Showman2009, Rauscher2012, Mayne2014, Agundez2014, Helling2016, Drummond2018_HD209458b, Drummond2020, Carone2020}, \hdeighteen{} \citep{Showman2008, Showman2009, Agundez2014, Helling2016, Drummond2018_HD189733b, Steinrueck2019}, and WASP-43~b \citep{Kataria2015, Mendonca2018_phasecurves, Mendonca2018_disequilibrium, Venot2020_wasp43b, Carone2020, Helling2020}. One major advantage of this practice is the ability to immediately measure the model against past simulation results, which serve as a benchmark. Such model hierarchy is crucial to investigate the complex physical and chemical processes in exoplanet atmospheres. 

In order to explore the diversity in exoplanet atmospheres, ensemble studies using GCMs with varying levels of complexity have been carried out. This is especially the case for atmospheric dynamics. \cite{Showman2015} have conducted a parameter study by varying the rotation period for nine 3D climate models of (non-)synchronously rotating giant exoplanets, and \cite{Kataria2016} have generated consistent climate models for nine known hot Jupiters. Furthermore, some studies have endeavoured to assimilate climate model ensembles into analytic theories \citep{Perez-Becker2013, Komacek2016, Zhang2017, Komacek2019}. These studies have established a variety of climate types and highlighted the potential variations in chemistry and cloud composition.
Regarding exoplanet atmospheric chemistry, however, due to the computation times involved in multi-dimensional chemical kinetics, most conducted ensemble or parameter studies are one-dimensional \citep[e.g.][]{Moses2013, Miguel2014, Heng2016}.
As such, they do not include the potentially important effects of zonal (i.e.~longitudinal) quenching or the zonal dependency of photochemistry. To date, one chemical kinetics ensemble study has incorporated zonal mixing and photochemistry, in order to assess the impact of atmospheric temperatures and chemistry on the phase curves of sub-Neptunes \citep{Moses2021}.  

In this work, we aim to explore the chemical diversity of exoplanet atmospheres over a wide range of temperatures, gravities and climate regimes, with a specific focus on disequilibrium chemistry due to atmospheric dynamics. 
In order to find a trade-off between simulation complexity and computation time, we employ a sequence of post-processed models: a 3D GCM with simplified radiative transfer, and a pseudo-2D chemical kinetics code with a reduced chemical network. With this methodology, we present a large grid of 144 climate and chemistry simulations, incorporating zonal and vertical mixing. In this work, we focus exclusively on the effects of zonal and vertical mixing, and we aim to investigate the additional effect of photochemistry in detail in a follow-up paper.

This paper is organized as follows. In Section~\ref{sec_methods} we present the sequence of modelling tools that is used for our atmospheric simulations. In Section~\ref{sec_grid}, the grid and the selected parameter space are discussed. The simulation results are presented in Section~\ref{sec_results} and further discussed in Section~\ref{sec_discussion}. In our discussion we focus on the impact of zonal advection (\ref{sec_zonal_context}), limb asymmetries (\ref{sec_limbs}), case studies of selected exoplanets (\ref{sec_case_studies}), observing windows for some molecules (\ref{sec_windows}), deep atmospheric quenching (\ref{sec_deep_quenching}), and brown dwarf--white dwarf binaries (\ref{sec_bd-wd}). Furthermore, some of the shortcomings of the models used in this work are discussed in Section~\ref{sec_limitations}. Finally, our conclusions are presented in Section~\ref{sec_conclusions}.

\section{Modelling Tools}
\label{sec_methods}

\begin{figure}
    \centering
    \includegraphics[width=0.9\columnwidth]{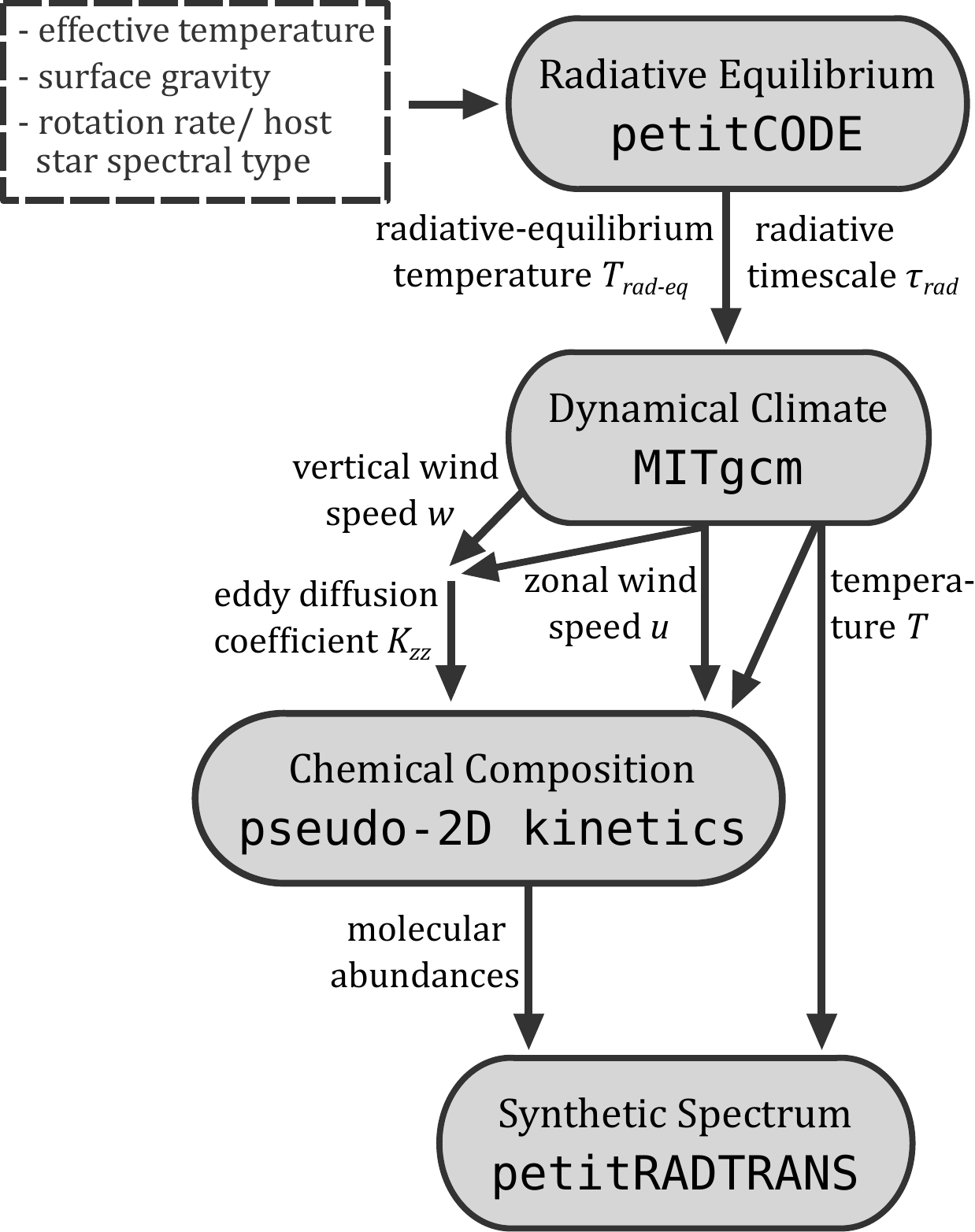}
    \caption{Schematic overview of the different models used in this work, and the post-processing sequence in which they are linked.}
    \label{fig_codes}
\end{figure}

For the forward modelling of our grid of planetary atmospheres, we use a range of modelling tools, further discussed below. Each tool relates to a different  
aspect of the atmosphere, namely radiative-convective equilibrium \citep[\textit{petitCODE},][]{Molliere2015, Molliere2017}, dynamical climate \citep[\textit{MITgcm},][]{Adcroft2004}, chemical composition \citep[pseudo-2D chemistry code,][]{Agundez2014}, and synthetic spectrum \citep[\textit{petitRADTRANS},][]{Molliere2019}. 
These codes are used in a post-processing sequence, meaning that there is no iteration or feedback between them. This is done in order to conserve the computational efficiency. We discuss the limitations of the post-processing approach in Section~\ref{sec_limitations_postproc}. To achieve cohesion in the modelling, all parameters are passed down the modelling chain consistently, as shown schematically in Fig.~\ref{fig_codes}. Hereby, the number of free parameters is restricted where possible. In Section~\ref{sec_parameter_dependence}, the parameter dependencies are discussed in more detail.

\subsection{Radiative-Convective Equilibrium}\label{sec_methods_petitcode}

The first step is the construction of several 1D pressure-temperature profiles for each planetary atmosphere. This is achieved with \textit{petitCODE}, a 1D plane-parallel radiative transfer code for exoplanet atmospheres \citep{Molliere2015, Molliere2017}. Starting from a gas mixture of solar metallicity and a carbon-to-oxygen (C/O) ratio of 0.55, the code converges the pressure-temperature profile to a valid solution, under the assumption of radiative-convective equilibrium and chemical equilibrium. We only consider opacities due to absorption, and neglect scattering processes. 

The molecular opacities of 
\chem{H_2}, \chem{C_2H_2}, \chem{H_2S} \citep{Rothman2013_HITRAN},
\chem{H_2O}, \chem{CO}, \chem{CO_2}, \chem{OH} \citep{Rothman2010_HITEMP}, 
\chem{CH_4} \citep{Yurchenko2014_CH4},
\chem{NH_3} \citep{Yurchenko2011_NH3},
\chem{HCN} \citep{Harris2006_HCN, Barber2014_HCN}, 
and \chem{PH_3} \citep{Sousa-Silva2015}
have been included in the computation, as well as opacities for \chem{Na} and \chem{K} \citep{Piskunov1995_VALD}, and collision-induced absorption of \chem{H_2}-\chem{H_2} and \chem{H_2}-He. We note that the opacities of \chem{TiO}, \chem{VO}, \chem{H^-} and several (ionized) metals have been excluded, despite their possible importance for the thermal structure of strongly irradiated planetary atmospheres \citep{Arcangeli2018, Lothringer2018, Parmentier2018}. The reason for this is twofold. First, the prevalence of these absorbers, especially TiO and VO, and their role in creating thermal inversions in the atmospheres of hot irradiated planets is still not fully understood (see e.g.~\cite{Piette2020} for a recent discussion). Since our goal is primarily to analyse the chemistry for a broad range of planets in a  consistent manner, we opt to leave out these absorbers, rather than complicate the picture. Second, the inclusion of these opacities tends to give rise to numerical difficulties when using \textit{petitCODE}, especially when the angle of incident radiation is very shallow. The omission of \chem{TiO}, \chem{VO}, \chem{H^-}, and several metal-opacities could lead to an underestimation of the day-side temperature and radiative forcing in our models for planets with $\Teff \geq 2200$~K. We return to the implications of excluding these opacities in Section~\ref{sec_limitations_ultrahot}.

For each planetary atmosphere in the grid, we follow a similar methodology in constructing a full radiative-equilibrium temperature view of the atmosphere as \cite{Showman2008} and \cite{Carone2020}. More specifically, for a fixed orbital distance multiple \textit{petitCODE} models are computed with varying values of $\mu = \cos{\theta}$, where $\theta$ is the angle of incident stellar irradiation. We have adopted values of $\mu = 0.1, 0.2, 0.3, 0.4, 0.5, 0.6, 0.7, 0.8, 0.9$ and 1.0. For the night side of the planet ($\mu = 0.0$) we computed a \textit{petitCODE} model without any irradiation from the host star. Finally, these models have been gathered so that their temperatures reach the same common deep adiabat, using the entropy-matching method (see Appendix~\ref{sec_appendix_adiabat}). This approach results in a 3D view of the radiative-equilibrium temperature of the planet.

Furthermore, we again use the same method as \cite{Showman2008} and \cite{Carone2020} for determining the radiative time-scales $\timescale{rad}$. We use \textit{petitCODE} to perturb each converged, pressure-dependent temperature profile. Next, the radiative time-scale at that particular temperature and pressure is computed with
\begin{equation}
    \label{eq_taurad}
    \timescale{rad}(p, T) = \Delta T \frac{\rho c_P}{\Delta F / \Delta z},
\end{equation} where the $\Delta T = 10$~K is the temperature perturbation, $\rho$ is the density, $c_P$ is the heat capacity and $\Delta F / \Delta z$ is the discretized vertical gradient of the flux perturbation at that atmospheric layer. Together with the radiative-equilibrium temperature profiles, the radiative time-scales are then used as input for the radiative forcing of the GCM. 

\subsection{Dynamical Climate}

The dynamical climate of the atmosphere is computed with a General Circulation Model (GCM), more specifically the \textit{MITgcm}\footnote{\url{mitgcm.org}} \citep{Adcroft2004}. While this GCM was originally developed for the Earth climate, it has been successfully applied to both rocky exoplanets \citep{Carone2014} and hot Jupiters \citep{Showman2009, Carone2020}. With \textit{MITgcm}, we solve the primitive hydrostatic equations on a cubed-sphere grid with a horizontal resolution of 192 $\times$ 32, which corresponds to a resolution of 128 $\times$ 64 on a latitude-longitude grid. This yields grid cells of about 2.8$\degrees$ wide. Furthermore, we use 45 vertical layers: forty layers are equally spaced in log-space between 0.1~mbar and 200~bar, and five additional layers with a thickness of 100~bar are added at the bottom of the atmosphere until 700~bar. This setup is the same as that of \cite{Carone2020} and yields a resolution of about three layers per atmospheric scale height.

Our climate model is forced using Newtonian relaxation of the local temperature towards a prescribed radiative-convective equilibrium temperature during the radiative time-scale:
\begin{equation}
    \frac{q}{c_p} = \frac{T_{\rm rad-eq}(\phi, \lambda, p) - T(\phi, \lambda, p)}{\timescale{rad}(p, T)},
\end{equation} where $q$ is the heating rate per unit volume, and $T$ and $T_{\rm rad-eq}$ are the local model and radiative-equilibrium temperatures respectively, with a dependence on the latitude $\phi$, longitude $\lambda$ and pressure $p$. The radiative-equilibrium temperatures and radiative time-scales are derived from the precomputed \textit{petitCODE} profiles. A sample of \textit{petitCODE} radiative-equilibrium temperature profiles, used as input for different climate models, is displayed in Appendix~\ref{sec_appendix_petitcode}. Rather than using the actual temperature, in the code the model is stepped forward using the potential temperature $\theta$, an adiabatically corrected temperature for a fluid parcel that is lifted from a reference pressure $p_0$ (in our model $p_0 = 700$~bar) to a new pressure $p$: 
\begin{equation}
    \theta = T \left( \frac{p_0}{p} \right)^\kappa.    
\end{equation}
Here, $\kappa = R / c_p$ is the ratio of the specific gas constant divided by the heat capacity at constant pressure. In the literature \citep[e.g.][]{Holton2013_ch4, Mayne2014}, the potential temperature can be found expressed using the Exner function $\Pi = \left( \frac{p}{p_0} \right)^\kappa$ as $\theta = T/\Pi$. The code is then stepped forward with   
\begin{equation}
    F_\theta = \frac{\theta_{\rm eq}(\mu, p) - \theta(\phi, \lambda, p)}{\timescale{rad}(p, T)},
\end{equation} where $F_\theta$ is the radiative forcing term in the energy equation ${c_p \Pi \frac{\rm{d} \theta}{\rm{d} t} = F_\theta}$. For each time step, the irradiation angle parameter $\mu = \cos{\phi} \cos{\lambda}$ is determined based on the horizontal location of the fluid parcel, assuming that the substellar point is located at $(\phi, \lambda) = (0\degrees, 0\degrees)$ and the day side at $-90\degrees < \lambda < 90\degrees$. The associated radiative-equilibrium temperature profile is the precomputed \textit{petitCODE} profile with that $\mu$-value, rounded up. For the whole night side, when $|\lambda| > 90\degrees$, the ${(\mu=0.0)}$-profile is used. The radiative time-scales $\timescale{rad}(p, T)$ are computed with an interpolation between temperatures, for a fixed pressure, on the precomputed \textit{petitCODE} grid using a third-degree polynomial. Finally, in order to prevent numerical problems such as unphysical temperatures, we apply a measure to set the temperature equal to the night-side equilibrium profile, whenever it would become negative during the temperature forcing time step. This numerical problem can sometimes arise on the night sides of highly irradiated and fast rotating planets, during the spin-up of the model.  
Overall, this setup is very similar to the one used by \cite{Showman2008} and especially \cite{Carone2020}. We note that the Newtonian relaxation scheme is a simplified method of radiative forcing, but one that has been extensively used to investigate the climates of hot Jupiters, because of its transparency and computational efficiency. This point is further discussed in Section~\ref{sec_limitations_postproc}.

Additionally, we applied friction to our climate simulations. Following the reasoning of \cite{Liu2013, Komacek2016, Carone2020}, we have used a basal drag that serves to anchor the bottom boundary of the simulation domain to the interior of the planet. This practise was demonstrated to reduce the dependence on initial conditions \citep{Liu2013} and numerically stabilize the simulation \citep{Carone2020}. We adopt the same description for basal friction as \cite{Carone2020}, using a Rayleigh-type friction proportional to the horizontal wind velocity $\mathbf{v_h}$:
\begin{equation}
    \label{eq_fric}
    \mathbf{F_v} = -\frac{1}{\timescale{fric}}\mathbf{v_h}.
\end{equation} Here, the friction time-scale $\timescale{fric}$ decreases linearly with pressure, from its maximal value of $\timescale{fric} = 1$~day at $p=700~$bar to a value of zero at $p < 490$~bar. The qualitative dynamical state is not sensitive to the exact value of $\timescale{fric}$ \citep{Liu2013, Komacek2016} and we adopt the value used by \cite{Carone2020}.
Furthermore, we applied a fourth-order Shapiro filter with a time-scale of 25~seconds, which acts as an explicit numerical diffusion mechanism (hyperviscosity) to smooth grid-scale noise, as is customary with (exoplanet) GCMs \citep[e.g.][]{Jablonowski2011, Showman2009, Heng2011_menou_phillips_I_suite, Carone2020}. 

All our simulations were run with fixed values for the specific gas constant and heat capacity, namely $R = 3589$~J kg$^{-1}$ K$^{-1}$ and $c_p$ = 1.3 $\cdot 10^4$~J kg$^{-1}$ K$^{-1}$ respectively. This is appropriate for a hydrogen-helium composition atmosphere with mean molecular weight 2.3. The chemical composition of the air does not come into play directly in the Newtonian-forced GCM, but only enters through the radiative temperatures and time-scales computed with \textit{petitCODE}. All simulations are performed for a fixed radius at the bottom of the model of 9.7$\cdot 10^4$~km (1.35~R$_{\rm Jup}$), the radius of a typical inflated hot Jupiter, so that the only differences between the grid simulations are the gravity, rotation rate and radiative forcing.

Following the recommendation of \cite{Sainsbury-Martinez2019}, we initialized our model with a hot adiabatic temperature profile. This allows for a comparatively fast convergence of the model to a steady state, even for the deep atmospheric layers, which are known to exhibit prohibitively long convergence times \citep[additionally, see the discussions in ][]{Mayne2017, Carone2020, Wang2020}. From there, all models were run for 1500 (Earth) days with time steps of 25~seconds, and the output has been averaged over the final 100~days of the simulation. The model convergence is further discussed in Appendix~\ref{sec_appendix_convergence}.

\subsection{Chemical Composition}\label{sec_methods_chemistry}

To compute the chemical composition of the atmosphere, we use the pseudo-two-dimensional chemical kinetics model developed by \cite{Agundez2014}. It involves an effectively one-dimensional vertical column of gas rotating along the equator of the planet. In this column, the coupled vertical continuity equations for each chemical species are integrated, incorporating vertical transport and production and loss rates from thermochemical kinetics \citep{Venot2012, Agundez2014}. The chemical reaction rates are computed with respect to a fixed background temperature which is discretized in longitude (zonally) and pressure (vertically). After a certain integration time, which is computed based on the zonal wind speed, the whole vertical column rotates along the equator as a solid body, thereby mimicking horizontal advection by a uniform equatorial jet stream. The integration of the coupled system of continuity equations \citep[eq.~(1) in ][]{Agundez2014} is then resumed for the new physical situation. This process is continued for a number of rotations around the planet, until the chemical mixing ratios reach a periodic steady state. For the most abundant atmospheric constituents this typically occurs after several tens of rotations. The result is a pseudo-2D, longitude- and pressure-dependent, chemical composition of the exoplanetary atmosphere. The pseudo-2D framework for chemical kinetics \citep{Agundez2014, Venot2020_wasp43b} represents an intermediate step between 1D photochemical kinetics models \citep{Moses2011, Venot2012} and fully coupled 3D chemical kinetics \citep{Drummond2020}. 

For the thermochemical reactions used to compute the production and loss rates, we have used the reduced chemical network by \cite{Venot2020_network}, which is available from the Kinetic Database for Astrochemistry \citep[KIDA, ][]{Wakelam2012}. It is an updated version of the original network used by \cite{Venot2012, Agundez2012, Agundez2014}, in order to resolve the existing discrepancies between different chemical kinetics schemes concerning the quenching level of \chem{CH_4} \citep[see also][]{Moses2014, Wang2016}. Furthermore, the number of chemical species and reactions has been reduced, with the aim of a numerically efficient coupling to dynamical models. The network includes 44 chemical species, consisting of C, H, N and O, and contains 582 reactions (288 reversible and 6 irreversible reactions). It is applicable to both cold and hot, hydrogen-dominated exoplanet atmospheres and brown dwarfs, and it has been shown to reproduce the dominant neutral species of those atmospheres (\chem{H_2O}, \chem{CH_4}, \chem{CO}, \chem{CO_2}, \chem{NH_3}, \chem{HCN} and \chem{C_2H_2}) accurately \citep{Venot2020_network}. The main advantage of the reduced network in this study is the numerical speed-up with respect to a full chemical network: we have measured that the model execution is a few tens of times faster, which allows us to employ it in a wide grid of chemical models. However, as a drawback, discrepancies with the full chemical network have been observed in the upper atmospheres of hot exoplanets \citep{Venot2020_network}. Furthermore, the network contains no photolysis reactions and has not been developed with photochemistry in mind, focusing instead on the accurate reproduction of thermochemical equilibrium and the dynamical quenching points. Therefore, it is not suitable for simulating the upper atmospheric chemistry. In this study, we do not use photochemistry, instead aiming to investigate the impact of dynamical disequilibrium chemistry over a wide range of parameters. We aim to address the impact of photochemistry in detail in a follow-up study, using the full version of the chemical network.

For each chemical kinetics simulation, the physical input data, such as temperatures and wind speeds, have been extracted from the corresponding GCM simulation.
In the pseudo-2D formalism, the zonal wind speed determines the rotation rate of the vertical gas column, and hence the horizontal advection rate. Additionally, information from the horizontal and vertical wind components is used to compute the degree of vertical mixing within this column, as detailed below. 
To obtain the longitude- and pressure-dependent temperature background, we extracted an equatorial band from the GCM simulation, averaged and weighted by $\cos{\phi}$ in latitude, for latitudes $\phi$ between $+20\degrees$ and $-20\degrees$. The zonal wind advection speeds have been derived from the GCM simulations as well, namely by averaging the local zonal wind speeds zonally, meridionally between $+20\degrees$ and $-20\degrees$, and vertically for pressures $p < 10$~bar, following the approach of \cite{Agundez2014}. Averaging zonally over all longitudes is justified for climate regimes dominated by a superrotating equatorial jet stream, the validity of which is further discussed in Section~\ref{sec_results_climates}. Most derived zonal wind speeds used in the chemical modelling are of the order of $1-2$~km/s. For some of the coldest planets, the derived zonal wind speeds approach zero, or are very slightly negative, i.e. the zonal mean flow is retrograde. In these cases, we have used a minimum wind speed of 10~m/s in the chemical models, in order to avoid numerical problems. The temperature and wind data resulting from the 3D GCM simulations are presented and discussed below, in Section~\ref{sec_results_MITgcm}.

Vertical mixing is incorporated in the chemical kinetics code as a diffusion process, the efficiency of which is given by a pressure-dependent eddy diffusion coefficient $\kzz(p)$. Although the vertical mixing in irradiated exoplanets is governed by the general atmospheric circulation, and thus it is not diffusive \citep{Zhang2018_and_showman_II_tidallylocked, Komacek2019}, it is still widespread practice to parametrize vertical mixing using $\kzz$. In this work, we compute $\kzz$ using the scaling relation derived by \cite{Komacek2019}:
\begin{equation}\label{eq_kzz}
    \kzz \sim \frac{w^2}{\left( \frac{1}{\timescale{chem}} + \frac{1}{\timescale{adv}} \right)},
\end{equation} where $w$ is the vertical wind speed, $\timescale{chem}$ the chemical time-scale, and $\timescale{adv}$ the horizontal advective time-scale of the wind flow. In this equation, we computed the vertical wind speed value by globally averaging the vertical wind from the corresponding GCM simulation on isobars. While it is conceivable that the vertical mixing efficiency varies with location, e.g.~between the day and night sides of a planet, it is difficult to assign a local $\kzz$-value to it, as it has been argued that the approximation of mixing as a diffusive process is mostly valid on the global scale \citep{Parmentier2013, Showman2020}. For this reason, the wind speeds involved in the $\kzz$ computation are globally averaged on isobars. 

The horizontal advective time-scale $\timescale{adv}$ represents the efficiency of chemical advection by the wind flow on isobars, and thus the extent to which horizontal chemical perturbations can be maintained. Here it is computed as a function of pressure $p$: 
\begin{equation}
    \timescale{adv}(p) = \frac{L}{v_h(p)},
\end{equation} where $L$ is a typical horizontal length scale, taken to be the planetary radius ($R_p = 1.35$~$R_{\rm Jup}$), and $v_h$ is the horizontal wind speed, extracted from the GCM simulation and globally averaged on isobars. 

Next, the appearance of the chemical time-scale $\timescale{chem}$ in equation~\eqref{eq_kzz} implies that a different $\kzz$-profile is associated with each individual chemical species \citep{Zhang2018_and_showman_I_fastrotating, Zhang2018_and_showman_II_tidallylocked, Komacek2019}. Although it is possible to derive chemical time-scales through careful analysis of the chemical network by determining the rate-limiting step as a function of pressure and temperature \citep{Tsai2018}, this is outside the scope of this work. Hence, we adopt here a constant chemical time-scale of $\timescale{chem} = 1.6\cdot10^5$~s, which is the intermediate value used by \cite{Komacek2019}, and we make an \textit{a posteriori} estimate of the chemical time-scale based on our chemical network, as a sanity check. \citet{Tsai2018} have performed a comparison of the chemical reaction pathway by which \chem{CH_4} is transformed into \chem{CO}, for three different chemical networks \citep{Tsai2017, Moses2011, Venot2012}. At temperatures between 1000K~and~1500~K and pressures higher than 1~bar, the rate-limiting reaction in the former two networks is the methanol (\chem{CH_3OH}) production reaction
\begin{equation}\label{eq_methanol}
    \chem{OH} + \chem{CH_3} + \chem{M} \longrightarrow \chem{CH_3OH} + \chem{M},    
\end{equation} where \chem{M} is any third species. In the \cite{Venot2012}-network this reaction is not the rate-limiting step, because a different reaction in this network accounts for efficient methanol production. However, the more efficient reaction was removed in the updated chemical network used here \citep{Venot2020_network}, making it very likely that the rate-limiting reaction in our network is indeed \eqref{eq_methanol}. We estimated the chemical time-scale by evaluating the reaction rate of \eqref{eq_methanol} for different converged chemistry models in our grid. We find values between $10^4$~s at 10~bar and $10^{10}$~s at 1~bar for warm to hot planets, while the cold ($\Teff < 1000$~K) planets tend to have longer chemical time-scales. Thus, our adopted value of $\timescale{chem} = 1.6\cdot10^5$~s seems suitable, although it must be remarked that the chemical time-scale can be expected to change by orders of magnitude depending on the planetary parameters or location inside the atmosphere.

Finally, we calibrated this scaling relation to the eddy diffusion coefficient profiles that have been determined using 3D simulations with passive tracers, for the planets \hdtwenty{} ($\kzz = 5\cdot10^4 \cdot p_{\rm bar}^{-0.5}$~m$^2$/s, \cite{Parmentier2013}) and \hdeighteen{} ($\kzz = 10^3 \cdot p_{\rm bar}^{-0.65}$~m$^2$/s, \cite{Agundez2014}). To achieve this simple calibration, the $\kzz$-profile obtained from eq.~\eqref{eq_kzz} was multiplied by a scale factor for the two models in our grid closest to the planets \hdtwenty{} and \hdeighteen{}. We found that a scale factor of 0.1 results in a good order-of-magnitude agreement for both cases. Therefore, we use 
\begin{equation}\label{eq_kzz_full}
    \kzz(p) = 0.1 \frac{w^2(p)}{\left( \frac{1}{\timescale{chem}} + \frac{v_h(p)}{R_p} \right)}
\end{equation} to compute vertical mixing profiles for our chemical kinetics simulations. The $\kzz$-profiles, and how they relate to the different planetary climate models in our grid, are presented and discussed below, in Section~\ref{sec_verticalwind_kzz}.

Finally, we remark that $\kzz$ can be expected to rise again in the deep atmosphere ($p > 1$~bar), more specifically when the atmosphere becomes convective and hence well-mixed. Our $\kzz$-profiles estimated via \eqref{eq_kzz_full} are based on atmospheric circulation theory, and do not scale well to the deep convective interior, because the low resolution of most exoplanet GCMs cannot capture convective motions. The uncertainty associated with $\kzz$ in the deep atmosphere is further discussed in Section~\ref{sec_deep_quenching}.

The initial state of the chemical kinetics integration is a converged, one-dimensional chemical kinetics simulation, assuming solar metallicity and C/O ratio, and a pressure-temperature profile corresponding to the substellar point of the planet. This choice of initial state results in an efficient convergence to a periodic steady state \citep{Agundez2012}.
Each chemical model was subsequently run for 100 rotations about the equator. Depending on the different zonal wind speeds used in each model, all chemical models have total simulation times between $10^7$~s and $10^{10}$~s. This is sufficient for the evolution from a vertically quenched initial state to a zonally and vertically quenched steady-state in the pseudo-2D framework. The model convergence to a steady state is demonstrated and discussed in Appendix~\ref{sec_appendix_convergence}.

\subsection{Synthetic Spectrum}\label{sec_methods_spectrum}

As a final step, the results from the GCM (temperatures) and the chemical kinetics simulations (molecular abundances) are assimilated to compute synthetic transmission spectra, using the radiative transfer package \textit{petitRADTRANS} \citep{Molliere2019}. In the spectral synthesis, we include line opacities for \chem{H_2}, \chem{C_2H_2} \citep{Rothman2013_HITRAN}, \chem{H_2O}, \chem{CO}, \chem{CO_2}, \chem{OH} \citep{Rothman2010_HITEMP}, \chem{CH_4} \citep{Yurchenko2014_CH4}, \chem{NH_3} \citep{Yurchenko2011_NH3} and \chem{HCN} \citep{Harris2006_HCN, Barber2014_HCN}, as well as Rayleigh scattering opacities for \chem{H_2} \citep{Dalgarno1962_H2_rayleigh} and \chem{He} \citep{Chan1965_He_rayleigh}, and continuum opacities in the form of collision-induced absorption from \chem{H_2}--\chem{H_2} and \chem{H_2} -- \chem{He} interactions.
For all species, the chemical abundance profiles are first computed with the pseudo-2D chemical kinetics code.  

Regarding species not included in the chemical network, such as Na and K, we take an agnostic approach. They are not incorporated in the spectral synthesis, despite their ubiquity in transmission spectroscopy \citep[e.g.][]{Sing2016}. Although it is possible to include these species with chemical equilibrium concentrations\footnote{One way to do this would be to use interpolation on a pre-calculated chemical equilibrium grid, at the pressure and temperature under consideration. See e.g.~\url{https://petitradtrans.readthedocs.io/en/latest/content/notebooks/poor_man.html}.}, this is against the spirit of this work, namely a consistent study of dynamical disequilibrium chemistry. Hence, we omit these opacities and indicate their absence wherever it is relevant. 

Using \textit{petitRADTRANS}, a synthetic transmission spectrum is then computed with a wavelength range between 0.3~$\mu$m and 30~$\mu$m, and with a spectral resolution of $\lambda/\Delta\lambda = 1000$. To this end, a pressure-grid of 100 layers is set up, evenly spaced in log-space between 100~bar and 10$^{-4}$~bar. An atmospheric structure is interpolated on to this pressure grid, consisting of a vertical temperature profile, extracted from the GCM simulation, and molecular abundances and mean molecular weight, derived from the chemical kinetics code. Finally, the effective radius is computed with a ray-tracing method (see \cite{Molliere2019} for more details), where a pressure of 10~mbar is used as a reference pressure for the planetary radius $R_p = 1.35$~$R_{\rm Jup}$.

In the computation of a synthetic transmission spectrum we take into account contributions from both the evening and morning limb of the exoplanet disc. We calculate the averaged effective transit radius of the planet as 
\begin{equation}\label{eq_morning-evening}
    R_{\rm ave} = \sqrt{\frac{R^2_{\rm m} + R^2_{\rm e}}{2}},
\end{equation} where $R_{\rm m}$ and $R_{\rm e}$ are the effective transit radii computed with \textit{petitRADTRANS} for a one-dimensional, vertical atmospheric structure extracted from the morning or evening limb respectively. We note that this approach in constructing a transmission spectrum is rather crude, since we do not incorporate contributions from the high latitudes of the projected disc, or variations in the atmospheric structure along the line of sight. Instead, the transiting area is here approximated as two half discs (morning and evening), which are assumed to each have a spherically uniform atmospheric structure. Thus, the resulting transmission spectrum will be biased towards the physical and chemical conditions of the equatorial region (between $\pm20\degrees$ latitude). The consequences of including a more realistic 3D approach to computing transmission spectra, including line-of-sight variations in the atmospheric structure, have been examined by \cite{Caldas2019}.

\section{Grid of Models}
\label{sec_grid}

\subsection{Free Parameters}

In our grid of exoplanet atmosphere models, we have three varying free parameters: the effective temperature $\Teff$ of the planet, the surface gravity $g$ of the planet, and the stellar type of the host star. Other necessary model parameters remain fixed (the atmospheric metallicity, C/O ratio), or they are changed self-consistently (rotation rate, Section~\ref{sec_parameter_dependence}). These parameters and their adopted values are described in what follows.

\subsubsection{Effective temperature}

The effective temperature $\Teff$ of the planet is defined as the blackbody temperature of an object with the same radius that would emit a flux equivalent to the real planet. Thus, it can be expressed as combination of the equilibrium temperature $\Teq$, resulting from energy balance with the stellar radiation, and the intrinsic temperature $\Tint$, resulting from radiation originating in the planet's interior \citep{Fortney2018}: 
\begin{equation}
    \Teff^4 = \Teq^4 + \Tint^4.
\end{equation}
The effective temperature is arguably the most important free parameter, affecting the radiative, dynamical and especially chemical time-scales. Therefore, we probe a wide range of temperatures, from fairly cold (400~K) to ultra-hot (2600~K) gaseous exoplanets. We choose a uniform and quite fine temperature spacing of 200~K, to properly diagnose climate and chemical regime changes in the atmosphere. This leads to 12 sampled temperatures: $\Teff = 400$~K, 600~K, 800~K, 1000~K, 1200~K, 1400~K, 1600~K, 1800~K, 2000~K, 2200~K, 2400~K, and 2600~K.

\subsubsection{Surface gravity}

The surface gravity depends on the planetary mass and radius, which display large variations, especially in the Neptune- and Jupiter-mass regime \citep[e.g.][]{Chen2017}. Indeed, hot Jupiters are often inflated and typically have surface gravities of $\sim$10~m/s$^2$ (or $\log g = 3$ in cgs units). However, some exoplanets have a very high density and thus surface gravity, such as WASP-18b \citep{Hellier2009_wasp18b}, which despite its high temperature of 2400~K has a $\log g$ of 4.3 \citep{Southworth2010, Stassun2017}. On the other side of the spectrum, extremely low density planets, so-called super-puffs, can have surface gravities as low as $\log g \approx 2$ \citep{Lee2016, Libby-Roberts2020}.

The effect of surface gravity is mainly twofold. First, through the density it affects the radiative forcing, see eq.~\eqref{eq_taurad}. Second, by changing the atmospheric scale height, the opacity of the atmosphere changes as well, which will have a noticeable effect on the planet's spectrum. In particular, the photospheric pressure scales with gravity, and the molecular features probed in transmission spectroscopy are small if the surface gravity is high. In this work, we opt for a coarse sampling of the surface gravity, choosing three values to represent the diversity exhibited by gaseous exoplanets: $\log g = 2, 3$ and 4.

\subsubsection{Stellar type}

The effect of the host star's stellar type on the planet atmosphere is twofold. First, as noted by \cite{Molliere2015}, the radiation coming from hotter stars is absorbed at shorter wavelengths than radiation originating from cooler stars. This causes a higher temperature gradient in the former, and a more isothermal atmosphere in the latter case. Second, by changing the stellar temperature in our grid, while keeping the effective temperature of the planet fixed, we effectively vary the distance to the host star, and thus also the rotation period of the planet. This is further detailed in Section~\ref{sec_parameter_dependence}.

Similar to \cite{Molliere2015}, we choose four values for the stellar type, namely M5, K5, G5 and F5 main sequence stars. The corresponding stellar effective temperatures are 3100~K, 4250~K, 5650~K and 6500~K respectively.
For the same stellar type, the stellar properties -- temperature, luminosity, mass and radius -- have been correlated based on table~2.4 of \cite{Hubeny2014}.

\subsection{Parameter Dependence}
\label{sec_parameter_dependence}

\begin{table}
	\centering
	\caption{Overview of the model properties in the grid, ordered according to their effective temperatures $\Teff{}$ (vertically) and host star type (horizontally).}
	\label{tab:grid}
	\begin{tabular}{c r cccc}
    \hline
    \hline
    $\Teff{}$ & & M5-star & K5-star & G5-star & F5-star \\
    \hline
    \noalign{\smallskip}\multirow{2}{*}{400~K} & $a$ (AU) & 0.049 & 0.186 & 0.421 & 0.859\\
                           & $\Porb{}$ (days) & 8.61 & 35.55 & 104.33 & 256.13\\
    \noalign{\smallskip}\multirow{2}{*}{600~K} & $a$ (AU) & 0.021 & 0.080 & 0.182 & 0.372\\
                           & $\Porb{}$ (days) & 2.45 & 10.13 & 29.73 & 72.98\\
    \noalign{\smallskip}\multirow{2}{*}{800~K} & $a$ (AU) & 0.012 & 0.045 & 0.102 & 0.208\\
                           & $\Porb{}$ (days) & 1.03 & 4.25 & 12.46 & 30.59\\
    \noalign{\smallskip}\multirow{2}{*}{1000~K} & $a$ (AU) & 0.008 & 0.029 & 0.065 & 0.133\\
                           & $\Porb{}$ (days) & 0.53 & 2.17 & 6.37 & 15.64\\
    \noalign{\smallskip}\multirow{2}{*}{1200~K} & $a$ (AU) & 0.005 & 0.020 & 0.045 & 0.092\\
                           & $\Porb{}$ (days) & 0.30 & 1.26 & 3.68 & 9.04\\
    \noalign{\smallskip}\multirow{2}{*}{1400~K} & $a$ (AU) & 0.004 & 0.015 & 0.033 & 0.068\\
                           & $\Porb{}$ (days) & 0.19 & 0.79 & 2.32 & 5.69\\
    \noalign{\smallskip}\multirow{2}{*}{1600~K} & $a$ (AU) & 0.003 & 0.011 & 0.025 & 0.052\\
                           & $\Porb{}$ (days) & 0.13 & 0.53 & 1.55 & 3.81\\
    \noalign{\smallskip}\multirow{2}{*}{1800~K} & $a$ (AU) & 0.002 & 0.009 & 0.020 & 0.041\\
                           & $\Porb{}$ (days) & 0.09 & 0.37 & 1.09 & 2.68\\
    \noalign{\smallskip}\multirow{2}{*}{2000~K} & $a$ (AU) & 0.002 & 0.007 & 0.016 & 0.033\\
                           & $\Porb{}$ (days) & 0.07 & 0.27 & 0.80 & 1.95\\
    \noalign{\smallskip}\multirow{2}{*}{2200~K} & $a$ (AU) & 0.002 & 0.006 & 0.013 & 0.027\\
                           & $\Porb{}$ (days) & 0.05 & 0.20 & 0.60 & 1.47\\
    \noalign{\smallskip}\multirow{2}{*}{2400~K} & $a$ (AU) & 0.001 & 0.005 & 0.011 & 0.023\\
                           & $\Porb{}$ (days) & 0.04 & 0.16 & 0.46 & 1.13\\
    \noalign{\smallskip}\multirow{2}{*}{2600~K} & $a$ (AU) & 0.001 & 0.004 & 0.010 & 0.020\\
                           & $\Porb{}$ (days) & 0.03 & 0.12 & 0.36 & 0.89\\
    \noalign{\smallskip}\hline
    \end{tabular}
\end{table}

Aside from the three free parameters, our grid of tidally locked atmospheric models is designed to be self-consistent in the adopted input parameters. For instance, the incident stellar flux, the equilibrium temperature and the rotation period of the planet are related under the assumption of synchronous rotation: two planets with the same equilibrium temperatures orbiting stars of a different stellar type will have different semi-major axes and orbital periods, and thus also rotation periods. More specifically, the orbital distance $a$ of a planet with an effective temperature $\Teff$ is given by
\begin{equation}
    a = \sqrt{ \frac{f (1-A_B) L_\ast }{ 16 \pi \sigma (\Teff^4 - \Tint^4)}},
\end{equation} where $A_B$ is the Bond albedo of the planet, $L_\ast$ is the stellar luminosity, $\sigma$ the Stefan-Boltzmann constant and $\Tint$ the intrinsic temperature of the planet. The heat redistribution factor $f$ equals 2 when the incoming radiation is distributed over the day-side, and 1 when it is distributed over the whole surface of the planet. In the calculations for our grid, we have assumed a global heat redistribution, and a Bond albedo of 0, i.e.~all incident radiation is absorbed by the planet. Furthermore, we adopted $\Tint = 200$~K (see Appendix~\ref{sec_appendix_adiabat} for a more detailed discussion of the intrinsic temperature). 

The rotation and orbital period is then computed via Kepler's third law:
\begin{equation}
    \Prot = \Porb = \sqrt{ a^3  \frac{4 \pi^2}{G(M_\ast + M_p)}},
\end{equation} with $M_\ast$ and $M_p$ the stellar and planetary mass respectively, and $G$ the gravitational constant. 
In the calculation of the orbital/rotational periods in our grid, we have kept the planetary mass fixed to 1~$M_{\rm Jup}$, so the rotation period does not vary as a function of gravity in our grid. Nevertheless, the planetary contribution to the total mass of the star-planet system is negligible. As a consequence, under the assumption of synchronous rotation, the rotation period of the planet scales with the planetary temperature as $\Prot \sim \Teff^{-3}$.

An overview of the computed orbital distance and orbital/rotation period for each model in the grid, as a function of planet temperature and stellar type, is given in Table~\ref{tab:grid}.

\section{Simulation Results}
\label{sec_results}

\subsection{Radiative-Convective Equilibrium}

The outcome of the \textit{petitCODE} simulations are temperature-pressure profiles, corresponding to different angles of incidence, for each simulation in the grid. For each combination of surface gravity and stellar type, the night-side temperature profiles are similar, while the profiles corresponding to different incident angles on the day side increase with the simulation's effective temperature. Each temperature profile is put onto a common adiabat in the deep atmosphere, equal to that of the planetary average temperature, using the method described in Appendix~\ref{sec_appendix_adiabat}. This results in the typical structure of irradiated exoplanets consisting of an irradiated zone, in which the temperature depends strongly on the incidence angle, connecting to a common, deep adiabatic temperature gradient \citep{Showman2008, Guillot2010, Carone2020}. A sample of the radiative-convective equilibrium temperature profiles, used as input for the GCM modelling, is displayed in Appendix~\ref{sec_appendix_petitcode}.

\subsection{Dynamical Climate}\label{sec_results_MITgcm}

\subsubsection{Overview of climate regimes}\label{sec_results_climates}

\begin{figure*}
    \centering
    \includegraphics[width=0.99\textwidth]{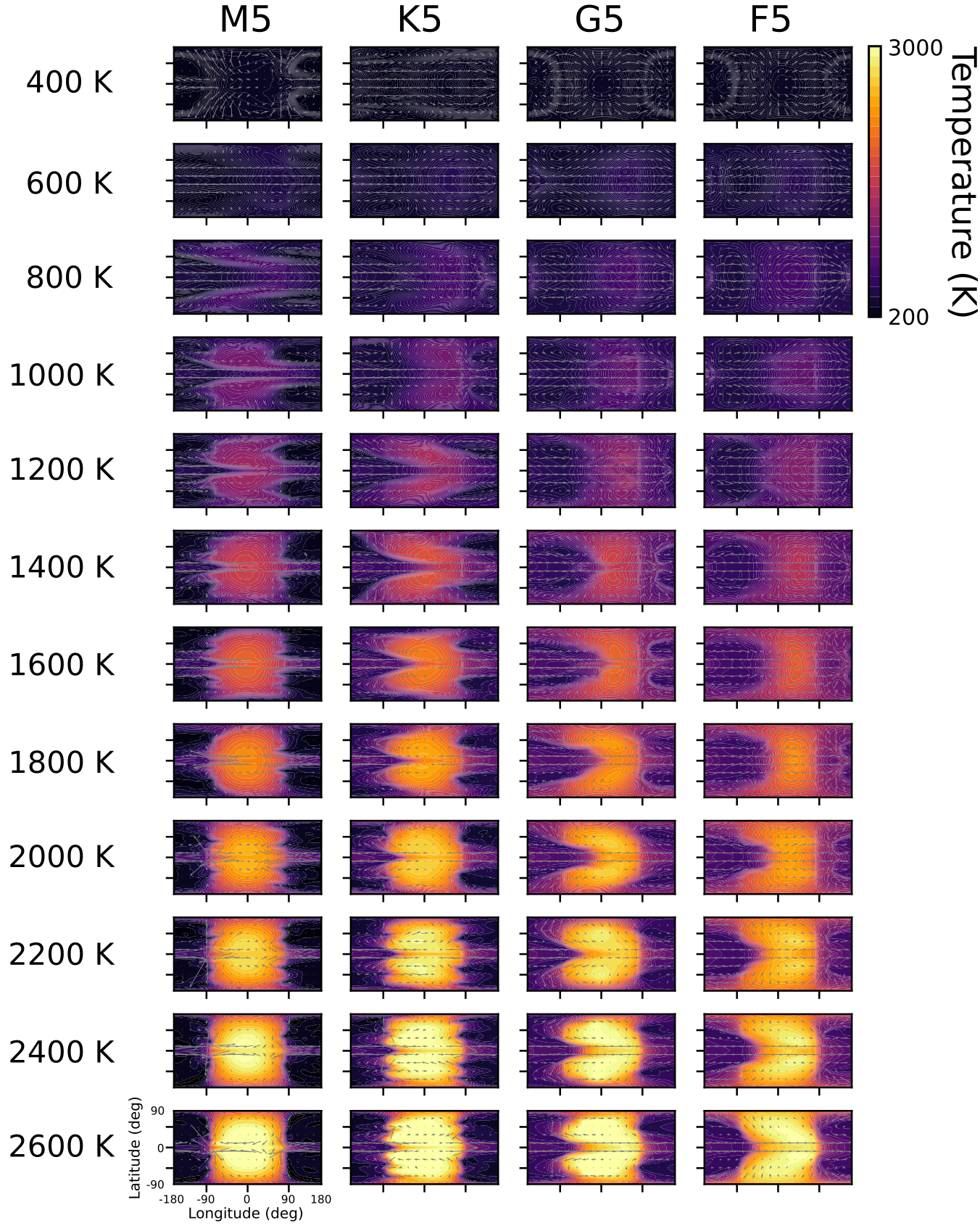}
    \caption{Temperature maps for the $g = 10$~m/s$^2$ models in our grid, plotted as isobaric slices at a pressure of 70~mbar. Grey arrows show the direction of the local horizontal winds, and the arrow length is scaled by the wind speed. The plots are ordered by host star type (horizontally) and planetary effective temperature (vertically). The substellar point is located at ($0\degrees, 0\degrees)$.}
    \label{fig_temperature_maps}
\end{figure*}

\begin{figure*}
    \centering
    \includegraphics[width=0.99\textwidth]{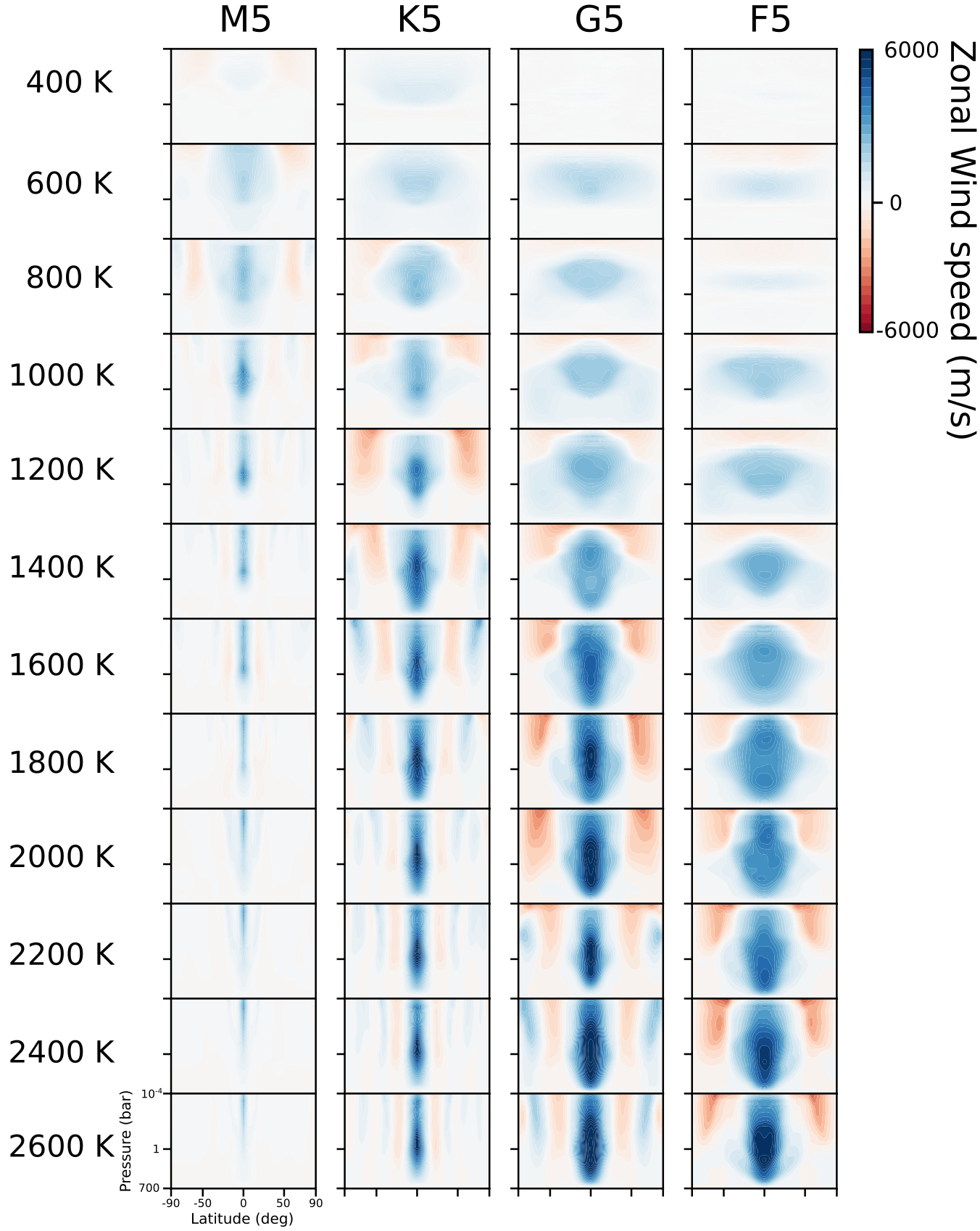}
    \caption{Zonally averaged zonal wind plots for the $g = 10$~m/s$^2$ models in our grid. Positive values of the wind speed (blue) correspond to prograde wind flow. The plots are ordered by host star type (horizontally) and planetary effective temperature (vertically).}
    \label{fig_zonalwind_plots}
\end{figure*}

The majority of the \textit{MITgcm} climate models in our grid of tidally-locked gaseous exoplanets have a climate regime dominated by equatorial superrotation, i.e.~a fast ($\sim$km/s) prograde jet stream around the equator; see Figures~\ref{fig_temperature_maps} and \ref{fig_zonalwind_plots} for $g = 10$~m/s$^2$. (The equivalent figures for $g = 1$~m/s$^2$ and $g = 100$~m/s$^2$ can be found in Appendix~\ref{sec_appendix_figures}.)\footnote{In this and following plots, 70~mbar was selected because it is a suitable pressure region to diagnose the dynamical-radiative coupling, and to facilitate comparisons with earlier work \citep[e.g.][]{Komacek2016, Komacek2017}.} This equatorial superrotation is a well-known and robust outcome of hot Jupiter GCMs \citep{Showman2011, Debras2020} and it has been established over a broad range of temperatures \citep{Komacek2016, Komacek2017}. We thus reconfirm it as the most prominent climate type in our grid.

Some notable deviations to the superrotating climate regime start to appear for very fast-rotating simulations in the grid. These simulations correspond to highly irradiated planets around cool stars, which reside in the lower left corner of Fig.~\ref{fig_temperature_maps}. Here, an equatorial retrograde flow starts to appear on the day side. A detailed discussion on this type of wind flow can be found in \cite{Carone2020}. Although the zonal mean wind still exhibits equatorial superrotation globally, as indicated by the prograde wind jets at the equator of all models (Fig.~\ref{fig_zonalwind_plots}), locally this retrograde flow is disrupting the superrotating jet stream. 
In doing so, it converges at the western terminator with the night-side prograde flow, resulting in locally high vertical wind speeds. The models in Fig.~\ref{fig_temperature_maps} that show equatorial retrograde winds at this pressure level all have rotation periods shorter than 8~hours. For the low-gravity models that critical rotation period is even shorter, for the high-gravity models, it is longer. The influence of gravity as additional parameter affecting retrograde flow has been established by \cite{Carone2020} as well. 

Additionally, the equatorial superrotation is not always present in the coldest planets of our grid. More specifically, out of all $\Teff=400$~K models in Fig.~\ref{fig_temperature_maps}, only the K5-orbiting planet shows signs of equatorial superrotation. The other models exhibit a more direct day-to-night-side flow (G5 and F5) or an equatorial retrograde flow on the day side together with a prograde flow on the night side (M5). The cold models corresponding to low (Fig.~\ref{fig_temperature_maps_appendix_g1}) and high surface gravity (Fig.~\ref{fig_temperature_maps_appendix_g100}) show analogous deviations from equatorial superrotation. This may indicate that the radiative forcing, caused by the permanent one-sided heating, for these cold models is insufficient to trigger the standing wave response that is necessary to excite and sustain the equatorial jet stream. The excitation of superrotation in this thermal regime seems to be less robust, and can be expected to depend on many other parameters such as the rotation rate, gravity, and numerical assumptions, like in the cases of Jupiter and Saturn \citep{Showman2020}. It can be questioned, finally, if these very cool, synchronously rotating models provide a proper representation of cool exoplanets, as their synchronization time-scales approach 1~Gyr and higher (see Appendix~\ref{sec_appendix_synchron}).

Another prominent trend is the qualitative change in the zonal wind flow with respect to changing effective temperature and rotation rate. Fig.~\ref{fig_zonalwind_plots} indicates that the strength of the equatorial jet stream is correlated with the effective temperature of the planet, as the zonal wind speed increases from almost zero for the coldest planets to about 6~km/s for the hottest. Interestingly, the fast-rotating planets (orbiting M5-stars) do not follow this pattern. Instead, the jet stream speed stays constant with effective temperatures increasing above 1200~K.
An additional effect of the rotation rate is the variation in the width of the equatorial jet stream for groups of models with the same effective temperature. Indeed, as the rotation period becomes shorter, the latitudinal extent of the jet decreases as well, as was previously found by \cite{Showman2011, Showman2015}, who showed that the jet width is proportional to the Rossby deformation radius. The latter quantity depends inversely on the $\beta$-parameter, and thus on the rotation rate. Equivalently, the jet width is reduced due to the exponential decay of the latitudinal extent of the standing equatorial Kelvin waves, needed in the superrotation excitation. The \textit{e}-folding decay length varies similarly with the $\beta$-parameter \citep{Holton2013_ch11}.
Finally, we find that the gravity does not have a big impact on the zonal mean wind flow, as indicated by the strong similarities between Figures~\ref{fig_zonalwind_plots}, \ref{fig_zonalwind_plots_appendix_g1} and \ref{fig_zonalwind_plots_appendix_g100}.

Finally, we can discuss the \textit{a priori} validity of adopting a single, zonally averaged wind speed to represent the zonal advection in the pseudo-2D chemistry code, considering the climate regimes displayed in Fig.~\ref{fig_temperature_maps} and~\ref{fig_zonalwind_plots}. As discussed above, the equatorial zonal wind speed is not always uniform. Indeed, longitudinal variations in the horizontal wind speed are noticeable. Most noteworthy are the cases with high rotation rates, where fast retrograde flow can be seen on the day-side. As was hypothesized by \cite{Carone2020}, this situation could result in a morning terminator composition that is affected by the day-side state, rather than the night-side state. Moreover, the fast vertical motions associated with the convergence of horizontal wind flow could cause locally high vertical mixing efficiencies, or a dynamical pile-up of clouds. None the less, in the zonal wind plots it can be seen that -- on average -- the superrotating jet stream dominates the atmospheric circulation, even in the cases with fast retrograde day-side flow. Hence, a uniform zonal wind speed should still be a good, first-order assumption for all cases, except the coolest ones ($\Teff = 400$~K).

\subsubsection{Heat redistribution}\label{sec_heat_redistribution}

\begin{figure*}
    \centering
    \includegraphics[width=0.76\textwidth]{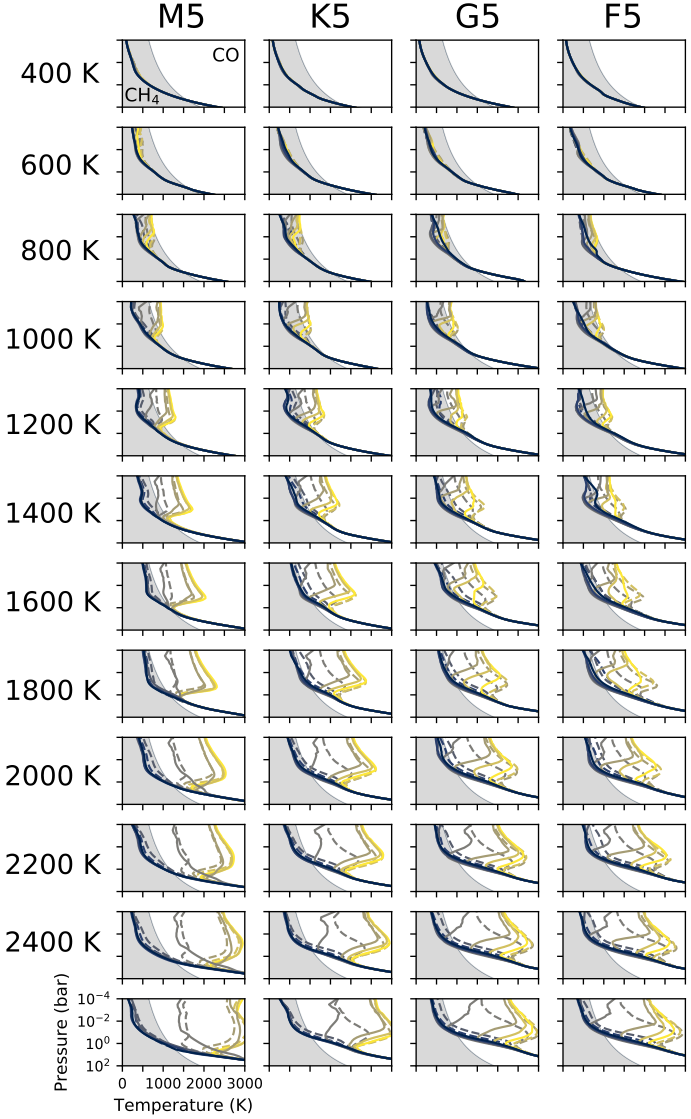}
    \caption{Temperature-pressure profiles from the $g = 10$~m/s$^2$ GCM models in our grid. Temperatures are meridionally averaged between $\pm20\degrees$, shown for longitudes spaced $30\degrees$ apart ($-180\degrees$, $-150\degrees$,\ldots , $150\degrees$), and colour coded with a gradual change from the anti-stellar (blue) to substellar point (yellow). Full and dashed lines mark locations west and east of the substellar point. The shaded background colour indicates the parameter space where methane is more abundant than CO in chemical equilibrium \citep{Visscher2012}. The plots are ordered by host star type (horizontally) and planetary effective temperature (vertically).}
    \label{fig_LPT_profiles}
\end{figure*}

\begin{figure}
    \centering
    \includegraphics[width=0.99\columnwidth]{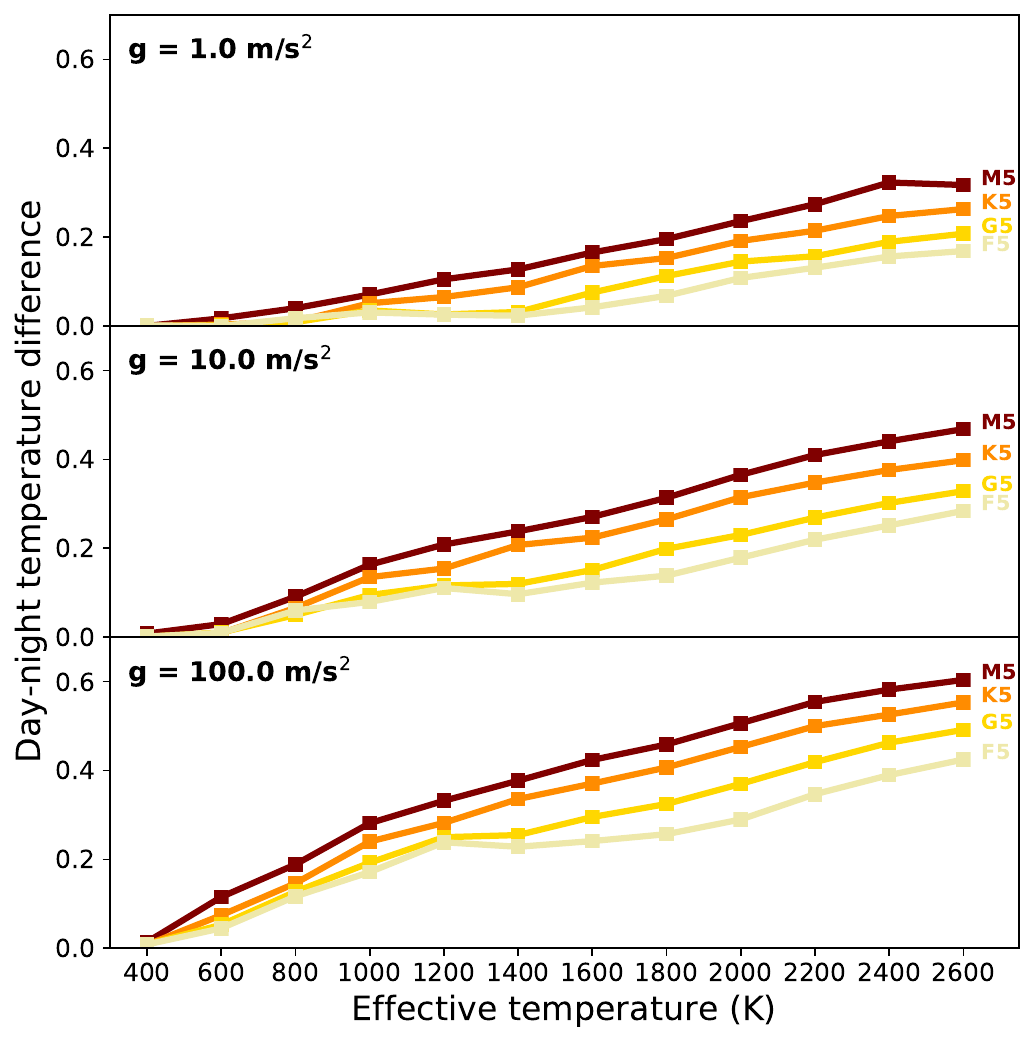}
    \caption{The relative mean temperature difference between the day- and night side is systematically larger for fast-rotating models (orbiting M5-stars) than slow-rotating models (orbiting F5-stars). Furthermore, it increases with planetary effective temperature. The relative mean temperature difference was computed by subtracting the average night-side temperature from the average day-side temperature, and dividing by the average day-side temperature. To compute the average day- or night-side temperature, we have taken the arithmetic mean of the temperature $T(\phi, \lambda, p)$ in our GCM model for all pressures $p$, latitudes (weighted with $\cos{\phi}$) and longitudes $\lambda$ of the day or night side respectively. The three panels show the low-, medium- and high-gravity models.}
    \label{fig_heat_redistribution}
\end{figure}

The temperature plots of the GCM models, as isobaric map (Figures~\ref{fig_temperature_maps}, \ref{fig_temperature_maps_appendix_g1} and \ref{fig_temperature_maps_appendix_g100}) or as a function of pressure (Figures~\ref{fig_LPT_profiles}, \ref{fig_LPT_profiles_appendix_g1} and \ref{fig_LPT_profiles_appendix_g100}), already indicate trends in the difference between the day- and night-side temperatures. To provide a good overview of the heat redistribution in all 3D GCM models, the temperatures of all models have been averaged, and the relative day-night temperature differences are plotted with respect to the planet's effective temperature for different surface gravities and host star types (Fig.~\ref{fig_heat_redistribution}). In all cases, a clear increasing trend with effective temperature is noticeable, which is in line with previous climate models \citep{Perna2012, Perez-Becker2013, Komacek2016} and observations \citep{Cowan2011, Komacek2017}. The reason often provided for this trend is that the radiative time-scale and heat advection time-scale both decrease with increasing irradiation, but the radiative time-scale more so. Thus, in these idealized GCM simulations, hotter planets can efficiently radiate away heat advected by the wind flow (see also \cite{Showman2020} for a more comprehensive discussion).

Another parameter influencing the atmospheric heat redistribution in irradiated exoplanets, is the rotation rate. Indeed, planets of the same effective temperature orbiting different stars, and hence having different rotation rates, show clear differences in their day-night temperature contrast (Figures~\ref{fig_temperature_maps} and \ref{fig_heat_redistribution}). More specifically, higher rotation rates inhibit the heat redistribution, resulting in larger day-night contrasts. This can be understood in terms of the force balance in the hydrodynamic momentum equation between the Coriolis force and the pressure gradient caused by the temperature difference between the day side and night side. In the case of fast rotation, the Coriolis force can support a bigger pressure gradient, resulting in larger day-night temperature contrasts \citep{Komacek2017}. This is clearly demonstrated in Fig.~\ref{fig_heat_redistribution}, where planets orbiting cooler host stars, with faster planetary rotation rates, exhibit a consistently larger day-night temperature contrast.

\subsubsection{Vertical wind speed and $\kzz$}\label{sec_verticalwind_kzz}

\begin{figure*}
    \centering
    \includegraphics[width=0.99\textwidth]{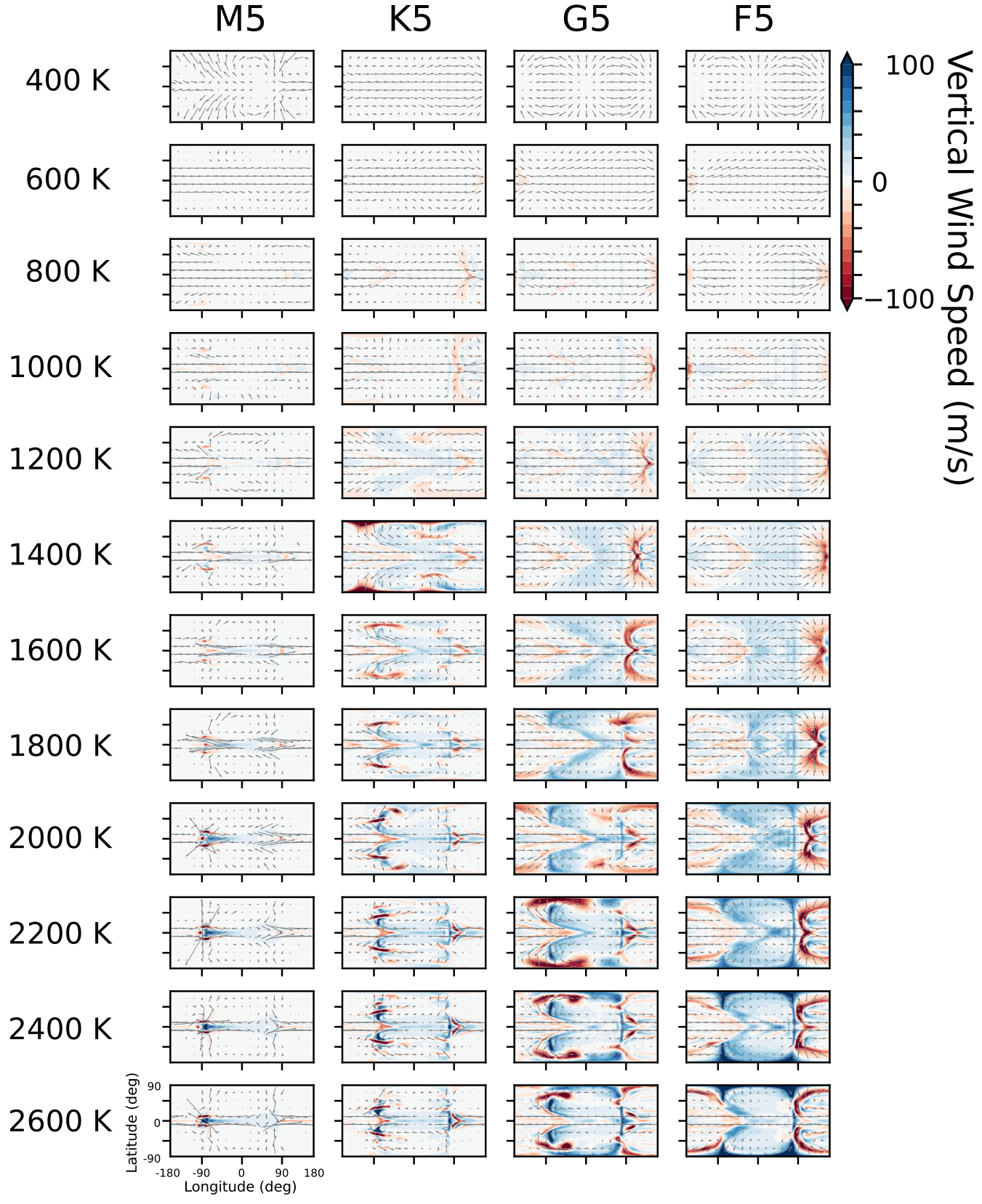}
    \caption{Maps of the vertical wind speed on a 70~mbar isobar for the $g = 10$~m/s$^2$ models in our grid. Upwelling motions (blue) are mostly present on the day side (longitudes between $-90\degrees$ and $+90\degrees$), whereas downwelling flow (red) is mostly visible on the night side, often as very localized fronts. The plots are ordered by host star type (horizontally) and planetary effective temperature (vertically). Thus, the rotation rate decreases along rows.}
    \label{fig_verticalwind_plots}
\end{figure*}

The vertical wind speeds in exoplanet atmospheres are a valuable diagnostic to constrain the amount of vertical mixing. As a consequence, understanding the effects of different physical parameters on the vertical wind speeds is key to predict disequilibrium chemistry. Analogous to the temperature maps of Fig.~\ref{fig_temperature_maps}, horizontal, isobaric maps of the vertical wind speeds are shown in Fig.~\ref{fig_verticalwind_plots} for the medium-gravity subsection of the grid. It can be seen that an increase in the effective temperature of the planet results in faster vertical wind speeds. This is to be expected, as the unilateral irradiation is the driver for the atmospheric circulation. The vertical motions at 70~mbar are characterized in general by upwelling flow at the day side, and downward flow occurring in more localized fronts. 

Among the most prominent fronts are those forming a `chevron' pattern at the equator, between $90\degrees$ and $180\degrees$ on the night side \citep[see also ][]{Rauscher2010, Zhang2018_and_showman_II_tidallylocked}. They are characterized by convergences in the horizontal wind field, where the wind from high latitudes flows equatorward and meets with the eastward advected air from the day side, resulting in very fast vertical flow in a `chimney'-like structure \citep{Parmentier2013, Komacek2019}. Both the location and speed of this vertical chimney seem to be influenced by the Coriolis force, as evidenced by their dependence on the planet's rotation rate (e.g. Fig.~\ref{fig_verticalwind_plots}, $\Teff=1200$~K, along row). Slowly rotating planets display a comparatively fast vertical wind chimney, which is located near the anti-stellar point. On the other hand, planetary atmospheres that rotate faster see these columns shifting westward to the evening terminator. The vertical wind speeds generally also decrease with faster rotation. For the fastest rotators, with $\Prot < 1~$day typically, the large-scale flow breaks up into small-scale substructures and the prominence of the chevron front gradually disappears. 

Beyond the presence of localized fronts, the vertical wind speeds in general show a clear dependence on the rotation rate, with fast rotators displaying slow vertical wind speeds and vice versa (Fig.~\ref{fig_verticalwind_plots} along rows). Indeed, the upward wind on the planet's day side reaches roughly 50~m/s in general (not including localized fronts) when the planet is rotating comparatively slow (G5 and F5). For faster rotators (K5), this is already reduced to 10~m/s, with the exception of the dynamically active equatorial band. Finally, in the case of the ultra-fast rotators (M5), vertical wind speeds become negligible in the regions outside the narrow equatorial band. Hence, the rotation rate is critical for a good estimation of the vertical motion in irradiated, tidally-locked exoplanets.

Upon comparing the vertical wind maps of models with different surface gravities, all sub-grids show similar trends with effective temperature and rotation rate/host star type. However, in absolute terms the values are very different (see Fig.~\ref{fig_verticalwind_plots} for medium, and Figures~\ref{fig_verticalwind_plots_appendix_g1} and \ref{fig_2dchem_appendix_g100} for low and high gravities). It is clear that a high surface gravity serves to reduce the vertical wind speeds overall, as the ($g=100$~m/s$^2$)-simulations generally do not have vertical wind speeds above 10~m/s. Regarding the low gravity ($g=1$~m/s$^2$)-simulations, on the other hand, the hot and slowly rotating planets display wind speeds that surpass $\pm100$~m/s on large parts of the isobaric surface plotted in Fig.~\ref{fig_verticalwind_plots_appendix_g1}. The inverse dependence between vertical wind speed and the surface gravity is evident, since the vertical wind speed scales with the atmospheric scale height $H=\frac{RT}{g}$ \citep{Komacek2019}. Hence, high-gravity planets can be expected to have lower eddy diffusion coefficients than their low-gravity counterparts. The vertical advection time-scale can then be estimated through $\timescale{vertical} \sim H^2/\kzz$. Given that $w\sim H$ and $\kzz \sim w^2$, first-order analytic theory would suggest that the scale height and vertical wind terms cancel out, and the vertical transport efficiency is not dependent on the gravity. However, in our GCM simulations, which span two orders of magnitude in $g$, we find that the models with higher gravity generally have lower vertical transport time-scales than models with low gravity (see also Section~\ref{sec_timescale_comparison}).

\begin{figure*}
    \centering
    \includegraphics[width=0.99\textwidth]{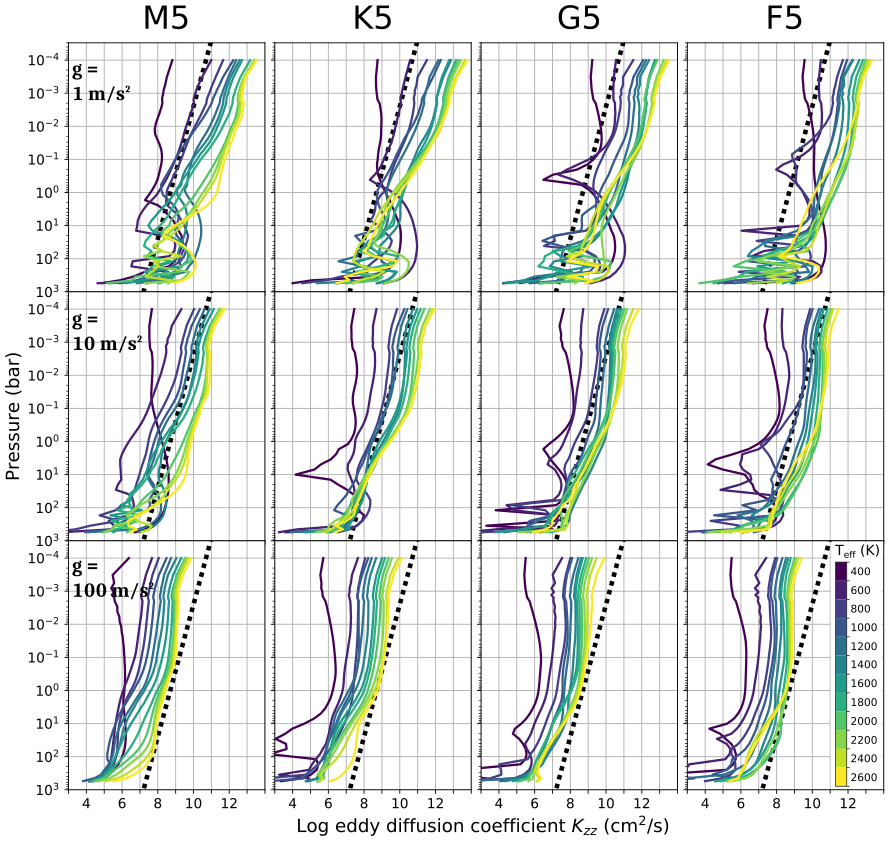}
    \caption{The eddy diffusion coefficient profiles, corresponding to all planetary climate simulations, show a clear increasing trend with effective temperature, and a decreasing trend with surface gravity. For each stellar type (\textit{columns}), models of different gravity are displayed (\textit{rows}). Dark blue and bright yellow correspond to the coolest and hottest planets in the grid, 400~K and 2600~K, respectively. The black dotted line is a reference profile, corresponding to the parametrization given by \protect\cite{Parmentier2013} for \hdtwenty{}.}
    \label{fig_kzz_profiles}
\end{figure*}

The eddy diffusion coefficients, computed using eq.~\eqref{eq_kzz_full} and based on the GCM wind maps, show the same trends as the vertical wind speeds, namely correlated with effective temperature and inversely dependent on the surface gravity (Fig.~\ref{fig_kzz_profiles}). Indeed, the value of the eddy diffusion coefficient changes by up to four orders of magnitude, depending on the effective temperature of the planet. This is roughly in both qualitative and quantitative agreement with the $\kzz$ computations by \cite{Komacek2019} (see their Fig.~7, for $\timescale{chem} = 1.6\cdot10^5$~s and an infinite drag time-scale). Furthermore, our $\kzz$-profiles show a similar pressure dependence as the power-law parametrization made by \cite{Parmentier2013} (dashed line in Fig.~\ref{fig_kzz_profiles}). Although the method employed in this work to derive $\kzz$ via \eqref{eq_kzz_full} is not as accurate as passive tracers in describing the vertical mixing through atmospheric circulation, our agreement with studies that do involve passive tracers \citep{Parmentier2013, Komacek2019} allows us to conclude that this description is sufficiently accurate for the purposes of this study.

While there are clear trends between the order of magnitude $\kzz$ value, and atmospheric parameters such as effective temperature and gravity, some of the derived $\kzz$-profiles have anomalous shapes, especially if the effective temperature is low ($400-600$~K). In these cases, $\kzz$ shows relatively little variation with altitude, rather than the steep monotonic increase that characterizes the hotter models. Moreover, in many low-temperature $\kzz$ profiles, a localized sharp drop can be seen (e.g.~K5, $g=10$~m/s$^2$, Fig.~\ref{fig_kzz_profiles}). Since the horizontal advection time-scale shows only small variations with pressure, the general morphology of the $\kzz$-profiles is attributed to the mean vertical wind speed. Thus, this sharp drop in $\kzz$ is associated with a sign-change in the vertical wind. It appears that the irradiation in the cold models is insufficient to drive a vertically coherent atmospheric circulation. The transition in climate regime is further exemplified in the wind maps (Fig.~\ref{fig_temperature_maps} and \ref{fig_zonalwind_plots}), and the absence of equatorial superrotation as we discussed in Section~\ref{sec_results_climates}. It has indeed been demonstrated that a low stellar irradiation results in climate regimes that strongly differ from typical hot Jupiters, with possible transitions between sub-regimes depending on rotation rate \citep{Showman2015}. We conclude that care should be taken when $\kzz$ is applied to cool ($\Teff < 600$~K) exoplanets, as extrapolations from hotter exoplanet atmospheres break down. As the focus in exoplanet science shifts from hot Jupiters to cooler planets, further investigations in the climate regimes of mildly irradiated gaseous planets would be beneficial.

Finally, we note that in the deepest parts of the atmosphere, any clear trends in the $\kzz$-profiles break down. Regardless, the vertical mixing strength in these layers is uncertain, as $\kzz$ is expected to increase again when the convective regime is reached (see Section~\ref{sec_deep_quenching}).

\subsection{Chemical Composition}\label{sec_results_chemistry}

\subsubsection{Overview}
\begin{figure*}
    \centering
    \includegraphics[width=0.97\textwidth]{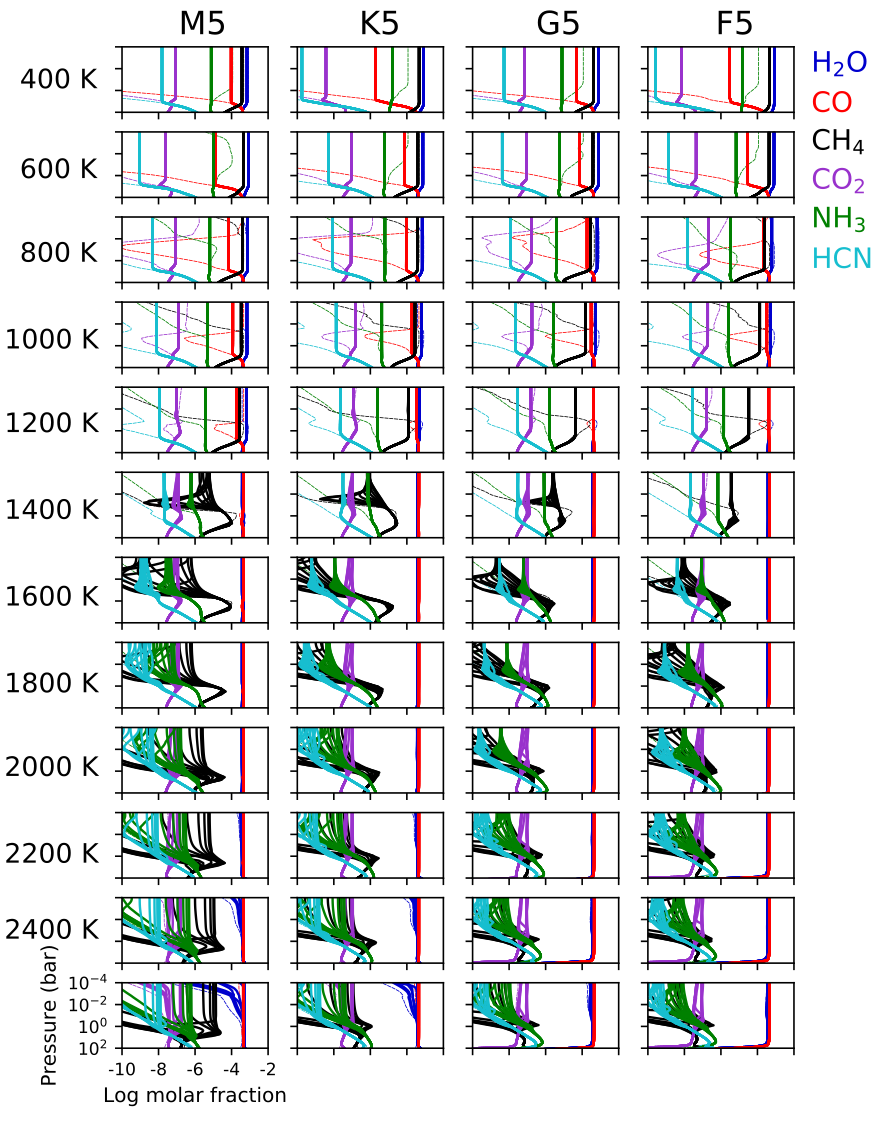}
    \caption{The abundances for the main molecular species, computed with the pseudo-2D chemistry code for $g = 10$~m/s$^2$, show a dichotomy of the zonal mixing regime.  On the one hand, hotter planets (bottom rows) sustain large chemical gradients between the hot day- and the cold night sides, as evidenced by the 12 vertical profiles that are plotted for different longitudes ($-180\degrees$, $-150\degrees$,\ldots , $150\degrees$). Colder planets (top rows), on the other hand, exhibit zonally quenched chemistry profiles, without any day-night variation. In these cases, all longitudinally sampled vertical profiles are plotted on top of each other. The dashed lines denote the chemical equilibrium composition at the substellar point.}
    \label{fig_2dchem}
\end{figure*}

The chemical abundances for some prominent molecules in exoplanetary atmospheres, computed with the pseudo-2D chemical kinetics code, are displayed in Fig.~\ref{fig_2dchem} for the simulations with medium surface gravity. The figure gives an overview of the dominant atmospheric constituents, besides \chem{H_2} and \chem{He}. In all cases, it can be seen that \chem{H_2O} is one of the most abundant molecules. This is complemented by \chem{CO} in the hot planets ($\Teff \geq 1400$~K), and by \chem{CH_4} in the cool planets ($\Teff \leq 1000$~K). In between these temperatures, coexistence of \chem{H_2O}, \chem{CO} and \chem{CH_4} is possible in equal abundance, but the rotation rate (captured in the host star type) can be seen to play a role in the balance between these three molecules. The transition between \chem{CO} and \chem{CH_4} as the main carbon-bearing species in exoplanet atmospheres is well understood in chemical equilibrium \citep{Lodders2002}. Additionally, it is crucial to map this transition in disequilibrium chemistry, since it serves as the basis for a number of simplifying assumptions in exoplanet chemistry studies \citep[e.g.][]{Cooper2006, Mendonca2018_disequilibrium, Drummond2018_HD209458b, Drummond2018_HD189733b, Steinrueck2019}.

Upon viewing the grid of pseudo-2D chemical kinetics results in Fig.~\ref{fig_2dchem}, a striking dichotomy can be seen. Cool planets with $\Teff < 1400$~K do not show longitudinal variations in their chemical composition. Hot planets with $\Teff > 1400$~K do show longitudinal variations. This is especially the case for temperature-sensitive molecules, such as \chem{CH_4}, \chem{NH_3} and \chem{HCN}, but even \chem{H_2O} shows longitudinal variations for those models where the temperature differences become very high, namely the ultra-hot, fast rotating planets. Thus, zonal quenching by the horizontal wind appears to be sufficiently strong to smear all chemical variations on isobars in the low-temperature regime. However, as the effective temperature increases, both the zonal wind speed (Fig.~\ref{fig_zonalwind_plots}) and day-night temperature differences (Fig.~\ref{fig_heat_redistribution}) increase. The former would keep the chemical composition horizontally homogeneous, whereas the latter would cause chemical variations, as the warm day side tends more towards a chemical equilibrium composition. Above the threshold of $\Teff \approx 1400$~K, it appears that the zonal temperature variation becomes the main cause for zonal chemistry variations. 
Finally, note that we did not incorporate photochemistry, which could be an additional catalyst for longitudinal chemistry variations, especially in the upper atmosphere \citep{Agundez2014, Venot2020_wasp43b}.

Vertical mixing ensures an almost constant chemical composition in the majority of the mid to high altitudes, especially in the atmospheric models of cool planets. We observe that, with our description of $\kzz$, the typical quenching pressure lies between 1~bar and 10~bar for the planets with $\Teff < 1600$~K. Here, the coldest planets have the deepest quenching points. Thus, although the eddy diffusion coefficient generally increases with effective temperature (Fig.~\ref{fig_kzz_profiles}), models with higher effective temperatures still have shallower quenching points, because of the faster reaction rates associated with higher temperatures. Furthermore, planets hotter than 1600~K show different turnoff points from their chemical equilibrium composition depending on the longitude probed. Species on the night side are typically still quenched at the 1~bar pressure level, but the quenching level increases as hotter longitudes are viewed. Consequently, for the hottest planetary atmospheres, species at the day-side longitudes remain nearly unquenched and in equilibrium. 

In the hottest atmospheres paired with G- and F-stars, the abundances of \chem{H_2O} and \chem{CO} decrease sharply in the deep atmosphere, around 100~bar. The temperatures at this depth reach about 5000~K, which is too hot for most molecules to exist. Hence, \chem{H_2O}, \chem{CO}, but also \chem{H_2} revert to their atomic components. A similar phenomenon is seen in the chemical abundances of hot, low-gravity planets (Fig.~\ref{fig_2dchem_appendix_g1}). We note that most NASA coefficients, used to calculate the chemical equilibrium, are only verified up to 5000~K \citep{Mcbride1993_NASAcoefficients}. Moreover, the temperature in the deepest parts of the atmosphere is rather uncertain (see Section~\ref{sec_deep_quenching}), and transitions in the equation of state can be expected at these conditions \citep{Saumon1995}.

The influence of the rotation/host star type on the chemical composition models manifests itself in three ways. First, a higher rotation rate is related to a slower and narrower equatorial jet (see Section~\ref{sec_results_climates}), and thus a lower degree of horizontal quenching in the chemistry models. Second, a higher rotation rate is associated with an inefficient heat redistribution (see Section~\ref{sec_heat_redistribution}), resulting in chemical equilibrium compositions that are highly longitudinally dependent. Third, a higher rotation rate correlates to slow vertical wind speeds and consequently lower eddy diffusion coefficients (see Section~\ref{sec_verticalwind_kzz}), leading to a reduced vertical mixing of the chemical species and a lower quenching pressure. 
In summary, the first and seconds effects complement each other, so that the fast rotating planets show more pronounced horizontal chemical variations, which is confirmed in Fig.~\ref{fig_2dchem} (see for instance along rows of $\Teff \geq 1400$~K). 
The third effect, of rotation leading to reduced quenching levels, can be discerned as well (Fig.~\ref{fig_2dchem}), in particular by comparing the departure of chemical species from their equilibrium track in the models with $\Teff=800$~K. Here, the quenching level strongly affects the \chem{CO}-content: a fast rotating planet (host star M5) has a low \chem{CO} content due to the relatively shallow quenching level, but a slowly rotating planet (host star F5) can be seen to have a deeper quench level, leading to a coexistence of \chem{CO}, \chem{CH_4} and \chem{H_2O}. Furthermore, the temperatures of the G5 and F5 models appear to be slightly higher near the quenching pressure, resulting in more CO in equilibrium (Fig.~\ref{fig_LPT_profiles}). By analogy, for planetary temperatures with $\Teff=1200$~K, \chem{CO}, \chem{CH_4} and \chem{H_2O} can coexist if they are fast rotating, but \chem{CO} and \chem{H_2O} dominate if they are slowly rotating.
We note that the set of 400~K-models also shows varying levels of CO-quenching for different host star types. In this case, however, the behaviour is not easily generalized into a trend with rotation rate, due to the more erratic $\kzz$-profiles of these models, highlighted earlier (Section~\ref{sec_verticalwind_kzz}).
For hotter planets, in general, the effect of rotation on the quenching level is limited.

Upon comparing the chemical models in Fig.~\ref{fig_2dchem} with their low (Fig.~\ref{fig_2dchem_appendix_g1}) and high (Fig.~\ref{fig_2dchem_appendix_g100}) surface gravity counterparts in the grid, agreement with the above trends is found. The dichotomy between horizontally homogeneous models ($\Teff < 1400$~K) and models showing horizontal chemical gradients ($\Teff > 1400$~K) is still present. However, the distinction is less clear for the low-gravity case, where a longitudinal dependence is never very pronounced. The high-gravity models, on the other hand, display very strong longitudinal chemistry gradients in the molar fractions of \chem{CH_4}, \chem{NH_3} and \chem{HCN}. The increasing chemical inhomogeneity for hot model atmospheres with increasing gravity, is most likely linked to a combination of two effects. First, the heat redistribution is relatively inefficient in high-gravity atmospheres (see Fig.~\ref{fig_heat_redistribution}), resulting in strong horizontal temperature gradients. Second, in this part of the parameter space, vertical mixing is more efficient than zonal wind advection. This point is discussed in more detail in Section~\ref{sec_timescale_comparison}. Thus, the compositional differences set by a longitudinally dependent chemical equilibrium, are maintained throughout the vertical extent of the atmosphere. Finally, the dichotomy threshold temperature of $\sim$1400~K appears to be independent of the surface gravity.   

\subsubsection{Time-scale comparison}\label{sec_timescale_comparison}

In order to disentangle and compare the advection effects of vertical mixing and zonal mixing, the typical time-scales associated with both processes can be computed. The vertical mixing time-scale is computed as $\tau_{\rm vertical} = H^2$/$\kzz$, where $H$ is the atmospheric scale height and $\kzz$ the eddy diffusion coefficient computed with \eqref{eq_kzz_full}. The horizontal mixing time-scale is computed as $\tau_{\rm zonal} = 2 \pi R_p$/$u$, where the planet circumference $2 \pi R_p$ is taken to be a typical length scale for the zonal wind advection, and $u$ is the zonal wind speed. Even though these time-scales are only zeroth order estimates of the mixing efficiencies, they can still provide insight in the relative importance of both mechanisms \citep[see e.g.~][]{Drummond2020}. 

\begin{figure}
    \centering
    \includegraphics[width=0.99\columnwidth]{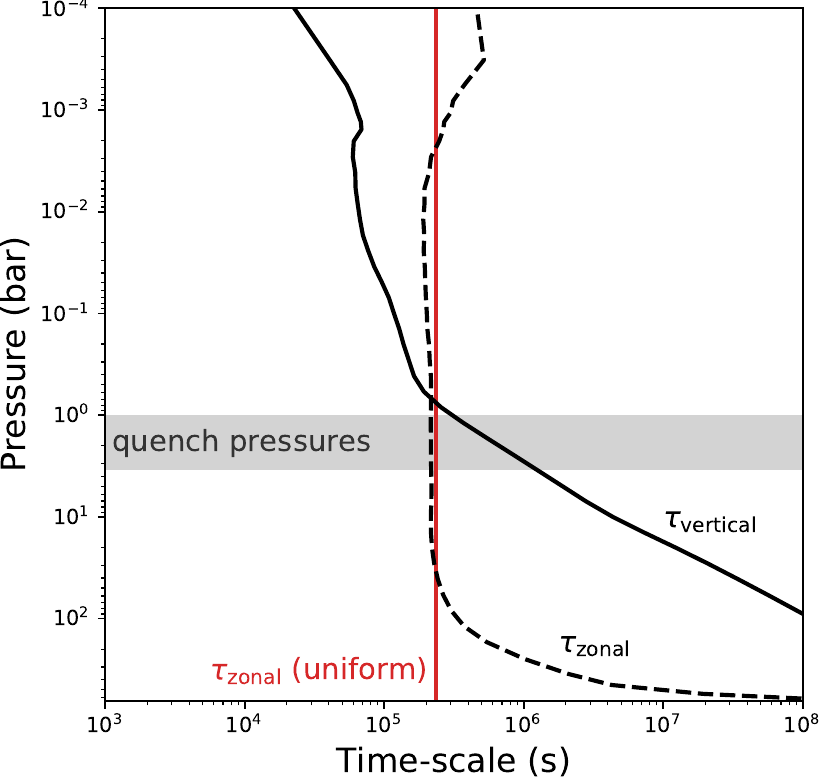}
    \caption{The vertical advection time-scale $\tau_{\rm vertical}$ is shorter than the zonal advection time-scale $\tau_{\rm zonal}$ for the upper part of this hot Jupiter model atmosphere (with $\Teff=1400$~K, $g=10$~m/s$^2$ and G5-star host). In the deeper atmosphere, the zonal mixing time-scale is much shorter. The pressure range in which the chemical abundances become quenched for this model, has been marked in grey. The time-scale associated with the uniform zonal wind used in the pseudo-2D computation, is marked with a solid red line.}
    \label{fig_timescales_hd20}
\end{figure}

In the relatively deep part of the atmosphere, up until $\sim$1~bar, the dominant advection type is zonal mixing ($\timescale{zonal} < \timescale{vertical}$, see Fig.~\ref{fig_timescales_hd20}). While the zonal advection time-scale stays almost constant with pressure, the time-scale associated with vertical mixing decreases strongly with altitude (see also Fig.~\ref{fig_kzz_profiles}), so that it becomes shorter than the zonal advection time-scale at lower pressures. The quenching pressures, at which chemical reactions no longer reach an equilibrium, are located in the zonally dominated regime. Consequently, chemical species are first horizontally quenched, and later vertically. This is in agreement with previous studies researching the quenching behaviour of \hdtwenty{} \citep{Agundez2014, Drummond2018_HD209458b, Drummond2020}.

\begin{figure}
    \centering
    \includegraphics[width=0.99\columnwidth]{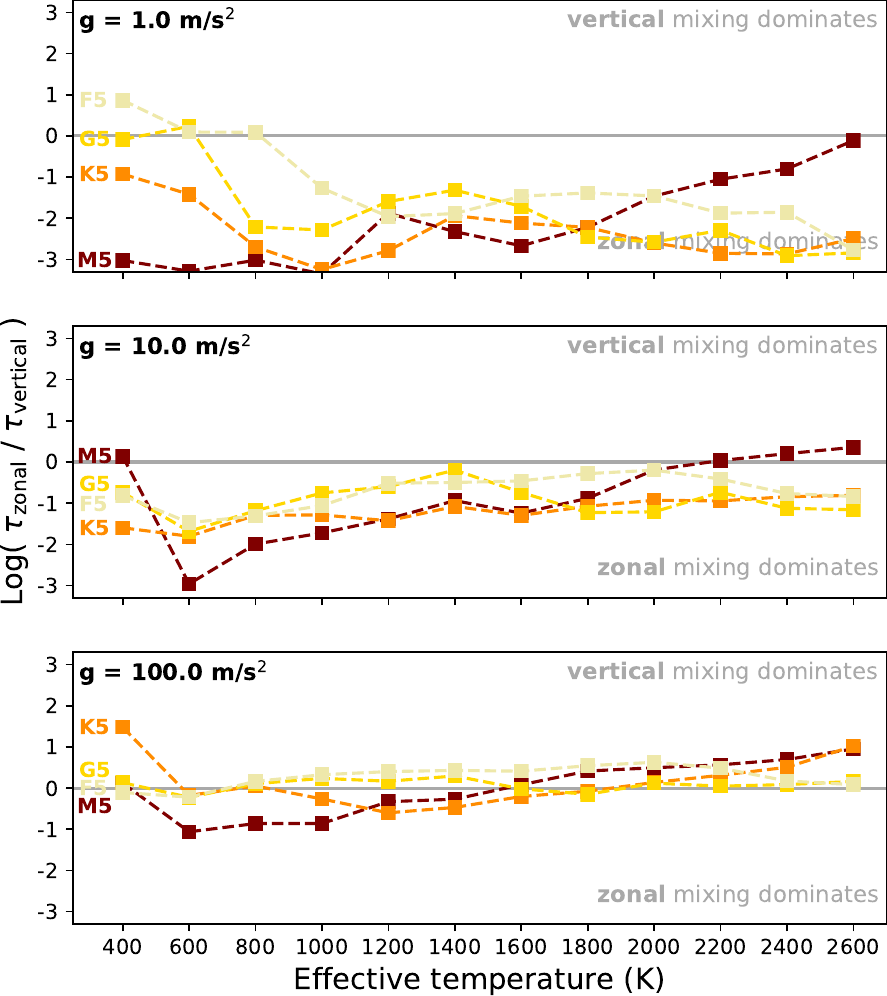}
    \caption{If the log ratio of the zonal advection time-scale $\tau_{\rm zonal}$ and the vertical mixing time-scale $\tau_{\rm vertical}$ is positive, the atmospheric mixing is more efficient in the vertical direction ($\tau_{\rm vertical}$ < $\tau_{\rm zonal}$). If it is negative, the atmospheric mixing is more efficient in the zonal direction ($\tau_{\rm zonal}$ < $\tau_{\rm vertical}$). In the ratio computation, the vertical mixing efficiency is sampled at 1~bar, and the zonal mixing efficiency is computed using the uniform zonal wind of the pseudo-2D framework. The three panels show the low-, medium- and high-gravity models.}
    \label{fig_timescales_all}
\end{figure}

Generalizing the time-scale analysis of Fig.~\ref{fig_timescales_hd20}, we have computed the ratio of the zonal and vertical time-scales at 1~bar, for each model in the grid, and plotted the result as a function of the effective temperature, for different model parameter combinations (Fig.~\ref{fig_timescales_all}). It becomes clear that, at 1~bar, the zonal advection time-scale is in most cases shorter than the vertical mixing time-scale. This is a consequence of the fast equatorial jet stream that is excited in almost all models. Thus, if the quenching pressures are located around 1~bar, as is the case for atmospheres with an effective temperature of $\sim$1400~K or lower, most atmospheres are primarily horizontally quenched, and thus do not exhibit strong chemical gradients on isobars. This is in agreement with our chemical models (Fig.~\ref{fig_2dchem}). On the other hand, if the quenching pressure is raised to lower pressures, as can be expected for hotter atmospheres, it can be expected that vertical mixing, rather than zonal, will gain importance, given the uniform zonal wind speed and the increase in $\kzz$ with pressure. 

The zonal-to-vertical mixing ratio seems to be sensitive to the host-star type/rotation rate, although it is difficult to find a clear trend spanning the parameter space. It could be envisaged that slowly rotating atmospheres have zonal-to-vertical mixing ratio's that are relatively high, since they tend to have a faster equatorial jet stream (see Fig.~\ref{fig_zonalwind_plots}). However, they also tend to have higher vertical wind speeds (see Fig.~\ref{fig_verticalwind_plots}), so the relative importance of zonal and vertical mixing is not trivially interpreted. Moreover, in our computation of $\kzz$ via eq.~$\eqref{eq_kzz}$, both effects are already taken into account. 

In some regions of the parameter space, vertical mixing is the more dominant process at 1~bar. This occurs in coldest models, for which the equatorial jet stream is insufficiently excited, as well as the slowest rotating (M5-typed), ultra-hot atmospheres. Finally, the models with high surface gravities ($100$~m/s$^2$) all exhibit a relatively high vertical mixing efficiency, with many of them having zonal-to-vertical time-scale ratio's above 1. The result is a lack of zonal advection for atmospheres with high surface gravity, which is either associated with chemical equilibrium for pressures higher than the quenching pressure, or associated with a zonal-to-vertical time-scale ratio above 1 for pressures below the quenching pressure. The corresponding chemical abundance plots likewise show this lack of zonal mixing as strong horizontal gradients in the molar fractions of most species (Fig.~\ref{fig_2dchem_appendix_g100}).

\subsection{Synthetic Spectrum}

\begin{figure*}
    \centering
    \includegraphics[width=0.99\textwidth]{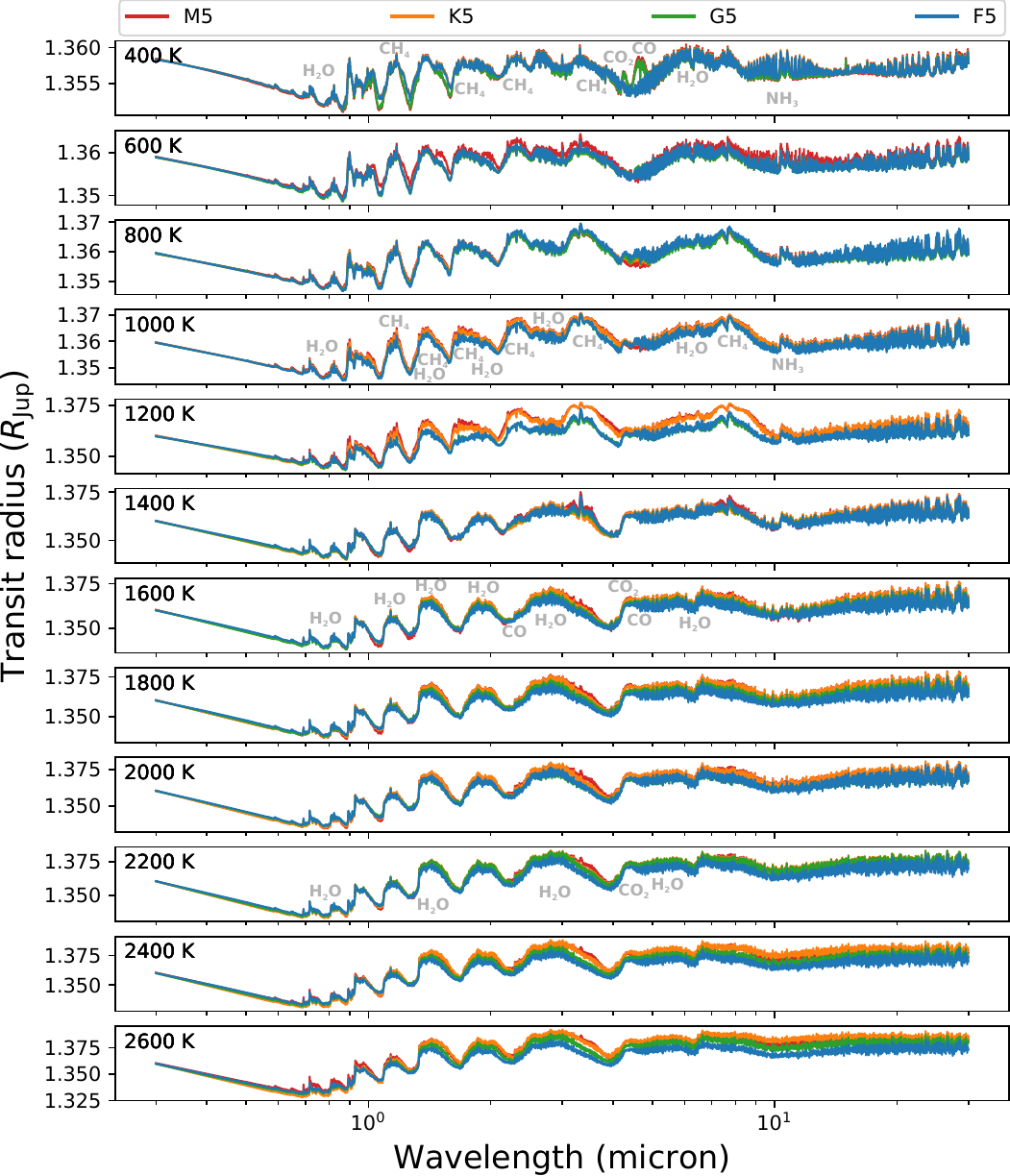}
    \caption{The synthetic transmission spectra computed with \textit{petitRADTRANS} for simulations with medium surface gravity ($g=10$~m/s$^2$), ordered by increasing effective temperature from top to bottom. Transmission spectra corresponding to different rotation rates/host star types are colour coded: (red, M5), (orange, K5), (green, G5) and (blue, F5). These simulations, differing in rotation period, have largely overlapping spectra, but in some molecular bands differences can be discerned. A selection of molecular bands and features has been annotated, either above or below the region of their highest opacity. Note that the vertical axes have differing scales. Moreover, the units are $R_{\rm Jup}$ for an easy comparison between planets orbiting different stars. As such, the actual transit depth, $R^2_{p}/R^2_\ast$,  will differ, depending on the stellar radius of the host star $R_\ast$.}
    \label{fig_many_spectra}
\end{figure*}

A suite of cloud-free transmission spectra, modelled with \textit{petitRADTRANS}, is shown in Fig.~\ref{fig_many_spectra}. The spectra are displayed according to effective temperature, with spectra corresponding to different rotation rates plotted together. We have opted here to plot the effective transit radius (in $R_{\rm Jup}$), computed from the morning and evening limb contributions (see eq.~\ref{eq_morning-evening}), in order to be able to compare simulations with different rotation rates/host star types directly. The transit depth, on the other hand, which is an observational parameter and is denoted by the ratio of the planetary and stellar radii $R^2_{p}/R^2_\ast$, will vary much more among spectra of different rotation rates, because the stellar radius varies self-consistently with the planetary rotation rate in the framework of synchronous rotation.

Upon inspecting the spectra in Fig.~\ref{fig_many_spectra}, from the coldest to the hottest atmospheres, molecular absorption bands can be distinguished for different species, which are in line with the most abundant species in the chemistry models. 
Water absorption is omnipresent. In the coldest atmospheres ($\Teff=400$~K), it is competing with Rayleigh-scattering at the short end of the spectrum (700 -- 900~nm). As the temperature increases and the contribution of methane diminishes, water absorption bands in the near-infrared become more pronounced, and at effective temperatures above $\sim$1400~K, water is the dominant opacity source for the majority of the spectral range. Furthermore, in the mid-infrared, the broad water absorption centred at $6.3$~$\mu$m is visible in all spectra.
Methane on the other hand, is only visible in the coldest planets. At $\Teff=400$~K, it is the dominant opacity source in the near-infrared, with very pronounced molecular bands. As the methane content decreases, however, it becomes more difficult to attribute spectral features to methane or water absorption, due to the large wavelength overlap of their absorption bands. This is illustrated for $\Teff=1000$~K, where both molecules contribute equally to the transmission spectrum.
Similarly to methane, ammonia is only visible in transmission spectra corresponding to the coldest atmospheres. It has a prominent absorption band at 10~$\mu$m, which is very pronounced for the $\Teff=400$~K models, but quickly diminishes as the temperature increases. Nevertheless, at this limited wavelength window, it seems to persist in atmospheres with effective temperatures of up to 1400~K. We discuss the outlook of observing ammonia in Section~\ref{sec_ammonia}.
Finally, molecular signatures of carbon monoxide and carbon dioxide are quite rare, as they both have a relatively poor absorption spectrum in the considered wavelength range. Still, at $4.3$~$\mu$m, the opacities of both \chem{CO} and \chem{CO_2} are at their maximum, and the resulting opacity bump can be seen in many spectra in Fig.~\ref{fig_many_spectra}. The potential of observing \chem{CO} and \chem{CO_2} is further discussed in Section~\ref{sec_co}. 

We note that the spectra corresponding to atmospheres with a different rotation rate/host star type, are generally very consistent. Nevertheless, a few noteworthy differences in the transmission spectra arise when the rotation rate -- and all associated climate and dynamical mixing properties -- are changed. Most notably, the coldest model with $\Teff=400$~K shows distinct absorption features at 4.3~$\mu$m and 4.6~$\mu$m, coming from \chem{CO_2} and \chem{CO} respectively, for stellar types M5 and G5, but these are not present for K5 and F5. This is a strong signature of disequilibrium chemistry for the two former cases. Indeed, the abundances of both \chem{CO} and \chem{CO_2} depend very strongly on the temperature in chemical equilibrium (Fig.~\ref{fig_2dchem}). Small changes in the quenching level can, consequently, induce order-of-magnitude-sized variations in the abundances of these species. This results in a spectrum which is very sensitive to the vertical mixing strength. 

A similar discrepancy between fast (M5, K5) and slow (G5, F5) rotators can be seen in the 1200~K-spectra. Here, the former show much more pronounced absorption bands of methane than the latter. From Fig.~\ref{fig_2dchem} it becomes clear that the methane content in the atmosphere depends critically on the chemical equilibrium composition at the quenching pressure. Indeed, the \chem{CH_4} equilibrium abundance slightly decreases as the host star gets hotter. This is caused by the gradual temperature increase with host star type in the deeper parts of the atmosphere (Fig.~\ref{fig_LPT_profiles}). Looking at the 1200~K-models in this figure, the atmosphere goes from a \chem{CH_4}-dominated equilibrium (M-star) to a CO-dominated equilibrium (G-, F-star) near the quenching pressure (a few bar). The eddy diffusion coefficient also shows a small gradual increase with host star type for these models (Fig.~\ref{fig_kzz_profiles}), so the quenching pressure is similar in all four cases. This results in a higher methane fraction throughout the atmosphere of fast rotating planets, and a stronger absorption signature in the spectrum. 

Finally, for transmission spectra corresponding to atmospheres of ultra-hot exoplanets ($\Teff > 2000$~K), the overall transmission depth can be seen to decrease with the rotation rate. Upon closer inspection, we have established that this decrease is mainly caused by the changing temperature structure with rotation rate, rather than the comparatively small influence of water dissociation, which can be seen in Fig.~\ref{fig_2dchem}.

Overall, we show that -- in certain cases -- the rotation rate of the planet can have a very strong impact on its transmission spectrum, even though it is negligible in most cases.
In the future, it could be interesting to see how taking into account the effect that the host star and rotation rate have on the climate and the chemical mixing efficiencies, can potentially bias exoplanet retrieval results, such as the inferred C/O ratio.
\section{Discussion}
\label{sec_discussion}

\subsection{Zonal Mixing In Context}\label{sec_zonal_context}

\begin{figure*}
    \centering
    \includegraphics[width=0.99\textwidth]{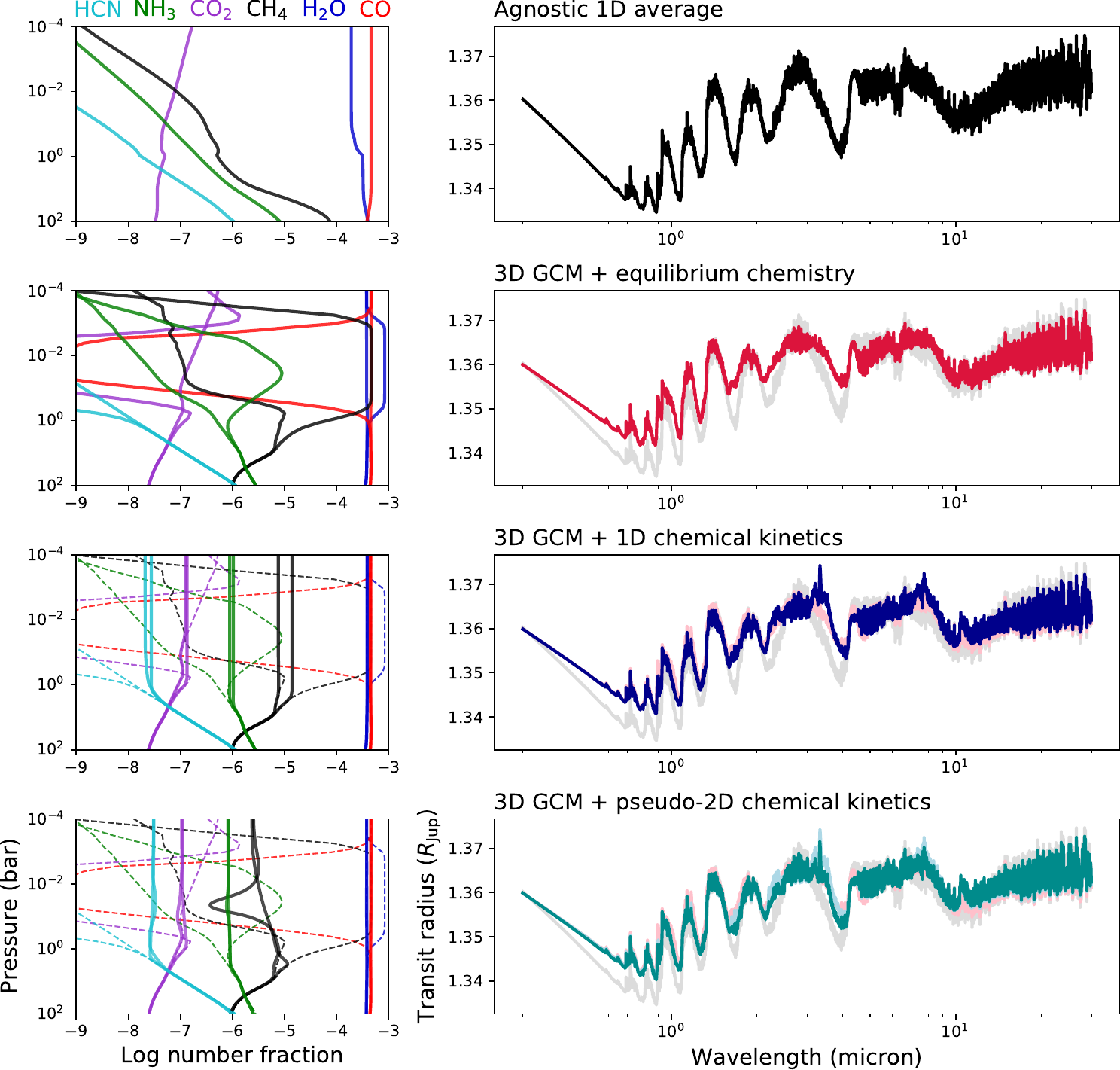}
    \caption{In this model hierarchy, the chemical composition of the morning and evening limbs (\textit{left}) and the corresponding synthetic transmission spectrum (\textit{right}) are shown for each modelling assumption. Four modelling assumptions are presented, from top to bottom: First, an `agnostic' one-dimensional model with equilibrium chemistry is displayed, corresponding to the planetary averaged flux distribution (see also Fig.~\ref{fig_different_adiabats}). Second, the pressure-temperature profiles are drawn from the 3D GCM, still assuming equilibrium chemistry. Third, the model is augmented with vertical mixing, resulting in potential departures (\textit{solid lines}) from equilibrium chemistry (\textit{dashed lines}). Finally, the introduction of the pseudo-2D framework incorporates horizontal advection of the chemical species, which represents the most sophisticated modelling step employed in this work. All hierarchical steps are based on a planet with $\Teff=1400$~K and $g=10$~m/s$^2$ orbiting a G5-star.}
    \label{fig_model_hierarchy}
\end{figure*}

\begin{figure*}
    \centering
    \includegraphics[width=0.99\textwidth]{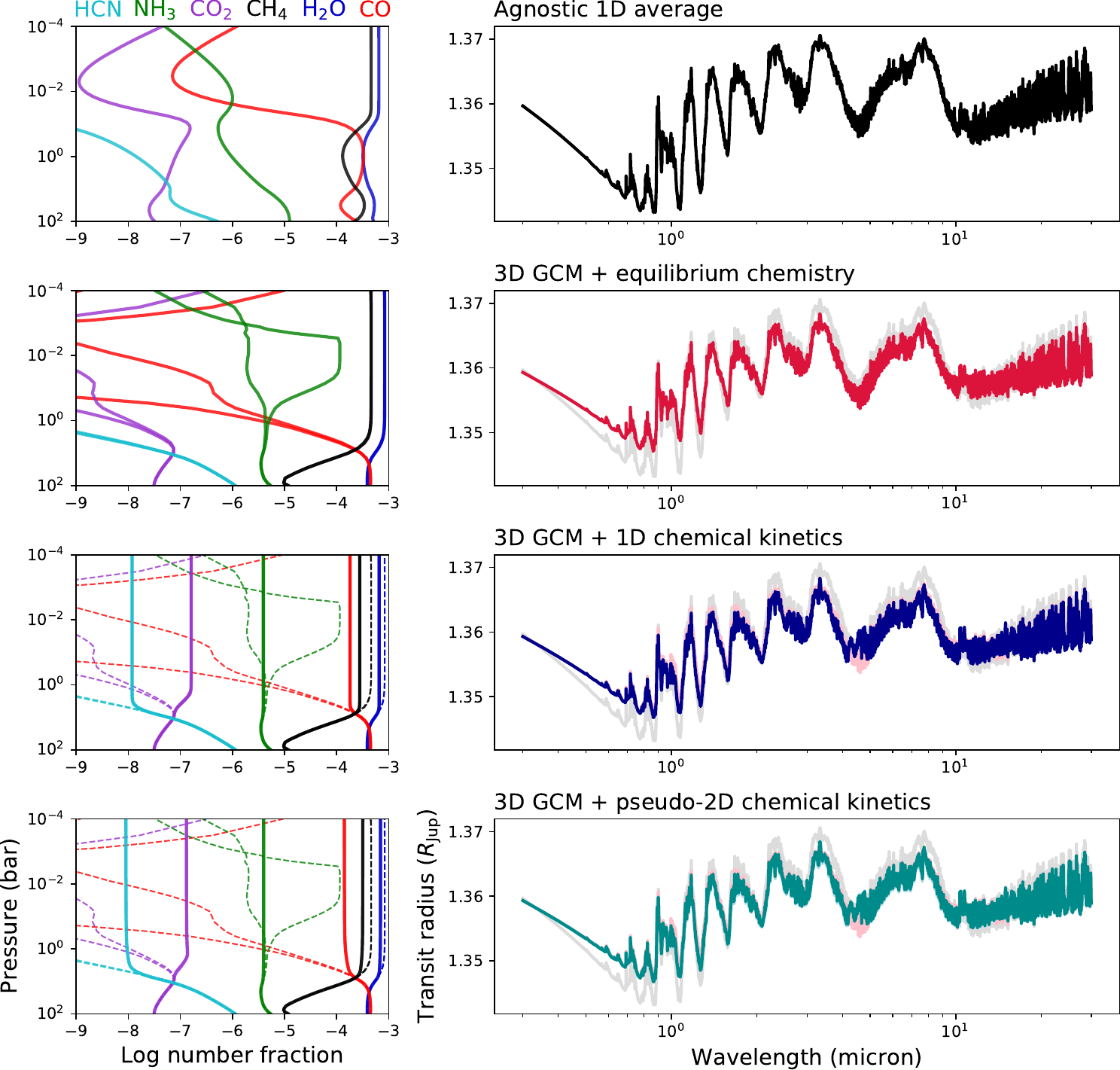}
    \caption{The same modelling hierarchy as in Fig.~\ref{fig_model_hierarchy} is shown, but here for a cooler planet with $\Teff=800$~K.}
    \label{fig_model_hierarchy_cool}
\end{figure*}

\begin{figure*}
    \centering
    \includegraphics[width=0.99\textwidth]{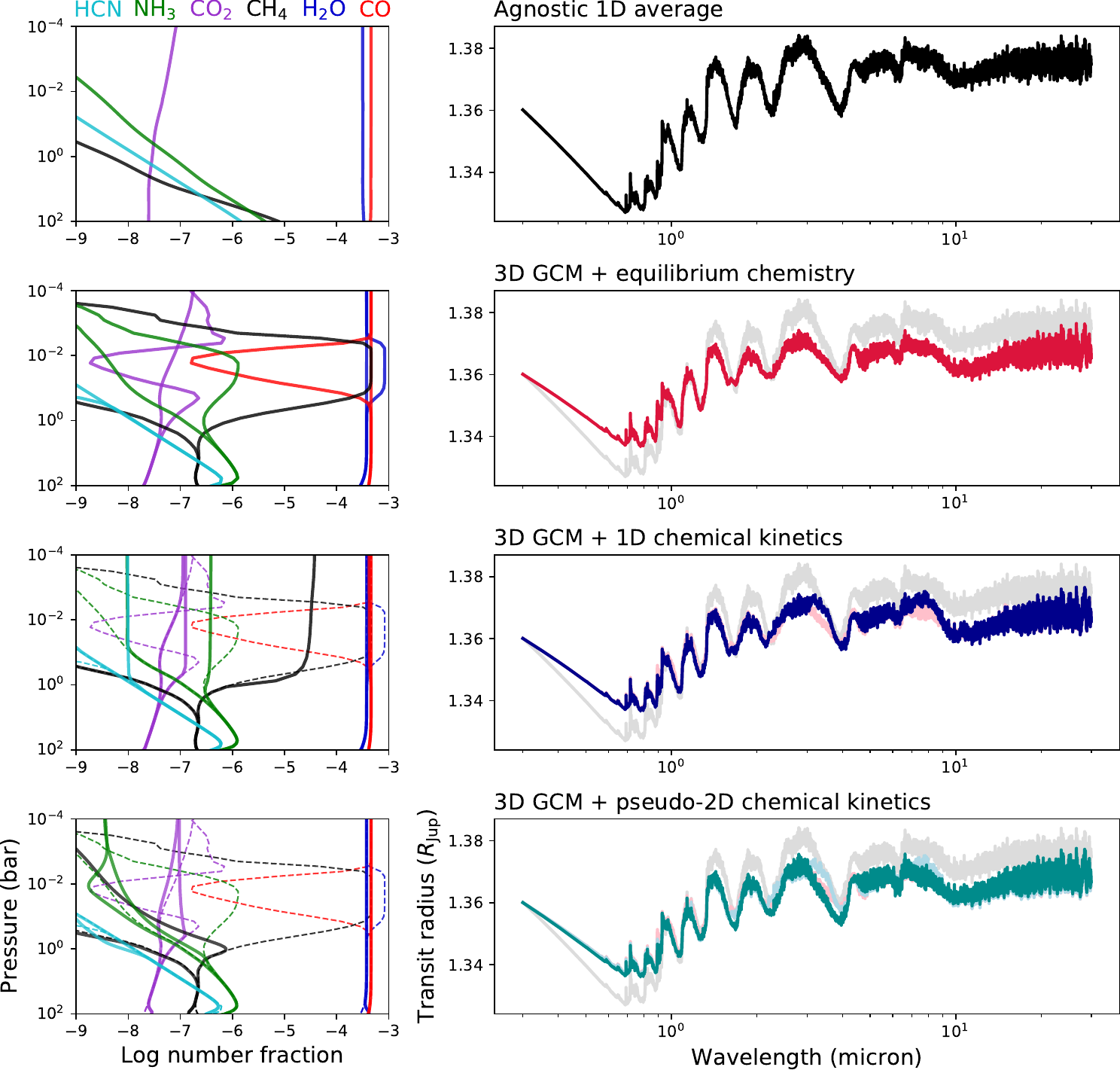}
    \caption{The same modelling hierarchy as in Fig.~\ref{fig_model_hierarchy} is shown, but here for a hotter planet with $\Teff=2000$~K.}
    \label{fig_model_hierarchy_hot}
\end{figure*}

Previous research into the coupling of chemical kinetics and dynamics, in pseudo-2D \citep{Agundez2014, Venot2020_wasp43b} or in 3D \citep{Mendonca2018_disequilibrium, Drummond2018_HD209458b, Drummond2018_HD189733b, Drummond2020}, has highlighted the potential of horizontal (zonal and/or meridional) advection to produce horizontal homogeneous chemical compositions. Indeed, upon comparison with 1D chemical kinetics models, many main molecules in the atmospheres of the exoplanets in consideration no longer show horizontal compositional gradients when horizontal mixing is taken into account. This is especially important for a species like \chem{CH_4}, that can show order-of-magnitude variations with longitude when horizontal mixing is not taken into account. Pseudo-2D chemical kinetics models have shown that in such case, contamination from the day side will reduce the overall methane content \citep{Agundez2014, Venot2020_wasp43b}. For the planet \hdtwenty{}, the same trend has been confirmed with a 3D GCM coupled to chemical kinetics \citep{Drummond2020}. For the planet \hdeighteen{}, this study also found a chemically homogeneous atmosphere, although in this cooler planet the methane abundance is expected to be enhanced due to meridional quenching from high latitudes. 

To assess the impact of zonal mixing in exoplanetary atmospheres, and disentangle it from other modelling assumptions, we analyse a hierarchy of simulations with increasing complexity. What follows is a short description of each modelling level.
\begin{enumerate}
    \item \textit{Agnostic 1D average:} a one-dimensional radiative-equilibrium temperature profile, computed for planetary averaged irradiation (see red line in Fig.~\ref{fig_different_adiabats}). With the denomination `agnostic', we stress that no preconceived notions about three-dimensional properties of the planetary atmosphere are used in the calculation (as opposed to averaging the results of a 3D GCM simulation). Chemical equilibrium is assumed. The synthetic spectrum is computed for this single vertical structure.\\
    
    \item \textit{3D GCM + equilibrium chemistry:} two vertical temperature profiles, corresponding to the morning and evening limbs, extracted from the GCM climate simulation. The profiles are weighted and meridionally averaged about the equator, in the same way as it was done for the pseudo-2D case (see Section~\ref{sec_methods_chemistry}). For each of these, the chemical equilibrium is determined. The synthetic spectrum is computed by taking into account equal contributions from the morning and evening limbs (see Section~\ref{sec_methods_spectrum}).\\
    
    \item \textit{3D GCM + 1D chemical kinetics:} same temperature profiles as above. For each of these, a 1D chemical kinetics model with vertical mixing is computed. The synthetic spectrum is computed as above.\\
    
    \item \textit{3D GCM + pseudo-2D chemical kinetics:} same temperature profiles as above. For the equatorial band, the pseudo-2D chemical kinetics framework is used to compute the chemical composition, incorporating both vertical and zonal mixing. The synthetic spectrum is computed as above. This is the nominal modelling approach for the grid models in this work.
\end{enumerate}

For a planetary atmosphere of 1400~K, the chemical composition and synthetic transmission spectrum for each of these four modelling steps are shown in Fig.~\ref{fig_model_hierarchy}. The overlapping transmission spectra demonstrate that the incorporation of the morning and evening limb temperatures, drawn from a 3D climate model, reduces the extent of the molecular features, compared to the simple agnostic model. It has been shown that the transit depth of molecular features can be reduced with respect to a one-dimensional mean temperature profile \citep{Caldas2019}. The reason is that the 3D approach results in colder temperatures for the morning terminator, due to horizontal advection from the night side. Furthermore, retrievals of exoplanet emission spectra, taking into account the 3D information of the day side temperature, result in retrieved temperature profiles that are close to the arithmetic mean temperature distribution \citep{Blecic2017}. Since the limbs will have lower temperatures than the dominating day side, it is reasonable to assume that the temperature probed with transmission spectroscopy will be lower than the arithmetic mean and also lower than the \textit{agnostic} temperature profile used here.

Additionally, the introduction of disequilibrium chemistry, through vertical and also zonal mixing, manifests itself in the appearance of methane absorption. This is most prominently seen at 3.3~$\mu$m. The case with only vertical mixing shows the strongest absorption by methane, and relatively high volume mixing ratios of about 10$^{-5}$ for both limbs. When horizontal quenching is taken into account, the mixing ratio and the prominence of the absorption band both drop, as methane is horizontally quenched to the day-side composition, similar to what was found by \cite{Agundez2014}.

When we apply the same hierarchical methodology to a planet with a cooler ($\Teff=800$~K, Fig.~\ref{fig_model_hierarchy_cool}) and a hotter ($\Teff=2000$~K, Fig.~\ref{fig_model_hierarchy_hot}) atmosphere, a trend starts to emerge. In the case of a cooler atmosphere, which shows methane absorption in the transmission spectrum, the switch from a 1D temperature profile to temperatures derived from a 3D GCM again has the biggest impact on the spectrum. However, subsequent levels of complexity very minorly or negligibly influence the transmission radius. The only change brought about by vertical mixing are the increased \chem{CO} and \chem{CO_2} mixing ratios, because of deep quenching, and the increased absorption between 4 and 5~$\mu$m accompanying it. Zonal mixing has no additional effect, since the abundances are all quenched quite deep, so that no longitudinal gradients are present. 

In the hierarchical case study of a hotter planetary atmosphere, each modelling level results in a distinct composition and spectrum (Fig.~\ref{fig_model_hierarchy_hot}). In line with the changes seen in the planets of 800~K and 1400~K, the use of 3D GCM temperatures changes the molecular absorption features strongly. Next, vertical mixing quenches the abundances at pressures of 100~mbar to 1~bar, which is higher up in the atmosphere than where it occurs in the planets with lower temperatures, despite more vigorous mixing. In this case, unlike the cool planetary atmosphere of 800~K, horizontal quenching has a large impact on the chemical composition. This is especially the case for \chem{CH_4}, reducing its fraction in the atmosphere from more than 10$^{-5}$ in the 1D kinetics case to below 10$^{-6}$ in the pseudo-2D case, for all pressures at the morning terminator. This change is reflected in the spectrum, where it can be seen that the transit radius reduces in the most prominent methane absorption bands.

In summary, from this discussion it appears that the consideration of more realistic temperature profiles, either drawn from a 3D GCM simulation \citep[e.g.][]{Venot2020_wasp43b, Lewis2020} 
or self-consistently parametrized \citep{Gandhi2020_daynight}, has a profound impact on both the exoplanet chemistry and the transmission signature. This illustrates the importance of taking into account horizontal heat redistribution and terminator asymmetries. Furthermore, the relevance of these effects applies to a wide range of atmospheric temperatures (Fig.~\ref{fig_model_hierarchy},~\ref{fig_model_hierarchy_cool},~\ref{fig_model_hierarchy_hot}). Where the temperatures modify the extent of the transmission features, disequilibrium processes act to increase or reduce absorption in specific wavelength ranges, since they have the potential to alter the concentration of some species by orders of magnitude. In relatively cool planets, vertical mixing can be seen to quench molecular abundances efficiently, as is expected for planets below 1200~K \citep{Line2013}. Because of this deep quenching, longitudinal gradients in the chemistry are almost non-existent. For this reason, horizontal mixing is quasi-irrelevant in relatively cool planets. In relatively hot planets, on the other hand, horizontal quenching is needed to avoid overpredicting the abundances of species like \chem{CH_4} and \chem{NH_3}, that tend to favour lower temperatures. The atmospheres of these planets are quenched at lower pressures, which -- in combination with their large day-night temperature differences -- results in strong horizontal chemical gradients in a 1D framework. Horizontal diffusion acts as a homogenization factor, which in general sets the chemical composition at the limbs closer to the day-side composition. 

\subsection{Limb Asymmetries}\label{sec_limbs}

\begin{figure}
    \centering
    \includegraphics[width=0.99\columnwidth]{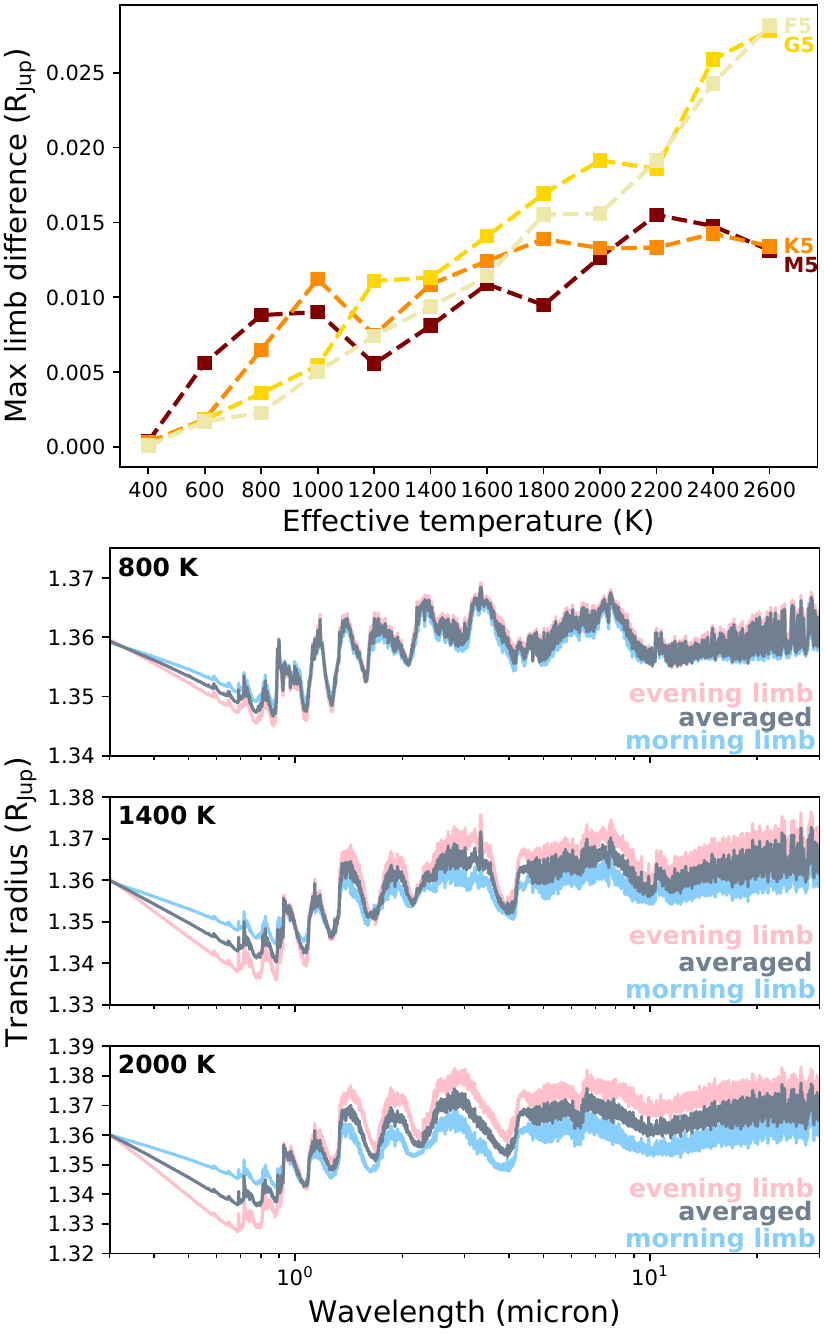}
    \caption{The maximal difference in transit depth between the morning limb ($-90\degrees$ longitude) and the evening limb ($90\degrees$~longitude) generally increases with the exoplanet's effective temperature (\textit{top panel}). The data displayed corresponds to grid models with $g=10$~m/s$^2$, and host star types/rotation rates as indicated with coloured lines. Next, the transmission spectra (\textit{grey}), along with those corresponding to the morning (\textit{blue}) and evening (\textit{pink}) limbs, illustrate in which wavelength ranges limb asymmetries result in the biggest differences (\textit{bottom three panels}). Spectra for three different grid models are shown: $\Teff=800$~K, 1400~K, and 2000~K. All of them have $g=10$~m/s$^2$ and a G5-star host.}
    \label{fig_limb_asymmetry}
\end{figure}

The modelling approach used in this work is well suited to explore the presence of morning and evening limb asymmetries in the diversity of exoplanet atmospheres, be it because the limbs differ in temperature, or in chemical composition. To this end, the transmission spectra corresponding to the morning and evening terminators are considered separately, where the atmospheric structure at $-90\degrees$ and $90\degrees$ longitude, respectively, is assumed to constitute the whole projected ring that is probed in transmission spectroscopy (see also Section~\ref{sec_methods_spectrum}).

\subsubsection{Detectability}

The maximal difference in the transit depths between the morning and evening terminators shows an increase with increasing temperature, for a subset of grid models with medium surface gravity (Fig.~\ref{fig_limb_asymmetry}). Indeed, considering a sample of models of increasing temperature -- 800~K, 1400~K, and 2000~K -- the maximal differences between both limbs are 0.004~R$_{\rm Jup}$, 0.011~R$_{\rm Jup}$, and 0.019~R$_{\rm Jup}$, respectively, for the G5-host star case. These values correspond approximately to 0.3, 0.8, and 1.5~per~cent of the radius of a typical inflated hot Jupiter (1.35~R$_{\rm Jup}$). 

Evidently, the potential for observing such a small difference will depend on the exact physical characteristics of the planet-star system. However, assuming a G5-type host star with a radius of 0.9~R$_\odot$ \citep[table~2.4,][]{Hubeny2014}, the estimated transit depth $R_p^2/R_\ast^2$ differences for the three cases shown would amount to 1, 2 and 4~ppm, respectively: too low to detect. Models corresponding to high ($g=100$~m/s$^2$) or low ($g=1$~m/s$^2$) surface gravities generally show a similar correlation with the effective temperature. The amplitude of the limb differences, however, changes, because of the change in scale height. In the low-gravity case, this results in limb asymmetries of 
38, 235 and 1951~ppm, which would be comfortably detectable with the \textit{James Webb Space Telescope} (\textit{JWST}), which would have a noise floor of 20--50~ppm \citep{Greene2016}. 

We note that this represents a very optimistic scenario, since the gravity of most transiting hot Jupiters will be higher than $1$~m/s$^2$. However, by doing this exercise, we illustrate that the detection of limb asymmetries with \textit{JWST} should be possible, and hinges on a combination of a comparatively low gravity, a medium to hot atmosphere, and a favourable host star. 

It is important to remark that neither clouds nor hazes are taken into account in our models. It has been shown that clouds and hazes can exhibit inhomogeneous distributions \citep{Parmentier2016, Powell2019, Steinrueck2020}, which add an additional layer of complexity in the comparison between the morning and evening limbs. On the upside, the observation of limb asymmetries could provide important information to break retrieval degeneracies \citep{Line2016, Kempton2017}. Finally, we note that, although our results suggest that ultra-hot ($\Teff > 2200$~K) Jupiters are the optimal candidates for limb asymmetries, we are careful with this conclusion. Ultra-hot exoplanets are expected to have additional opacity sources \citep{Lothringer2018, Arcangeli2018, Parmentier2018}, not included here (see Section~\ref{sec_methods_petitcode}). Moreover, hydrogen gas dissociation on the day side and recombination on the night side has been shown to reduce the day-night temperature contrast in these planets \citep{Tan2019}. Hence, it can be expected that the limb asymmetries associated with these planets are not as large as presented in Fig.~\ref{fig_limb_asymmetry}. This point is discussed further in Section~\ref{sec_limitations}.

\subsubsection{Thermal asymmetry, but chemical similarity}

Upon examining the causes of limb transit depth differences in our models, we find that temperature asymmetries between the morning and evening limbs are much more important than differences in the chemical composition. 
The morning and evening limbs are at different temperatures because of eastward heat advection. Hot, day-side air and cold, night-side air are both transported eastward, so that the morning limb ($-90\degrees$ longitude) is generally colder than the evening limb ($90\degrees$ longitude). This can be readily seen in Fig.~\ref{fig_temperature_maps}. Likewise, the separated transmission spectra show that the morning limb has a lower transmission signal in the infrared, being the colder of the two (Fig.~\ref{fig_limb_asymmetry}). We conclude that limb asymmetries are maximized when the heat transport is sufficiently efficient for heat redistribution to take place, but not too efficient, so as to remove all horizontal temperature gradients.

Chemical differences between the two limbs, on the other hand, are either not present, or appear to be of secondary importance in the interpretation of limb transit depth asymmetry. As we established in Section~\ref{sec_results_chemistry}, planets in our grid with $\Teff < 1400$~K tend to have a chemical composition that is horizontally homogeneous. Thus, for these cases, any differences in the transit depth of the limb are caused by the temperature structure. Also for hotter cases, which can have horizontal chemical gradients, the chemical composition is not the deciding factor in setting the transit depth. One scenario is that chemical differences between both terminators remain small due to efficient horizontal quenching (see Fig.~\ref{fig_model_hierarchy_hot}). In another scenario, the atmosphere is too hot for efficient horizontal mixing, and the chemical composition remains close to equilibrium, as would be the case for the hottest planets. In both cases, any difference in the limb compositions is directly attributed to temperature differences. Moreover, the dominant absorbing species, \chem{H_2O} and \chem{CO}, remain approximately homogeneously distributed, even in hot atmospheres. It is not expected that three-dimensional advection will introduce an additional asymmetry in the equatorial regions of the morning and evening limbs \citep{Mendonca2018_disequilibrium, Drummond2020}.

The conclusion that thermal and not chemical limb differences are causing the transit depth asymmetry, has been formulated earlier by \cite{Agundez2014} for the hot Jupiters \hdtwenty{} and \hdeighteen{}. Here, we confirm and widen it to a large parameter space. That temperature variations drive the limb asymmetry is visible in the individual transmission spectra (Fig.~\ref{fig_limb_asymmetry}). The amplitude of all molecular features is affected, rather than a select number of absorption bands related to a single species, which would be the case when chemical variations were the main cause of the spectral variation. Interestingly, the limb variation appears to be the smallest around 1~$\mu$m, roughly the wavelength range of \textit{HST}/WFC3. This additionally highlights the need for \textit{JWST} in order to be able to study the exoplanet atmosphere limbs separately with ingress--egress spectroscopy.

\subsection{Implications for Benchmark \textit{JWST} Targets}\label{sec_case_studies}

For a selection of planets, we sample the nearest atmospheric model from our grid of climate and chemistry models, and compare it with previous literature, theoretical models and observations. In the comparison with observations, mostly \textit{Hubble Space Telescope} (\textit{HST}) transmission spectroscopy, we apply manual vertical shifts in the transit depth, to correct for the fixed planetary radius in our models. We choose to focus on planets that have been heavily studied with 3D GCMs before, and/or will be observed with the \textit{James Webb Space Telescope} (\textit{JWST}). Note that we do not aim to provide quantitatively accurate atmospheric models for these specific exoplanets; this would be impossible given the wide grid spacing and model assumptions (see also Section~\ref{sec_limitations}). Rather, we aim to put our multi-dimensional modelling approach in a broader context by connecting the generic grid models back to the known diversity of exoplanets. 

\begin{figure*}
    \centering
    \includegraphics[width=0.99\textwidth]{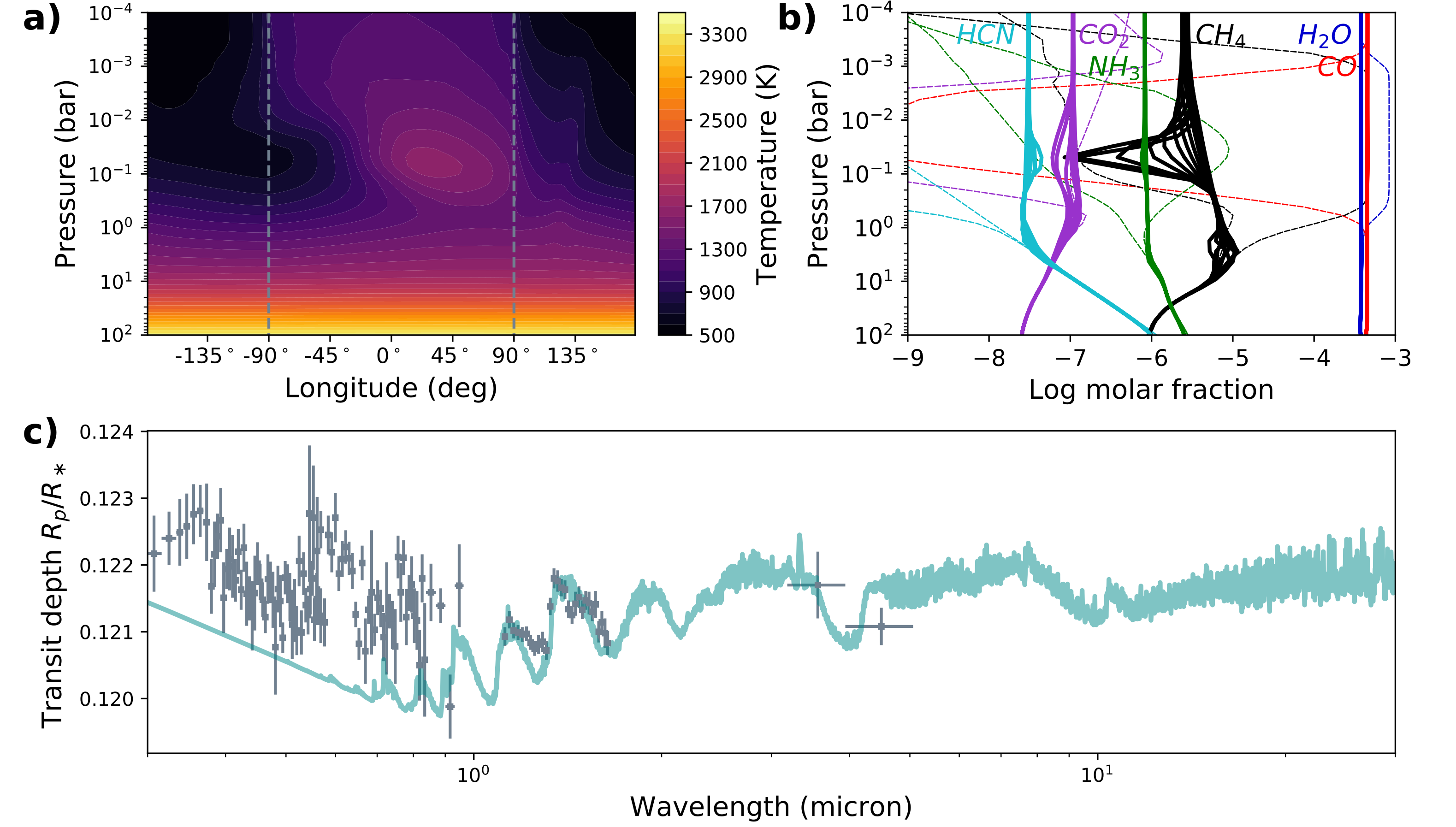}
    \caption{Physical diagnostics for an \hdtwenty{}-like planet. This grid model has $\Teff=1400$~K, $g=10$~m/s$^2$ and host star type G5. Plotted are the meridional mean temperature used in the pseudo-2D chemistry code (\textit{a}), the chemical composition (\textit{b}), and the synthetic transmission spectrum, in a comparison with observations (\textit{c}). The observational data (\textit{grey}) are from \citet{Sing2016}. The transmission spectrum is cloud-free, and does not contain Na/K opacities. In the chemistry plot, the equilibrium compositions of the morning and evening limb are plotted as thin dashed lines.}
    \label{fig_hd20}
\end{figure*}

\subsubsection{HD~209458~b}\label{sec_hd20}

The hot Jupiter \hdtwenty{}, with its mass of 0.73~$M_{\rm Jup}$ and radius of 1.39~$R_{\rm Jup}$ \citep{Stassun2017}, has a gravity of 9.4~m/s$^2$. It further has an equilibrium temperature of $\sim$1450~K \citep{Barstow2017} and a rotation period of 3.5~days. Going by these measurements, the grid model it corresponds most to, has $\Teff=1400$~K, $g=10$~m/s$^2$ and rotation period 2.32~days/host star type G5 (see Table~\ref{tab:grid}).\footnote{In determining the best matching model to the physical planetary parameters, we first match the temperature and gravity as close as possible. Next, we match the rotation/host star type. In doing so, we aim to obtain a good match for the planetary rotation rate, rather than the actual stellar type of the host star, since the rotation rate has a much bigger impact on the atmospheric modelling.}
Diagnostics for the temperature, chemical composition and spectrum for this model are shown in Fig.~\ref{fig_hd20}. The synthetic transmission spectrum in this figure has been computed assuming a stellar radius of 1.19~M$_{\odot}$, and a manual vertical shift has been applied to match the data of \cite{Sing2016}. This is necessary because all grid models have been computed with the same radius ($1.35$~R$_{\rm Jup}$), so the synthetic transmission spectrum differs from the data by a constant value.

The transmission spectrum of our \hdtwenty{}-like model agrees reasonably well with the observational data (Fig.~\ref{fig_hd20}, panel~c). The \textit{Spitzer} data points at 3.6~$\mu$m and 4.5~$\mu$m, and the water molecular feature at 1.4~$\mu$m are reproduced, although the amplitude of the latter is a bit overestimated, probably because we did not include clouds \citep{Deming2013}. There is no good match between our model spectrum and the \textit{HST}/STIS data in the optical wavelength range. The optical scattering slopes are determined by a variety of haze particles with non-grey opacities, as well as broad-base sodium and potassium absorption lines \citep{Sing2016}, both of which are not included in our opacities (see Section~\ref{sec_methods_spectrum}).  

Regarding the chemical composition (Fig.~\ref{fig_hd20}, panel~b), our results for \hdtwenty{} show some quantitative differences with the pseudo-2D model obtained with the same chemical kinetics code by \cite{Agundez2014}. The upper atmosphere ($p < 10^{-3}$~bar) modelled by \cite{Agundez2014} shows distinct longitudinal gradients in many species, because of photochemistry. This effect is not taken into account in this work, and we plan to include it in a future study. Furthermore, we do not find the same strong vertical gradients in the molar fractions of \chem{CH_4}, \chem{NH_3} and \chem{HCN}. Instead, we observe an efficient vertical mixing with almost constant molecular abundances. The discrepancy can be explained by the different temperatures employed. The atmospheric temperatures used by \cite{Agundez2014} have been derived from a 3D GCM with a temperature inversion \citep{Showman2009}, resulting in a much hotter day side, closer to equilibrium. This results in a strong decrease in \chem{CH_4}, \chem{NH_3} and \chem{HCN}. It is now generally accepted that \hdtwenty{} does not have a strong thermal inversion \citep{Diamond-Lowe2014, Schwarz2015}, so we expect the temperatures (Fig.~\ref{fig_hd20}, panel~a) and chemical abundances to be more vertically uniform. An additional potential for discrepancies is the chemical scheme employed. In this work, we use the chemical network of \cite{Venot2020_network}, which is an updated and reduced version of the \cite{Venot2012}-network used by \cite{Agundez2014}. We note that, for \hdtwenty{}, the variations due to the chemical network are expected to be small \citep{Venot2020_network}. Despite the methodology differences, we obtain similar abundances for the dominant molecules, \chem{H_2O} and \chem{CO}, as \cite{Agundez2014}, and we arrive at the same conclusion that horizontal advection efficiently mixes the atmosphere toward the day-side composition.

The 3D chemistry-coupled GCM simulation of \hdtwenty{} by \cite{Drummond2020} predicts horizontally homogeneous chemistry, similar to what we find, although there is a difference in the vertical quenching pressures. Molar fractions for \chem{CH_4}, \chem{NH_3} and \chem{HCN} become vertically constant around 100--200~mbar in the model of \cite{Drummond2020}, whereas this happens deeper in the atmosphere in our work (a few bar). Again, the different chemical network employed by these authors, namely that of \cite{Venot2019}, and without \chem{C_2H_2}, could result in slightly different quenching pressures compared to this work, although this effect is expected to be minor for \hdtwenty{} \citep{Venot2020_network}. 

Rather, when comparing the vertical advection time-scales (Fig.~\ref{fig_timescales_hd20} in this work, figure~E.1 in \cite{Drummond2020}), we find that our time-scales are about 2 orders of magnitude shorter near 1~bar, which could explain the discrepancy. The reason for the time-scale difference is not easily unravelled, since \cite{Drummond2020} have estimated it directly based on the vertical wind flow $w$ as $\timescale{\rm vertical} = \frac{H}{|w|}$, whereas we have based our estimation on the eddy diffusion coefficient as $\tau_{\rm vertical} = H^2$/$\kzz$. Upon closer comparison, it seems the 1~bar-temperatures in our model ($\sim$1400~K) are slightly lower than those in \cite{Drummond2020} (1600~K--1800~K), resulting in a shorter scale height and thus potentially more efficient vertical advection of chemical species. Note that the 1~bar-temperatures do not depend on the state of the deep atmosphere \citep{Drummond2020}, and that uncertainties of $\sim$100~K are typical in hot Jupiter GCMs.

Another possible reason for the comparatively high quenching pressure in our work compared to \cite{Drummond2020} could be that the vertical mixing efficiencies are in fact similar, but zonal mixing is more efficient in the work of \cite{Drummond2020}, resulting in an efficient homogenisation of the chemistry to the day-side composition between 0.1~bar and 1~bar. Indeed, the zonal wind speed in our model (2.6~km/s) is somewhat lower than in \cite{Drummond2020} ($>4$~km/s), and the vertical and zonal advection time-scales are comparable (Fig.~\ref{fig_timescales_hd20}).

It appears that temperature and wind speed variations between models -- even if they are small -- can influence the quenching pressures so as to cause order-of-magnitude quantitative changes in the atmospheric chemistry.
Other than the quantitative differences associated with the vertical quenching, we find a good qualitative agreement with the more sophisticated 3D chemical kinetics model of \cite{Drummond2020}. 
A planet such as \hdtwenty{} seems to be rather sensitive to variations between models, because the chemical and advection time-scales (Fig.~\ref{fig_2dchem}), as well as the vertical and horizontal advection time-scales (Fig.~\ref{fig_timescales_hd20} and \ref{fig_timescales_all}) are similar. A hierarchy of modelling approaches, in tandem with benchmarking of GCMs and chemistry codes, should be used to alleviate this issue.

\begin{figure*}
    \centering
    \includegraphics[width=0.99\textwidth]{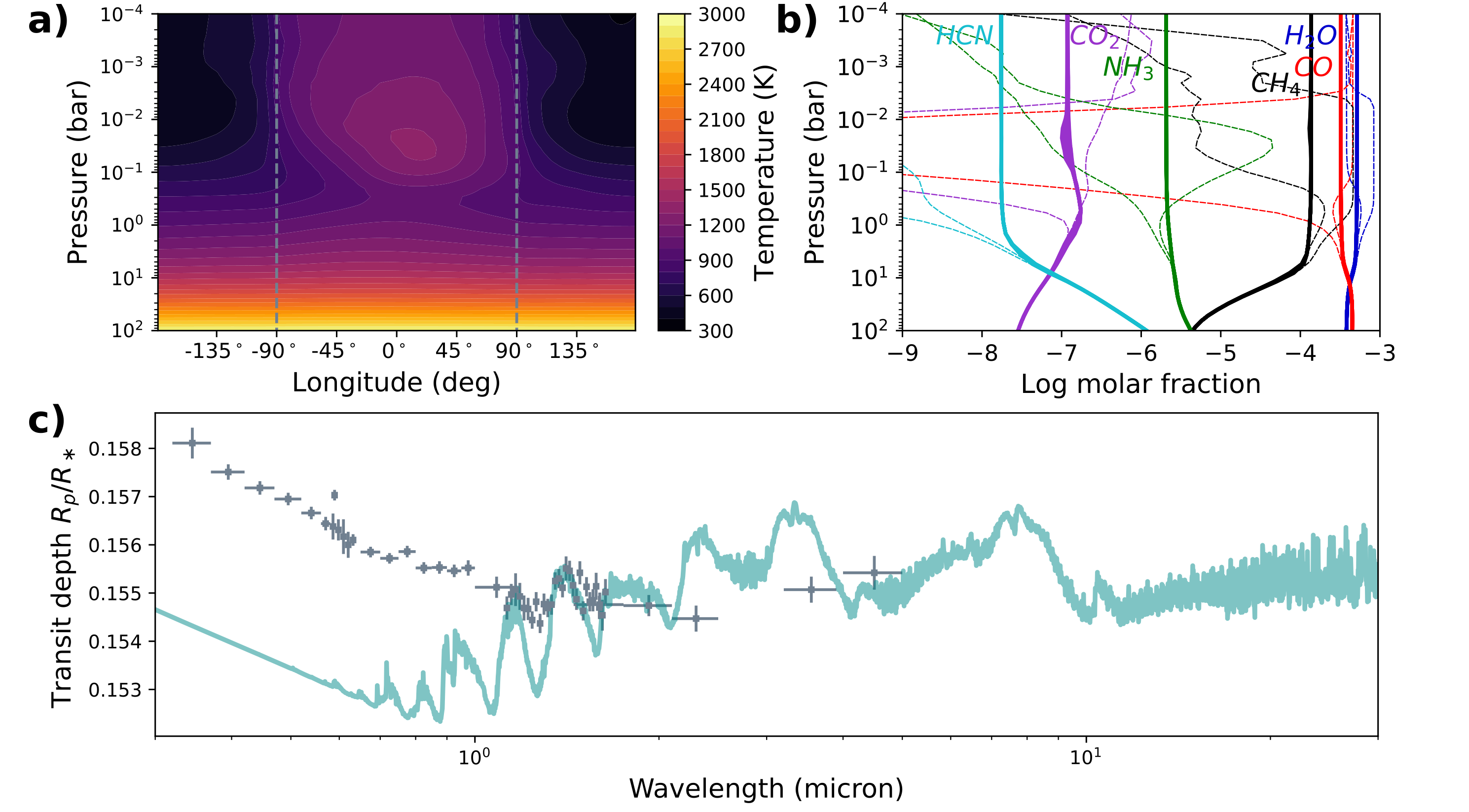}
    \caption{Physical diagnostics for an \hdeighteen{}-like planet. This grid model has $\Teff=1200$~K, $g=10$~m/s$^2$ and host star type K5. Plotted are the meridional mean temperature used in the pseudo-2D chemistry code (\textit{a}), the chemical composition (\textit{b}), and the synthetic transmission spectrum, in a comparison with observations (\textit{c}). The observational data (\textit{grey}) are from \citet{Sing2016}. The transmission spectrum is cloud-free, and does not contain Na/K opacities. In the chemistry plot, the equilibrium compositions of the morning and evening limb are plotted as thin dashed lines.}
    \label{fig_hd18}
\end{figure*}

\subsubsection{HD~189733~b}\label{sec_hd18}

The irradiated exoplanet \hdeighteen{} has a mass of 1.13~$M_{\rm Jup}$ and a radius of 1.13~$R_{\rm Jup}$ \citep{Stassun2017}, resulting in a surface gravity of 21.9~m/s$^2$. Furthermore, it is slightly cooler than \hdtwenty{}, with an equilibrium temperature of $\sim$1200~K \citep{Barstow2017}, and a rotation period of 2.2~days. Our best matching grid model has $\Teff=1200$~K, $g=10$~m/s$^2$ and rotation period 1.26~days/host star type K5.

In analogy with \hdtwenty{}, after applying a vertical offset to our synthetic transmission spectrum, the observational data from the \textit{HST} and \textit{Spitzer} \citep{Sing2016} are matched, except wherever clouds or hazes are concerned (Fig.~\ref{fig_hd18}, panel c). This includes the very prominent slope at optical wavelengths, which has been attributed to Rayleigh scattering on high altitude haze \citep{Pont2013, Sing2016, Steinrueck2020}. Furthermore, since our best-match model has a lower gravity than \hdeighteen{}, the amplitudes of the synthetic transmission features are too high for the observational data, even without taking into consideration that \hdeighteen{} is very cloudy. 

We expect \hdeighteen{} to be chemically homogeneous (Fig.~\ref{fig_hd18}, panel b), as we have established for planets with $\Teff < 1400$~K. This is in agreement with \cite{Agundez2014}, except for the rising \chem{HCN} they find with altitude, which is a product of photochemistry on the day side. A notable distinction with their work, are their homogenized values for methane ($10^{-6}$) and ammonia ($5\cdot10^{-6}$), which are higher in our model ($10^{-4}$ and $5\cdot10^{-5}$ respectively). We attribute this difference to the climate model: whereas the day side and evening limb in \cite{Agundez2014} are still CO-dominated in equilibrium, our model appears to be cooler at the quenching pressure ($\sim1$~bar), allowing for a considerable equilibrium fraction of methane at all wavelengths. This strong dependence of the overall chemical composition on the deeper atmospheric temperature is indicative of a potential source of uncertainty in atmospheric models. This issue is further discussed in Section~\ref{sec_methane} for \chem{CH_4} and Section~\ref{sec_deep_quenching} in general.

We further find a good agreement with the work by \cite{Drummond2020} for most species. They report a deep meridional quenching of methane, which increases the methane content before it is vertically mixed throughout. This mechanism is a 3D-effect that is not present in our work. Nevertheless, we obtain a similar methane fraction, since the chemical equilibrium in our deep atmosphere already has a sizeable methane fraction before quenching. We conclude that the temperatures between 1 and 10~bar play an important role in the chemical composition. Although the temperature discrepancies between our work, and those of \cite{Agundez2014} and \cite{Drummond2020}, can likely be attributed to physical model parameters, such as the low gravity of 10~m/s$^2$, care must be taken so that the temperatures in these layers are well converged \citep{Sainsbury-Martinez2019, Drummond2020}, and unaffected by other physical processes that raise or lower the internal temperature \citep{Fortney2020}. We elaborate on this point further in Section~\ref{sec_methane}.

\begin{figure*}
    \centering
    \includegraphics[width=0.99\textwidth]{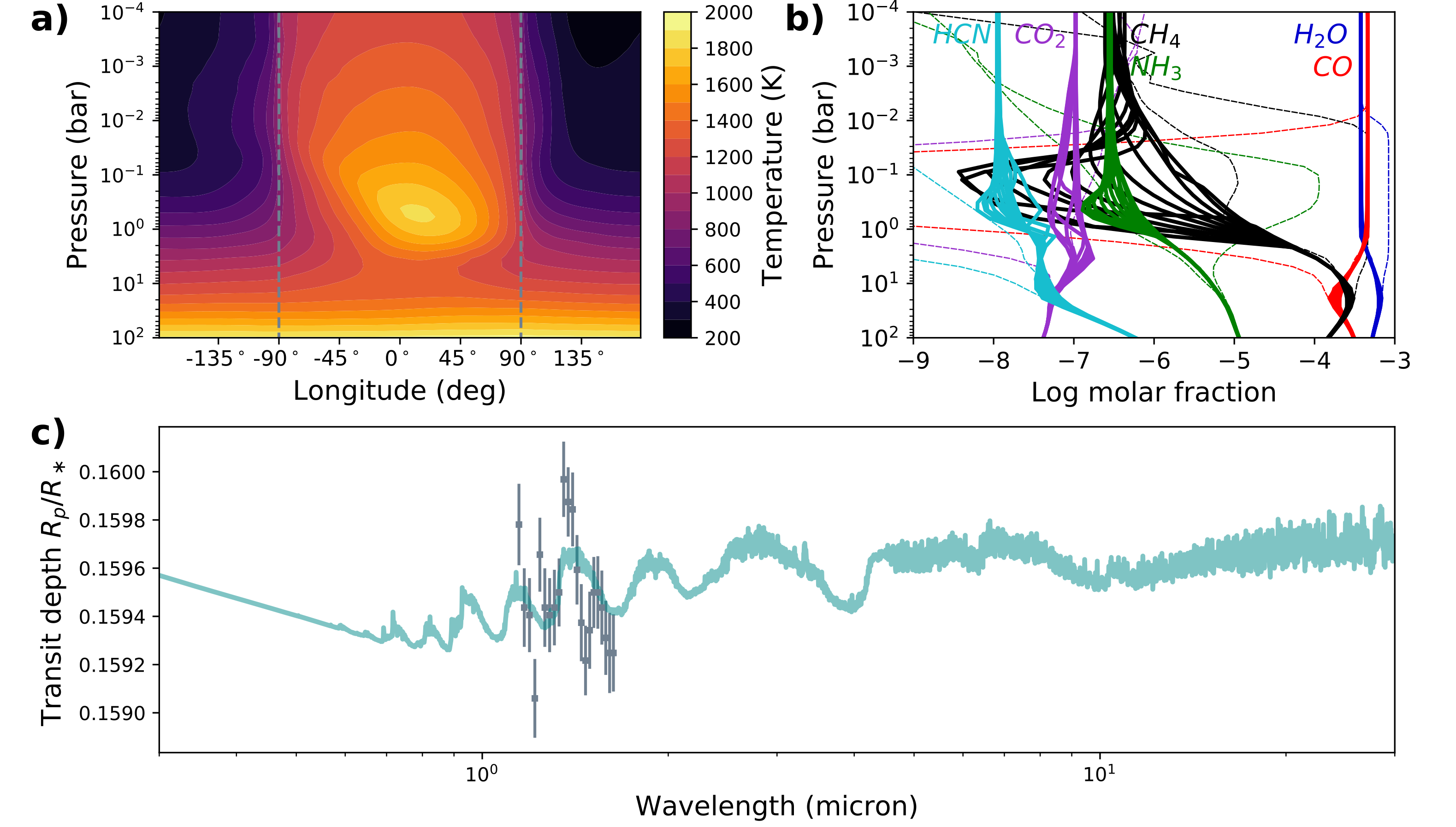}
    \caption{Physical diagnostics for an WASP-43~b-like planet. This grid model has $\Teff=1400$~K, $g=100$~m/s$^2$ and host star type K5. Plotted are the meridional mean temperature used in the pseudo-2D chemistry code (\textit{a}), the chemical composition (\textit{b}), and the synthetic transmission spectrum, in a comparison with observations (\textit{c}). The observational data (\textit{grey}) are from \citet{Kreidberg2014}. The transmission spectrum is cloud-free, and does not contain Na/K opacities. In the chemistry plot, the equilibrium compositions of the morning and evening limb are plotted as thin dashed lines.}
    \label{fig_wasp-43b}
\end{figure*}

\subsubsection{WASP-43~b}\label{sec_wasp43b}

WASP-43~b is a dense and fast rotating hot Jupiter with a mass of 1.998~$M_{\rm Jup}$ and radius 1.006~$R_{\rm Jup}$ \citep{Esposito2017}. This results in a gravity of 48.9~m/s$^2$. It has a very similar temperature to \hdtwenty{}, namely $\sim$1400~K \citep{Esposito2017}. Finally, it has a short rotation period of only 0.81~days. We match it to a grid model with  $\Teff=1400$~K, $g=100$~m/s$^2$ and rotation period 0.79~days/host star type K5.

The synthetic spectrum of our WASP-43~b-like simulation (Fig.~\ref{fig_wasp-43b}, panel c) shows a reasonable match with the observational transmission spectrum \citep{Kreidberg2014}, although the high surface gravity in our best-match model (100~m/s$^2$) results in a lower scale height, and thus, low-amplitude molecular features. Nevertheless, the correspondence between the data and our forward model is acceptable within the error bars, which is aided by the fact that this planet is expected to have a solar composition and is cloud-free within the observable pressure range \citep{Kreidberg2014}.

Because of its comparatively fast rotation, WASP-43~b has a limited heat redistribution, resulting in a relatively cold night side and morning terminator (Fig.~\ref{fig_wasp-43b}, panel a, and \cite{Carone2020}). This large day-night contrast results in some chemical gradients around 100~mbar, especially of methane. In general, we find a very good agreement between the chemical composition for a WASP-43~b-like planet (Fig.~\ref{fig_wasp-43b}) and another pseudo-2D chemical kinetics model for this planet \citep{Venot2020_wasp43b}. In the pressure range where the two models overlap, the only big difference is that the horizontal transport seems to be more efficient in the work by \cite{Venot2020_wasp43b}. This is explained by the horizontal wind speed they adopt: 4.6~km/s, compared to 2.9~km/s in this work. This substantial difference stems from different modelling assumptions in the 3D GCM simulation \citep[see][for a detailed comparison]{Carone2020}. An additional reason for discrepancies could be the different chemical networks used, namely the reduced \cite{Venot2020_network}-network used in this work and the \cite{Moses2013}-network used by \cite{Venot2020_wasp43b}.

In comparison with a 3D chemical advection model with simplified chemistry \citep{Mendonca2018_disequilibrium}, we find good agreement for \chem{H_2O}, \chem{CO} and \chem{CO_2}, and somewhat less for \chem{CH_4}. A possible reason could be that the night-side temperatures are higher in the study of \cite{Mendonca2018_disequilibrium}, resulting in a methane abundance that is already horizontally uniform, and lower than seen in Fig.~\ref{fig_wasp-43b}, panel b. No strong meridional advection is expected for this planet \citep{Mendonca2018_disequilibrium}.

\begin{figure*}
    \centering
    \includegraphics[width=0.99\textwidth]{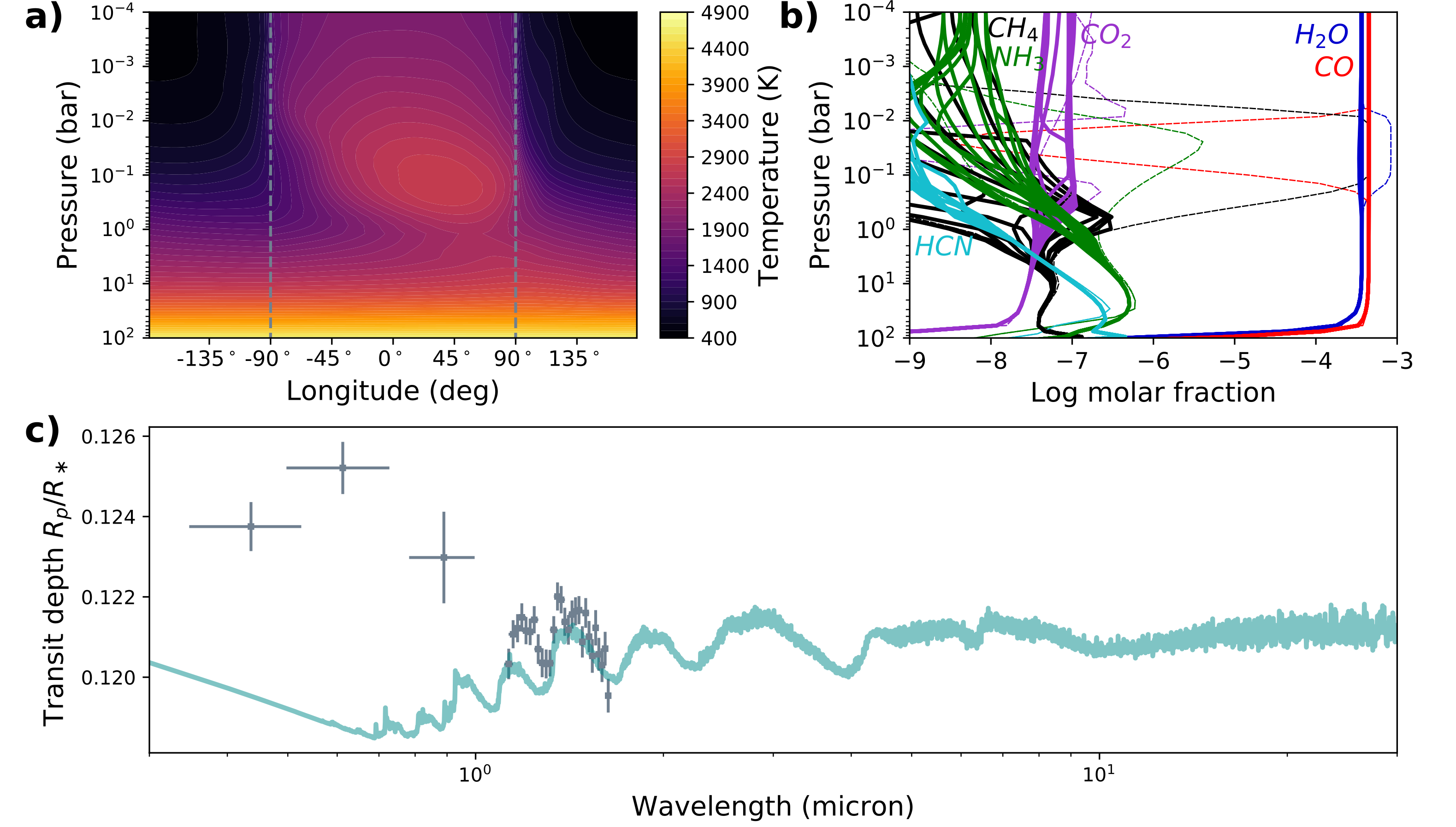}
    \caption{Physical diagnostics for an WASP-121~b-like planet. This grid model has $\Teff=2400$~K, $g=10$~m/s$^2$ and host star type F5. Plotted are the meridional mean temperature used in the pseudo-2D chemistry code (\textit{a}), the chemical composition (\textit{b}), and the synthetic transmission spectrum, in a comparison with observations (\textit{c}). The observational data (\textit{grey}) are from \citet{Evans2016}. The transmission spectrum is cloud-free, and does not contain Na/K opacities. In the chemistry plot, the equilibrium compositions of the morning and evening limb are plotted as thin dashed lines.}
    \label{fig_wasp-121b}
\end{figure*}

\subsubsection{WASP-121~b}

The ultra-hot Jupiter WASP-121~b has a mass of 1.183~$M_{\rm Jup}$ and a radius of  1.865~$R_{\rm Jup}$ \citep{Delrez2016}, and hence a surface gravity of 8.4~m/s$^2$. It is orbiting an F-type star on a comparatively short orbit of 1.3~days, giving rise to its high equilibrium temperature of $\sim$2400~K. In our grid, it is best matched by a model with $\Teff=2400$~K, $g=10$~m/s$^2$ and rotation period 1.13~days/host star type F5.

When the spectrum of our WASP-121~b-like atmosphere (Fig.~\ref{fig_wasp-121b}, panel c) is compared to observations \citep{Evans2016}, a discrepancy is seen in both the shape of the water band and the optical wavelengths, despite our good match to the physical temperature and gravity of WASP-121~b. The excess absorption has been interpreted as TiO- and VO-opacity, with potential contributions from FeH \citep{Evans2016} or a partial cloud cover at the morning terminator \citep{Parmentier2018}. These opacities -- typical for ultra-hot planets -- are not included in our homogeneous modelling approach, and we discuss the applicability of our grid to ultra-hot atmospheres in Section~\ref{sec_limitations}. 

Our atmospheric models for a WASP-121~b-like planet display a hot atmosphere, with a day-side temperature of about 2000~K, a poor heat redistribution, and cold night sides (Fig.~\ref{fig_wasp-121b}). This results in very low abundances of \chem{CH_4}, \chem{NH_3} and \chem{HCN} on the day side, and somewhat increased fractions at the night side. Horizontal quenching is, however, sufficiently quick, and these molecules never attain number fractions over $10^{-6}$. 

Since we did not include TiO, VO, or ionized-metal opacities in our models, we do not retrieve a strong day-side thermal inversion, and thus, the day-side temperatures in our model are too low to cause water dissociation, which occurs typically at temperatures above 2500~K \citep{Lothringer2018, Parmentier2018}. Hence, we find a constant water abundance throughout the atmosphere. Previous models mapping the 2D or 3D chemical composition of ultra-hot Jupiters have either invoked chemical equilibrium \citep{Parmentier2018} or only vertical mixing \citep{Molaverdikhani2020_hatp7b}, so the question whether water can exhibit horizontal gradients, or limb asymmetries, in ultra-hot atmospheres, remains to be answered. We note that in our hottest models with $\Teff=2600$~K and inefficient heat recirculation (M5-star host), the day-side does reach temperatures high enough for \chem{H_2O} dissociation. This is also visible in the chemistry models (Fig.~\ref{fig_2dchem}). In these cases, it appears that horizontal circulation is not efficient enough to completely mix the atmosphere to the water-depleted, day-side state. It is important to understand the modifying role of dynamical disequilibrium chemistry in molecular dissociation, as it has been shown that 3D effects near the limbs can bias retrievals of ultra-hot Jupiters \citep{Pluriel2020}.

\begin{figure*}
    \centering
    \includegraphics[width=0.99\textwidth]{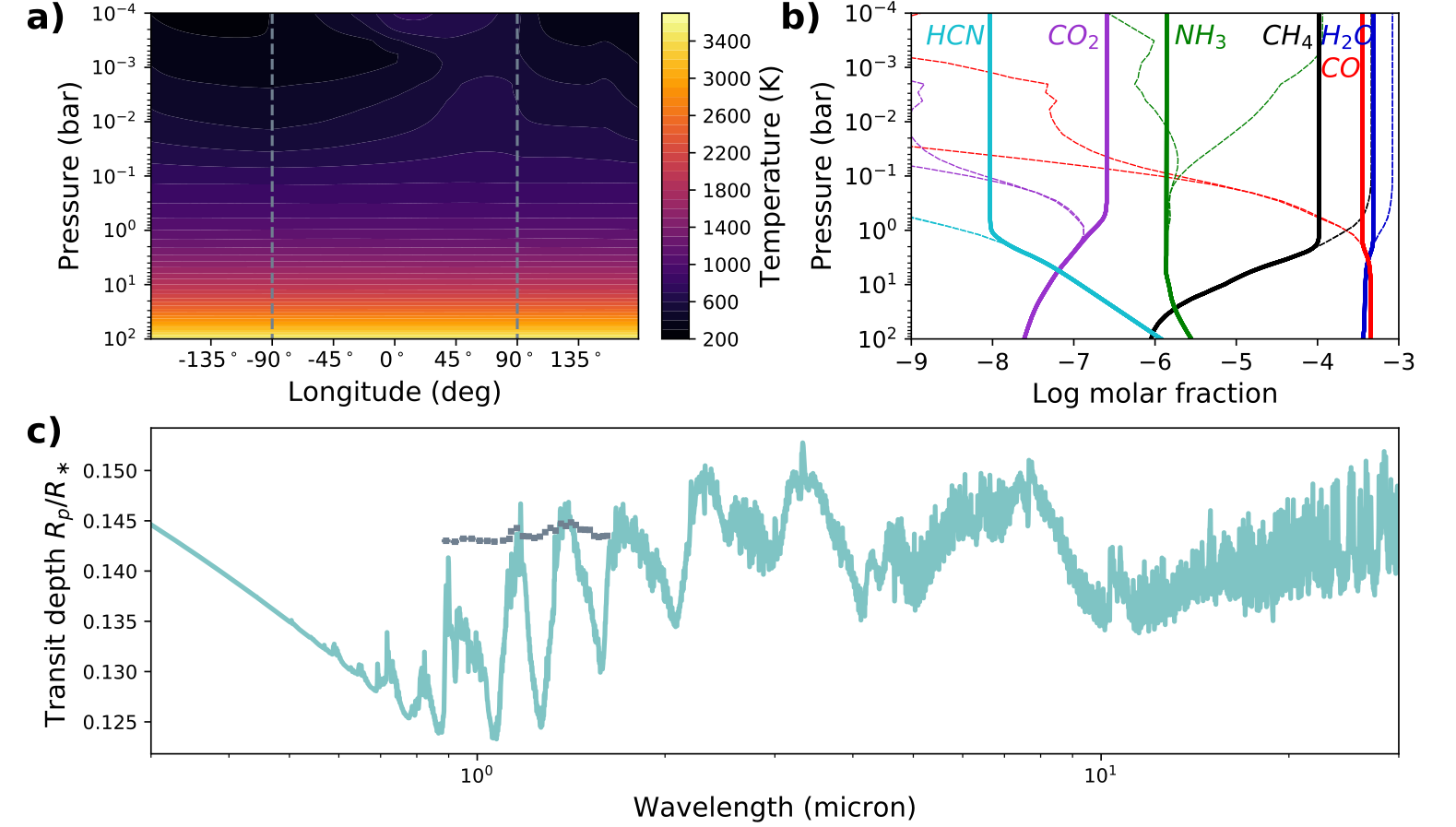}
    \caption{Physical diagnostics for an WASP-107~b-like planet. This grid model has $\Teff=800$~K, $g=1$~m/s$^2$ and host star type K5. Plotted are the meridional mean temperature used in the pseudo-2D chemistry code (\textit{a}), the chemical composition (\textit{b}), and the synthetic transmission spectrum, in a comparison with observations (\textit{c}). The observational data (\textit{grey}) are from \citet{Spake2018}. The transmission spectrum is cloud-free, and does not contain Na/K opacities. In the chemistry plot, the equilibrium compositions of the morning and evening limb are plotted as thin dashed lines.}
    \label{fig_wasp-107b}
\end{figure*}

\subsubsection{WASP-107~b}

The warm sub-Saturn mass planet WASP-107~b has a mass of 0.096~$M_{\rm Jup}$ and radius of 0.96~$R_{\rm Jup}$ \citep{Anderson2017, Piaulet2020}. This results in a very low gravity of 2.6~m/s$^2$. Furthermore, its equilibrium temperature is about $\sim$750~K, with an orbital period of 5.7~days. Hence, we opt to match it to a grid model with $\Teff=800$~K, $g=1$~m/s$^2$ and rotation period 4.25~days/host star type K5. Its low gravity and effective temperature make it a suitable case study for our low-gravity models with disequilibrium chemistry.

The synthetic transmission spectrum of our WASP-107~b-like model (Fig.~\ref{fig_wasp-107b}, panel c) does not correspond well to the \textit{HST} transmission spectrum presented by \cite{Spake2018}. The main reason for the discrepancy is a thick, opaque cloud or haze layer at altitudes of at most a few mbar \citep{Kreidberg2018}. This severely mutes the water absorption, which would normally be very extended because of the low gravity and high scale height of this planet. 

The chemical composition is homogeneous, in accordance with the dichotomy established in Section~\ref{sec_results_chemistry}. Additionally, the climate does not exhibit strong temperature variations (Fig.~\ref{fig_wasp-107b}). Because of vertical mixing, the atmospheric chemistry is CO-dominated under our assumption of solar metallicity and C/O ratio, although a considerable methane fraction is still present. The transmission spectrum still shows very prominent methane absorption. 

Interestingly, retrieval modelling by \cite{Kreidberg2018} concluded a depletion of methane for this planet. This could indicate a composition that is far removed from solar values, and/or a combination of a high internal temperature with deep vertical mixing \citep{Kreidberg2018, Fortney2020}. Our calculation of $\kzz$ yields $10^8$--$10^9$~cm$^2$/s for this WASP-107~b-like model, which should be high enough to quench methane at low abundances, as demonstrated for the comparable planet WASP-117~b \citep{Carone2021}. However, a low intrinsic temperature of 200~K has been adopted for all models in this work. Hence, we do not observe the reducing behaviour of deep quenching very strongly here. 

\subsection{Notable Detection Windows}\label{sec_windows}

\subsubsection{Observing methane}\label{sec_methane}

Methane is an important molecule to observe in exoplanetary atmospheres, since its abundance -- together with that of water -- can provide a handle on the C/O ratio, which can be linked to planet formation processes \citep{Oberg2011, Notsu2020}. For this reason, several attempts have been made to constrain the methane abundance in exoplanet atmospheres, using \textit{Hubble Space Telescope} WFC3 transmission spectroscopy. However, many of these studies have resulted in an absence of methane absorption, even in the atmospheres of cool planets with equilibrium temperatures below 1000~K \citep{Kreidberg2018, Benneke2019_GJ3470b, Carone2021}. In one instance, transmission spectroscopy has provided significant evidence for both water and methane absorption \citep{Chachan2019}. Additionally, in a different planet, with an estimated equilibrium temperature of at most 850~K, a methane detection has been reported using high-resolution spectroscopy \citep{Guilluy2019}, although a re-analysis of the data with a different line list has not resulted in a detection \citep{Gandhi2020_highres}. 

The difficulty of observing methane with transmission spectroscopy could be in part due to the largely overlapping absorption bands of water and methane in the \textit{HST}/WFC3 wavelength range \citep[see][for a discussion]{Bezard2020}. Moreover, since methane formation is more favourable in relatively cold environments, future methane observations with transmission spectroscopy will likely be plagued by cloud obscuration \citep{Crossfield2017, Gao2020, Taylor2020}. Nevertheless, it is instructive to discuss the detectable presence of methane in the cloudless grid presented in this work, taking into account disequilibrium chemistry through horizontal and vertical mixing.

In the medium-gravity atmospheric models of our grid, methane absorption bands become prominent at effective temperatures of 1200~K or lower. Methane is not necessarily the dominant carbon-bearing molecule in these models, but the \chem{CH_4} volume mixing ratio approaches or exceeds 10$^{-4}$ (Fig.~\ref{fig_2dchem}), which should indeed be above the detection threshold \citep[e.g.][]{Fortney2020}. All rotation rates at 1200~K show a high methane abundance with prominent absorption bands (Fig.~\ref{fig_many_spectra}), but the fast rotating models more so, because the quench point is situated at slightly lower pressures. In this temperature regime, meridional quenching, not considered in this work, could result in an additional increase in the methane abundance \citep{Drummond2020}. Even at 1400~K, methane has an opacity comparable to that of water in some distinct methane absorption bands at 2.3~$\mu$m, 3.3~$\mu$m and 7.8~$\mu$m. The former of those, however, is located between two water absorption bands, and may thus be difficult to distinguish from clouds. Hence, the latter appear to be more favourable for methane detections. Furthermore, we note that the distinct methane absorption bands diminish at effective temperatures of 600~K or lower (Fig.~\ref{fig_many_spectra}). At these low temperatures, the spectrum consists of numerous overlapping absorption bands of \chem{H_2O}, \chem{CH_4} and \chem{NH_3}. 

When we compare the transmission opacities of methane with our low- and high-gravity atmospheric models, we find that the low-gravity models display the same general trend as above, with methane becoming the dominant opacity source at temperatures equal or below 1200~K. At 1400~K, however, no methane absorption peaks at 2.3~$\mu$m, 3.3~$\mu$m or 7.8~$\mu$m are present. The high-gravity models, on the other hand, can show methane opacities comparable to those of water for relatively hot atmospheres even. The main cause is the comparably low heat redistribution (Fig.~\ref{fig_heat_redistribution}), and thus cool temperatures near the limbs (Fig.~\ref{fig_temperature_maps_appendix_g100}), which can give rise to a strong longitudinal dependence of the chemical composition, and methane formation on the night side and near the limbs (Fig.~\ref{fig_2dchem_appendix_g100}). Thus, for these very high-gravity cases, methane could be detectable at effective temperatures of 1400~K or hotter, depending on the rotation rate and horizontal mixing efficiency of the planet. However, it is unlikely that the effect of an increased temperature on the atmospheric scale height is enough to offset the adverse effect of gravity.

Thus, our pseudo-2D, cloud-free chemical kinetics models indicate that it
may be worthwhile to investigate planets with effective temperatures between 600~K and 1200~K for the presence or absence of methane absorption. The 3.3~$\mu$m absorption band, which will become accessible through \textit{JWST}, appears to be the most suitable. For some hot Jupiters with high densities ($g > 10$~m/s$^2$), it may even be valuable to employ transmission spectroscopy to hunt for methane originating from the night side. If night-side clouds and photochemical hazes are not too dense, methane could exhibit a relatively strong signal. We find that the rotation rate has some influence on the detectability of methane in the atmosphere, by raising or lowering the vertical mixing quench pressure. In this context, rapidly rotating planets are more prone to contain appreciable methane concentrations. 

We caution, however, that the \chem{CH_4} abundance is strongly impacted by the vertical mixing efficiency, which is not straightforward to determine accurately. Moreover, the internal temperature of the planet plays an important role in this analysis. In this work, an intrinsic temperature of
200~K is adopted for all planets, which is likely too low for inflated exoplanets, by several 100~K \citep{Thorngren2019, Sarkis2020, Fortney2020}.
The deep-atmosphere temperatures determine the vertical chemical equilibrium gradient. If a species' equilibrium abundance increases with altitude, a lower eddy diffusion will result in a higher disequilibrium abundance, and vice-versa. Furthermore, a steeper gradient will result in a higher sensitivity to the eddy diffusion coefficient. Indeed, it has been shown that atmospheric chemistry can be altered by the internal temperature \citep{Agundez2014_GJ436b, Fortney2020}. On top of that, the convergence of deep atmospheric layers is a well-known problem in GCM simulations \citep{Amundsen2016, Mayne2017, Carone2020, Wang2020, Showman2020}. As input temperatures for chemical disequilibrium studies are often derived from GCM simulations \citep[e.g.][and this work]{Agundez2014, Venot2020_wasp43b, Helling2019}, care must be taken that the deep ($p > 1$~bar) atmospheric temperatures are well-converged, or do not feed back into the chemistry through deep vertical mixing.  In this work, we have applied the recommendation laid out by \cite{Sainsbury-Martinez2019} to ensure that a suitable temperature profile is attained in the deep atmosphere \citep[see also the discussion in][]{Drummond2020}. We conclude that a detailed study of the planetary interior, and its impact on the chemistry in a 3D context, would be beneficial. Furthermore, \textit{JWST} observations of transiting exoplanets with effective temperatures below $\sim$1200~K will be instrumental in constraining the processes of methane quenching and interior heating, through vertical mixing of the deep atmosphere, or horizontal transport from the night side.

\subsubsection{Bump in CO$_2$ opacity}\label{sec_co}

\begin{figure}
    \centering
    \includegraphics[width=\columnwidth]{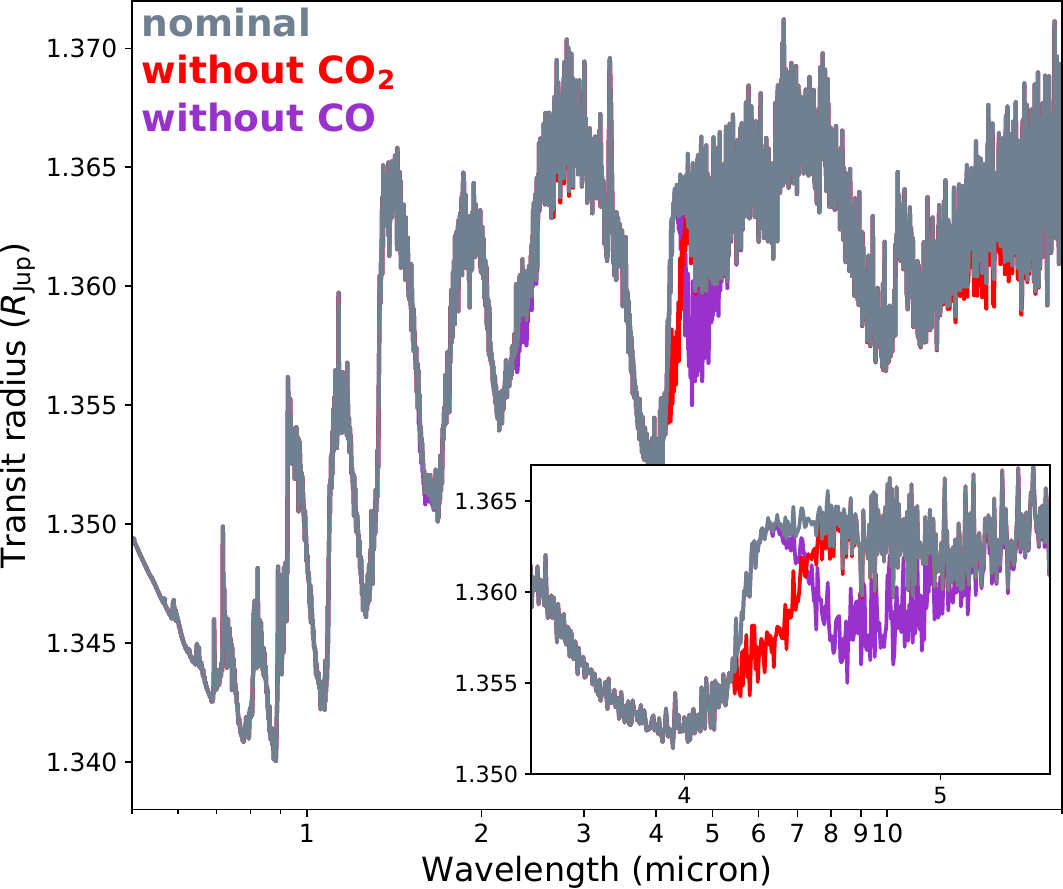}
    \caption{This transmission spectrum, corresponding to a simulation with $\Teff=1400$~K, $g=10$~m/s$^2$ and stellar type K5, shows a distinct opacity bump at 4.3~$\mu$m because of absorption by \chem{CO_2}. The inset shows the \chem{CO_2} absorption band in more detail.}
    \label{fig_CO2_opacity}
\end{figure}

Carbon dioxide is an important molecule in exoplanet atmospheres, because of its impact on habitability of rocky planets, as a greenhouse gas \citep[e.g.][]{Schwieterman2018}. In irradiated gaseous exoplanets, it can be used as a probe for the atmospheric metallicity and C/O ratio \citep{Zahnle2009, Moses2013, Fleury2020}. Furthermore, unlike methane, it is fairly unaffected by chemical disequilibrium (Fig.~\ref{fig_2dchem}). 

Although carbon dioxide only has a few absorption bands, the opacity in these bands can be relatively high, so that \chem{CO_2} dominates the opacity at 4.3~$\mu$m, even at low mixing ratios of $\sim$10$^{-7}$. This is illustrated for a medium-gravity model with $\Teff=1400$~K in Fig.~\ref{fig_CO2_opacity}. The opacity bump at 4.3~$\mu$m by \chem{CO_2}, and to a lesser extent \chem{CO}, is also very distinctive in hotter atmospheres (Fig.~\ref{fig_many_spectra}). When assuming a G5-host star radius, we estimate the transit depth difference associated with this \chem{CO_2} transmission feature to be about 65~ppm or 1~ppm, for our low ($g=1$~m/s$^2$) and medium ($g=10$~m/s$^2$) gravity models respectively, making it an intriguing, albeit challenging, observing window for this molecule.

\subsubsection{Outlook of observing ammonia}\label{sec_ammonia}

\begin{figure}
    \centering
    \includegraphics[width=\columnwidth]{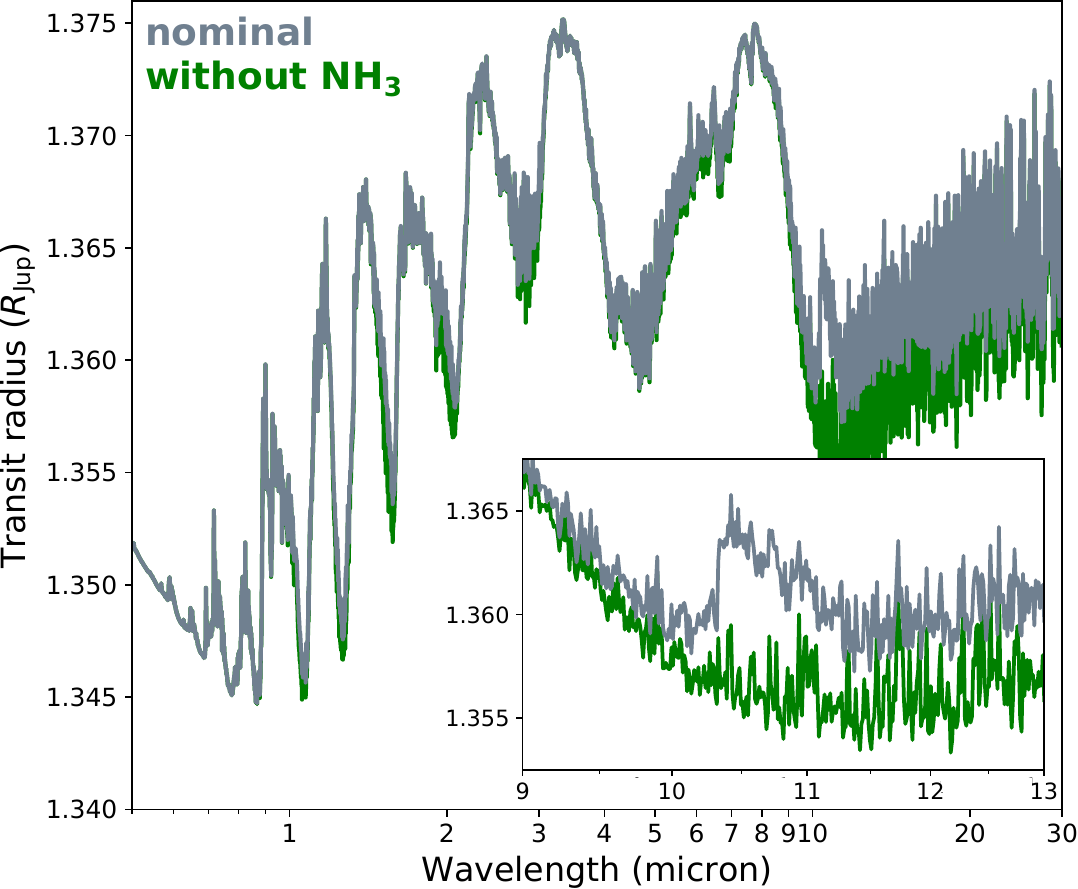}
    \caption{In this transmission spectrum, corresponding to a simulation with $\Teff=1200$~K, $g=10$~m/s$^2$ and stellar type M5, a clear opacity band of \chem{NH_3} at 10~$\mu$m can be seen. The inset shows a detail of the \chem{NH_3} absorption feature.}
    \label{fig_NH3_opacity}
\end{figure}

Ammonia (\chem{NH_3}) has an interconversion relation with nitrogen gas (\chem{N_2}), so that, in chemical equilibrium, the former is expected to occur in colder, and the latter in hotter exoplanet atmospheres \citep[e.g.][]{Venot2012, Moses2014}. Together with the photochemically produced \chem{HCN}, these molecules are expected to be among the main nitrogen-bearing molecules in exoplanet atmospheres. Up to now, detections of nitrogen-bearing molecules in transiting exoplanet atmospheres have been scarce and tentative \citep{Tsiaras2016, Macdonald2017}. Still, there is much potential in nitrogen-detections, since these molecules should be good indicators of disequilibrium chemistry (both through vertical mixing and through photochemistry). Furthermore, information on the atmospheric C/N or N/O ratio would provide valuable constraints for planet formation theories \citep[e.g.][]{Mordasini2016, Piso2016, Cridland2020}. 

Some of the spectra in the cool parameter space of our grid feature an \chem{NH_3} opacity peak in the mid-infrared, at 10$\mu$m (Fig.~\ref{fig_NH3_opacity}), making it a potential target for spectroscopy with \textit{JWST}/MIRI. Because the ammonia peak at 10~$\mu$m is quite distinctive, and located in a region with only pseudo-continuous water opacity, it is visible in a large range of synthetic spectra, from 1400~K to 400~K (Fig.~\ref{fig_many_spectra}). In a behaviour analogous to methane (Section~\ref{sec_methane}), the relative amplitude of the 10~$\mu$m ammonia feature appears to increase with gravity. However, the absolute amplitude of the feature, and thus its detectability, decreases with gravity, so medium- to low-gravity planets would still be more suitable to detect ammonia. For the model shown in Fig.~\ref{fig_NH3_opacity}, we estimate a transit depth difference of 1~ppm between the nominal spectrum and one without \chem{NH_3}. For the low-gravity case, the feature amplitude is estimated at 106~ppm. Finally, we did not find a strong dependence of ammonia absorption on the rotation period.

\subsection{Impact of Deep Atmospheric Quenching}
\label{sec_deep_quenching}



In the case of deep quenching ($p>1$~bar), both the thermal structure of the atmosphere at depth and the vertical mixing strength at depth are critical in setting the disequilibrium composition higher up in the atmosphere. This is highlighted through our comparison of \hdtwenty{} (Fig.~\ref{fig_hd20}) and \hdeighteen{} (Fig.~\ref{fig_hd18}) with earlier simulations using a different thermal structure and $\kzz$-profile \citep{Agundez2014}. As pointed out in Section~\ref{sec_case_studies}, discrepancies in these cases are often caused by a different temperature structure at depth, resulting in a different chemical equilibrium composition at the vertical quenching pressure. As such, any uncertainties in the deep thermal profile could be propagated to the chemical composition in general.

If the thermal evolution in the deep part of the atmosphere is adiabatic, the associated vertical temperature gradient tends to be steep, and a small shift in the quenching pressure could result in a big change to the chemistry. Moreover, if the quenching happens near a chemical transition, like in our \hdeighteen{} simulation (Fig.~\ref{fig_hd18}), the impact on the chemical composition is exacerbated. Therefore, it appears that great care should be taken in the selection of both the temperature profile, and the vertical mixing strength at depth, if the quenching pressure is high. We now discuss the determination of these two factors.

\subsubsection{Vertical Mixing in the Deep Atmosphere}

Giant irradiated exoplanets are expected to consist of a stably stratified radiative atmosphere, overlying a deep convective atmosphere \citep[e.g.][]{Marley2015}, with a radiative-convective boundary situated roughly between 1~bar and a few 100~bar, depending on the planet's gravity and incident irradiation \citep{Thorngren2019}. In the radiative part, which includes the photosphere, vertical mixing is expected to be dominated by atmospheric circulation or atmospheric waves, which can be approximated as a 1D diffusion process with coefficient $\kzz$ \citep{Parmentier2013, Charnay2015, Komacek2019}. The eddy diffusion coefficient generally increases with altitude (Fig.~\ref{fig_kzz_profiles}), and hence reaches its lowest values when approaching the deep atmosphere ($p \gg 1$~bar). Convection in the deep atmosphere, however, ensures efficient mixing, and the value of $\kzz$ can be expected to increase again as the vertical mixing becomes dominated by small-scale turbulent convection, which is not captured by GCMs.

In the deep convective region, mixing length theory can be used to estimate the eddy diffusivity ($\kzz \sim wL$) \citep[e.g.][and references therein]{Wang2015, Zhang2020}, resulting in values around $10^9$~cm$^2$/s for Jovian planets. However, knowledge of the vertical velocities and/or the thermal structure, in addition to a good estimate of the mixing length \citep{Smith1998, Bordwell2018} are still required to determine $\kzz$ in the deep regions of giant exoplanets.

It should thus be clear that is not trivial to determine a vertical mixing profile, especially in the 3D context of an irradiated exoplanet. The pressure of the radiative-convective boundary is location-dependent \citep[][and our Fig.~\ref{fig_different_adiabats}]{Rauscher2014_andShowman}, which makes it more likely that on the cold night side quenching pressures fall within the convective zone. Compressible hydrodynamic simulations suggest that convective overshooting is limited \citep{Freytag2010}, so $\kzz$ could change quickly close to the radiative-convective boundary, as some previous chemical kinetics studies have assumed \citep[e.g.][]{Moses2011, Venot2020_wasp43b}. Furthermore, convectively driven, vertically propagating gravity waves could provide an additional source of vertical mixing above the radiative-convective boundary \citep{Medvedev2019, Bordwell2020}, but resolutions of typical exoplanet GCMs are too low to capture these waves. The potentially rapid increase in vertical mixing efficiency in the deep interior is not taken into account in our $\kzz$-profiles, but could contribute to the prevalence of deep quenching, and the associated chemical uncertainties. Future hydrodynamic studies focusing on the transition between the deep convective zone and the stratified radiative layers above, would provide valuable constraints for exoplanet disequilibrium chemistry.

\subsubsection{Temperatures in the Deep Atmosphere}

The assumed temperature profile in the deep part of the atmosphere is crucial to the chemistry, not only locally, but throughout the vertical extent of the atmosphere if quenched. However, unlike the 3D temperature structure at lower pressures, which can be relatively well constrained with a GCM, the deep thermal structure is often ill-constrained by GCMs, because it evolves very slowly \citep{Carone2020, Mendonca2020, Wang2020, Showman2020}. This can result in temperature-profiles that are only partially converged, featuring a small thermal inversion in the deep ($p\approx10$~bar) layers \citep[e.g.][]{Agundez2014, Amundsen2016, Venot2020_wasp43b, Drummond2020}. Although it can be argued in some cases that the observable atmosphere is not affected by the deep thermal evolution, this does not appear to be true in general \citep{Carone2020}. Additionally, we stress that care should be taken when unconverged temperature profiles are used as input for disequilibrium chemistry studies, because of the potential for deep quenching.

Simulations by \cite{Sainsbury-Martinez2019} suggest that GCMs tend to converge to a hot, deep adiabatic temperature profile when evolved up to a full steady-state. Our temperature plots indeed show that an adiabatic gradient is reached at depth (Fig.~\ref{fig_LPT_profiles}). Although this fact is reassuring, there is no guarantee that this adiabat is physically suitable. \cite{Fortney2020} have demonstrated the sensitivity of the chemical disequilibrium abundances to the interior temperature. Irradiation \citep{Thorngren2019}, tidal heating \citep{Agundez2014_GJ436b, Fortney2020}, radius inflation \citep{Tremblin2017}, Ohmic heating \citep{Batygin2011}, or the planet age \citep{Fortney2020} all impact the interior temperature, and thus potentially also the chemical abundances if quenching occurs deeply.

It this context, a self-consistent model coupling the atmospheric circulation with the deep interior evolution of giant exoplanets in a 3D framework would be very beneficial. Such a coupling is needed in order to study the interface near the radiative-convective boundary, and gauge the impact of the interior temperature on the (observable) atmospheric layers above. 
Finally, we note that the deep interior is not totally inaccessible to observations. Indeed, the detection of certain refractory species in the gas-phase at high altitudes could place valuable constraints on the physical conditions below \citep{Sing2019}.


\subsection{Implications for Brown Dwarf -- White Dwarf Binaries}\label{sec_bd-wd}

By systematically applying our grid parameter combinations of effective temperature and stellar type, under the assumption of synchronous rotation, we have produced numerous atmospheric models with rotation rates below 10~hours (residing in the `lower left' corner of the grid space, Table~\ref{tab:grid}, Fig.~\ref{fig_temperature_maps} and following). These increasingly short rotation periods are not typical for giant exoplanets, of which the fastest rotators have orbital periods of $\sim$1~day, whereas these models have rotation periods of the order of hours. However, because of their high equilibrium temperatures and short rotation periods, these ultra-fast rotating models have interesting real-life analogues, namely brown dwarf -- white dwarf (BD--WD) binaries, a rare class of objects with similarities to both brown dwarfs and ultra-hot Jupiters \citep{Lothringer2020}. This is especially the case for our high gravity set of models, with $g=100$~m/s$^2$. In this section, we highlight the applicability of our models to BD--WD binaries.

Two important distinctions between our modelling approach, stemming from a self-consistent extrapolation of hot Jupiter modelling, and BD--WD binaries are: 
\begin{itemize}
    \item the high intrinsic temperatures in irradiated brown dwarfs;
    \item the ionizing (UV-)radiation from the white dwarf host.
\end{itemize} Indeed, whereas we have adopted an intrinsic temperature of 200~K in our models, values of up to 2000~K have been retrieved for irradiated brown dwarfs \citep{Lothringer2020}. This could impact both the circulation \citep{Carone2020}, as well as the chemistry \citep{Fortney2020} when the vertical quench level is sufficiently deep. Furthermore, 
as explained in Section~\ref{sec_limitations}, the incorporation of longitude-dependent photochemistry could be important for these objects in the future. 

Regarding the atmospheric circulation, we compare our results with recent previous studies of BD--WD binary climates \citep{Lee2020, Tan2020_WD-BD}. In general, we find qualitatively agreeing circulation regimes with a large day-night temperature contrast, a very thin, eastward zonal-mean equatorial jet stream, and multiple smaller jet structures at high latitudes. Similarly to \cite{Lee2020}, we obtain retrograde flow at the equatorial day-side, resulting in high local wind speeds near the morning terminator, which is typical for fast rotation and high gravity \citep{Carone2020}.

From our chemistry modelling, it appears that BD--WD binaries should feature a chemical regime that is only partially zonally quenched. A certain contamination from the day side can be expected, but our models suggest that advection by the zonal mean circulation is not efficient enough to remove all horizontal chemical gradients. Depending on local thermal conditions, these objects could exhibit zonally changing water absorption, if horizontal quenching of thermally dissociated \chem{H_2O} is inefficient. Thus, besides the temperature itself, it may be needed to take into account the local chemical composition to interpret phase curve observations \citep{Casewell2015}. We stress that our models do not include a stratosphere or photochemistry, both of which could be present in BD--WD binaries \citep{Lothringer2020}, and could impact the circulation and chemical composition, as well as their zonal dependence. 

\section{Methodology Limitations and Future Work}\label{sec_limitations}

\subsection{Post-Processing Models}\label{sec_limitations_postproc}

The connection between the radiative transfer calculations with \textit{petitCODE} and the 3D climate model \textit{MITgcm} is made with the so-called Newtonian cooling (also called Newtonian relaxation) framework. It comprises a parametrized atmospheric forcing, lacking self-consistent feedback of dynamical processes to the radiative heating and cooling terms. This simplified radiation treatment is ideal for dynamical studies, but less suitable for a direct comparison with observational fluxes \citep{Showman2020}.
Comparisons between GCMs coupled with Newtonian cooling or more robust radiative schemes yield qualitatively similar outcomes, but with quantitative differences. The day--night temperature contrast was found to be reduced (compare \cite{Showman2009, Amundsen2016} with \cite{Showman2008, Mayne2014}), especially at high altitudes, where radiative time-scales are short.
Furthermore, the lack of a self-consistent radiation can sometimes result in the appearance of a weak, dynamically induced temperature inversion. This can be attributed to an insufficient coupling between radiation and dynamics at the hot day side of irradiated exoplanets \citep{Carone2020}. Indeed, such feature disappears when the output of a GCM with Newtonian relaxation \citep{Showman2008} and with self-consistently coupled radiation \citep{Showman2009} are compared.

Nevertheless, Newtonian cooling has been used widely for parameter and ensemble studies of exoplanet climates, as well as detailed dynamical studies that benefit from a parametrized radiative time-scale, since it is faster and more readily interpreted than a full radiation-coupled climate model \citep[see e.g.][]{Showman2008, Rauscher2010, Mayne2014, Zhang2018_and_showman_I_fastrotating, Zhang2018_and_showman_II_tidallylocked, Sainsbury-Martinez2019, Carone2020}. Moreover, in a comparison made between a suite of models with Newtonian relaxation \citep{Komacek2016, Zhang2017} and with a more sophisticated double-grey radiation scheme \citep{Komacek2019}, it was noted that the results yielded by those simulations agree qualitatively, as well as quantitatively by up to a factor of $\sim$2 over a large parameter space. This motivates our choice for Newtonian relaxation as a mechanism to consistently investigate a large grid of planetary atmospheres.

Furthermore, employing a fixed background temperature for the chemical kinetics calculations neglects feedback which could have taken place in a coupled dynamical-chemical model. Two of such feedback mechanisms that could be important, come to mind. First, chemical heat sources or sinks could impact the temperature because of the endo-/exothermic nature of chemical reactions. Second, the presence of chemical elements can influence the temperature via the radiation field through radiative heating and cooling. The impact of this feedback mechanism has been studied in 1D for two hot Jupiters \citep{Drummond2016} and a hot Neptune \citep{Agundez2014_GJ436b}. 
Both of these studies concluded that temperature differences of up to 100~K are possible for cases of strong chemical disequilibrium, in comparison with post-processed chemistry models. A similar magnitude in temperature variation was found by \cite{Steinrueck2019}, who used parametrized disequilibrium chemistry to investigate the feedback to the radiation and temperature field in a 3D GCM. The uncertainties associated with the Newtonian-cooling GCM simulations of this work are about $\sim$100~K as well. Therefore, we do not expect the post-processing methodology that was used in this work to undermine our conclusions. Still, it would be beneficial if the impact of consistent chemistry-, dynamics- and radiation-coupling could be investigated in a broader parameter space in future studies.

\subsection{Ultra-Hot Jupiters}\label{sec_limitations_ultrahot}

There are many indications that ultra-hot Jupiters, highly irradiated exoplanets with day-side temperatures above $\sim$2200~K, are emerging as a distinct class of planets \citep{Baxter2020}, featuring a range of physical processes that are not included in this study. Indeed, ultra-hot Jupiters are expected to have thermal inversions (stratospheres) on the day side, caused by atomic and ionized metal absorption, and/or TiO and VO absorption \citep{Lothringer2018, Arcangeli2018, Parmentier2018}. These opacities can give rise to temperature excesses of 1000~K. Thermal inversions, along with the detection of species that cause them, is still a highly active topic of research \citep[e.g.][]{Piette2020}.
Furthermore, molecular dissociation on the day side and recombination on the night side of these planets is expected to constitute a substantial part of the heat transport budget \citep{Bell2018}, thereby heating the night side of these planets and reducing the day-night temperature \citep{Tan2019, Keating2019}.

These physical effects, which are now in the process of being understood, make it difficult to interpret atmospheric models that span a large range of equilibrium temperatures. More concretely, by excluding thermal inversions and hydrogen recombination, our 3D climate models are expected to have both cooler day sides and cooler night sides than what would be considered realistic, based on the current known ultra-hot Jupiter population. As of yet, it is unclear whether the inclusion of strongly absorbing opacities and hydrogen recombination would increase or decrease the day--night temperature contrast compared to the models presented here. However, recent phase curve observations of ultra-hot Jupiters suggest that the overall effect would be a day--night contrast that is lower than obtained with the simplifications assumed in this work \citep{Mansfield2020}. Nevertheless, these simplifying assumptions were made in favour of having a more continuously varying suite of atmospheric models to investigate, and to build up a hierarchical framework from which more complex models can draw. 

\subsection{Pseudo-2D Framework}
The pseudo-2D framework used in this work has the obvious disadvantage that dynamics is only taken into account in a parametrized or simplified way. First, in the vertical direction, the widely-used but ill-determined $\kzz$ parameter is used. By using scaling relations \citep{Komacek2019}, we have applied informed estimates of the eddy diffusion coefficient, based on the global vertical wind speed and zonal wind advection, over a large parameter space. However, the vertical mixing could have local efficiency extrema, such as vertical advection `chimneys' \citep{Parmentier2013, Zhang2018_and_showman_II_tidallylocked, Komacek2019, Carone2020, Steinrueck2020}, which are not captured in this approach. Since we have established that the chemical composition, in some cases, depends strongly on an accurate determination of the quenching pressure, the ability to make precise predictions of the eddy diffusion coefficient over a wide range of exoplanets will remain important in the future.

Next, in the zonal direction, a uniform, unidirectional wind advection is assumed. This assumption is valid for most of the climates in this grid, as discussed in Section~\ref{sec_results_climates}. However, the assumption of a uniform zonal wind restricts the area under scrutiny to the equatorial region, since it is expected to break down in the high-latitude regimes of synchronously rotating exoplanets. As observational data becomes more accurate, taking into account the full 3D geometry of exoplanet transits may become necessary \citep{Caldas2019, Pluriel2020}. 

Last, the meridional direction is absent in our framework. Meridional chemical advection has been studied with a GCM coupled to simplified chemical relaxation \citep{Mendonca2018_disequilibrium, Drummond2018_HD209458b, Drummond2018_HD189733b}, as well as a fully coupled 3D chemical kinetics models \citep{Drummond2020}. We have discussed these research results in our case studies of \hdtwenty{}, \hdeighteen{} and WASP-43b (Section~\ref{sec_hd20}, \ref{sec_hd18} and ~\ref{sec_wasp43b} respectively). In only one case -- \hdeighteen{} -- meridional advection has been found to significantly impact the chemical composition. Incidentally, this is the planet with the lowest equilibrium temperature. Thus, it remains an open question if this finding may be extrapolated, and if meridional quenching plays an important role in the chemistry of all cool exoplanet atmospheres.  

Despite these simplifying assumptions, solving chemical kinetics in a pseudo-2D framework has the obvious advantage of being reasonably fast compared to 3D GCMs with fully coupled chemistry. This advantage, combined with its more parametrized nature, allows for broad parameter studies, such as this work and \cite{Moses2021}. Potentially, detailed microphysical cloud modelling \citep[e.g.][]{Helling2019} could also benefit from a pseudo-2D framework to incorporate simplified atmospheric dynamics, since cloud formation and gas-phase chemistry are intrinsically coupled. Considering that more observations of phase curves and potential limb asymmetries will become available in the \textit{JWST} and \textit{Ariel} era, pseudo-2D chemical kinetics will remain a useful tool to interpret and investigate these spatially resolved data.

\subsection{Photochemistry}

Photochemistry is a potentially important chemical disequilibrium effect that is currently not taken into account in this work. Photochemistry can cause photolysis, molecular dissociation in the upper atmospheres of exoplanets and brown dwarf--white dwarf binaries, leading to the reduction of some molecules, such as \chem{CH_4}, and the increase of others, such as \chem{HCN} \citep{Moses2011, Venot2012}. Chemical kinetics in combination with the pseudo-2D framework applied in this work is well equipped to take into account the changing irradiation angle with longitude \citep{Agundez2014, Venot2020_wasp43b}.
This is especially the case for the self-consistent variation of the host star type employed in the grid setup of this work. Recently, the need for such photochemical modelling has been demonstrated for the hot Jupiter HAT-P-41~b \citep{Lewis2020}. We thus aim to include photochemistry in a detailed follow-up study.

\section{Conclusions}
\label{sec_conclusions}

In this work, we have synthesized a large grid of atmospheric models for synchronously rotating, irradiated giant exoplanets. Our cloud-free models include the three-dimensional climate of the planet, together with two-dimensional information about the chemical composition of the atmosphere. From these models, we have synthesized transmission spectra to assess the potential impact on observational data. These simulations were conducted for temperatures ranging from 400~K to 2600~K, gravities between 1~m/s$^2$ and 100~m/s$^2$, and for different rotation rates. We have included chemical disequilibrium effects in the form of vertical mixing, parametrized by the eddy diffusion coefficient $\kzz$, and horizontal mixing, approximated as uniform eastward advection. Our aim was to explore the chemical diversity over a wide range of effective temperatures, gravities and rotation rates, with a specific focus on the role of vertical and horizontal mixing.

Our main finding is that a dichotomy is expected in the chemical compositions of exoplanet atmospheres. Cool planets with effective temperatures below 1400~K are expected to be chemically homogeneous, whereas hot planets with effective temperatures above 1400~K show longitudinal changes in their composition. The established homogeneity in cool planets is caused by deep vertical mixing, where species are quenched before chemical equilibrium leads to strong longitudinal gradients. Furthermore, through a time-scale analysis we found that for most cases, the zonal advection is more efficient than vertical mixing at a pressure of 1~bar. In hotter planets with effective temperatures above 1400~K, zonal advection is not efficient enough to completely remove longitudinal chemical gradients, and colder regions, like the night side and morning terminator, tend to be chemically contaminated by air advected from the day side. These longitudinal gradients, mainly of \chem{CH_4}, \chem{NH_3} and \chem{HCN}, are exaggerated with increasing gravity.

The rotation rate was found to impact the chemical composition in multiple ways. A high rotation rate leads to a reduced day-to-night heat redistribution, a slower horizontal wind advection, and lower vertical mixing efficiencies. This results in bigger horizontal variations of the chemical composition, and in some cases, a shift in the \chem{CO}--\chem{CH_4} balance. In the latter case, the impact of the rotation rate on the transmission spectrum could be considerable, but in general, it is minor.

By adopting a hierarchical approach in composing the synthetic transmission spectrum, we established that the inclusion of a three-dimensional temperature, informed by a GCM, has the biggest impact on the chemical composition at the exoplanet limbs, and on its transmission signature. Chemical disequilibrium effects are of secondary importance. For cool planets ($\Teff < 1400$~K), horizontal advection appears to be unimportant if vertical quenching occurs deep enough. For hotter planets, on the other hands, horizontal advection plays an important homogenizing role, as the chemical composition is set close to that expected on the day side of the planet.

Differences in the atmospheric structure at the morning and the evening terminator lead to different contributions of both limbs to the exoplanet transit spectrum. For a wide parameter space, we confirm the conclusion of \cite{Agundez2014} that the temperature, rather than the chemical composition, is the main cause for transit depth asymmetries between the morning and evening limbs. Furthermore, the maximal transit depth difference between the two limbs was found to increase with the effective temperature of the atmosphere. We estimate that, for a cloud-free planet, featuring a low gravity, a hot atmosphere and a slow rotation rate, limb differences could be detectable with \textit{James Webb Space Telescope} ingress and egress transit observations.

Finally, we note some of the limitations of our modelling setup, which could be improved upon in future studies. It is likely that the day--night temperature contrast in our hottest planets ($\Teff$ > 2000~K) is overestimated, since we did not include hydrogen dissociation and recombination feedback in our climate models. GCMs with detailed radiative transfer, including some high-opacity species omitted in this work, 
would be needed to more accurately constrain the thermal structure for these ultra-hot planets. Two other potentially important effects, that are not included in our setup, are meridional advection and photochemistry. Regarding the former, it remains to be seen to what degree, and in which planets, meridional quenching can impact the chemical composition. Regarding the latter, we aim to investigate the impact of photochemistry over a broad parameter range in a detailed follow-up study.
Despite these modelling limitations, we generally find reasonable agreement between the atmospheric models presented in this work, and previous studies \citep{Agundez2014, Mendonca2018_disequilibrium, Venot2020_wasp43b, Drummond2020}. We find that the existing discrepancies are often caused by differences in the temperature structure and $\kzz$. This sensitivity highlights that care should be taken in the derivation of the atmospheric thermal structure, especially when vertical mixing is efficient, and the quenching pressures are high \citep{Fortney2020}. In the future, it would be beneficial to investigate the interior temperature impact on the observable atmosphere in a 3D context.  

\section*{Acknowledgements}

We wish to thank Maria Steinrueck, William Pluriel and Tiziano Zingales for helpful discussions. We also thank the anonymous referee for a very helpful review that led to the improvement of this manuscript.
RB is a PhD~fellow of the Research Foundation -- Flanders (FWO). 
LD acknowledges support from the FWO research grant G086217N. 
LC acknowledges support by the DFG grant CA 1795/3.
OV thanks the CNRS/INSU Programme National de Plan\'{e}tologie (PNP) and CNES for funding support.
MA acknowledges support from Spanish MICIU through grants RyC-2014-16277, AYA2016-75066-C2-1-P, PID2019-106110GB-I00, and PID2019-107115GB-C21.
PM acknowledges support from the European Research Council under the European Union's Horizon 2020 research and innovation program under grant agreement No.~832428.
The computational resources and services used in this work were provided by the VSC (Flemish Supercomputer Center), funded by the Research Foundation Flanders (FWO) and the Flemish Government -- department EWI.


\section*{Data Availability Statement}

The data underlying this article will be shared on reasonable request to the corresponding author.



\bibliographystyle{mnras}




\appendix

\section{Assessment of Model Convergence}\label{sec_appendix_convergence}

In order to ensure that the models presented in this work are converged to a steady-state, in which they do not exhibit substantial, non-cyclic evolution, diagnostic parameters have been plotted as a function of simulation time for a typical general circulation model (Fig.~\ref{fig_GCM_convergence}) and chemical kinetics simulation (Fig.~\ref{fig_chem_convergence}).

\subsection{GCM Convergence}

\begin{figure}
    \centering
    \includegraphics[width=\columnwidth]{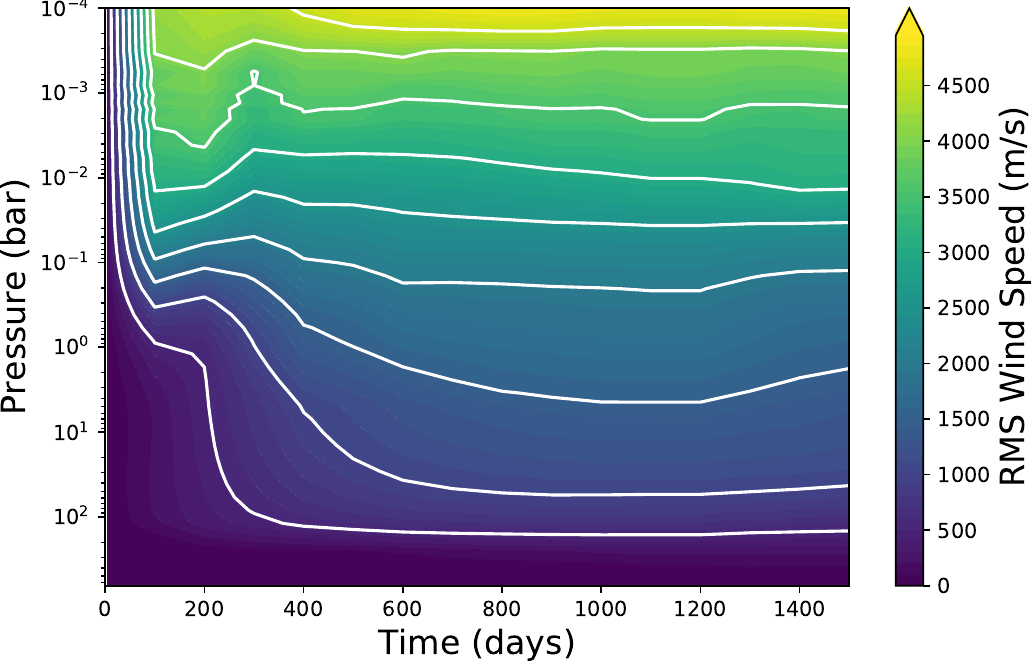}
    \caption{The root mean square (RMS) wind speed on isobars of this 3D GCM simulation ($\Teff=1400$~K, $g=10$~m/s$^2$, G5) shows a quick evolution in the first 400 days of the numerical integration, before attaining a steady-state. The RMS wind speed is weighted with the cosine of the latitude, and horizontally averaged on isobars. The simulation is evaluated after 1500~days and averaged over 100~days.}
    \label{fig_GCM_convergence}
\end{figure}

The convergence rate of irradiated exoplanet GCMs is a long-standing subject of scrutiny. Especially the deep atmospheric layers ($p > 10$~bar), which have a comparatively long radiative time-scale, have been noted to evolve very slowly \citep[e.g.][]{Amundsen2016}, leading to impractically long computation time requirements \citep{Wang2020, Mendonca2020}. Moreover, it has been shown that, in some cases, the deep atmospheric evolution can have a dynamical feedback on the observable atmospheric layers above \citep{Carone2020}. In order to combat the slow spin-up times of the deep atmospheric layers, \cite{Carone2020} have prescribed an accelerated forcing for these layers on to a pre-computed adiabatic temperature profile. A different method, that has been proposed by \cite{Sainsbury-Martinez2019}, is to initialize the temperature structure with a hot adiabat. We have employed the latter method in this work. In addition to the basal drag applied in our models, this serves to stabilize and spin up the atmosphere on a time-scale of $\sim$1000 (Earth) days, as can be seen in Fig.~\ref{fig_GCM_convergence}.

\subsection{Chemical Kinetics Convergence}

The pseudo-2D chemical kinetics model used in this work also relies on a convergence of the initial profile to a steady-state. The model is initialized with a vertically quenched chemistry profile corresponding to the substellar point, following the recommendation of \cite{Agundez2012}, after which it is run for 100 rotations. We demonstrate that 100 rotations are sufficient to ensure the convergence for the models in this grid. 

The total simulation time of the models in our grid varies depending on the zonal wind speed $u$ (see Fig.~\ref{fig_zonalwind_plots}, \ref{fig_zonalwind_plots_appendix_g1}, \ref{fig_zonalwind_plots_appendix_g100}) as $t = 100 \times \frac{2 \pi R_p}{u}$, with $R_p$ the fixed planet radius of 1.35~R$_{\rm Jup}$. Hence, models with fast equatorial jet streams are run for shorter times than models with slow equatorial flow. As a consequence, generally, models with high effective temperatures and slow rotation rates are run for $\sim$10$^7$~s, while the coolest models are run for $\sim$10$^{10}$~s. A formal requirement for convergence is that the total simulation time is higher than the vertical transport time-scale of the model. Since $\kzz$ scales with the effective temperature (Section~\ref{sec_verticalwind_kzz}), the simulation times roughly match the maximal vertical advection times needed for convergence, namely $10^6$--$10^8$~s for the high-, $10^6$--$10^{10}$~s for the medium-, and $10^7$--$10^{10}$~s for the low-gravity models. However, since we start the pseudo-2D model from a converged 1D model at substellar point, the initial composition is already close to the expected steady-state, and thus, the actual convergence should be faster.

\begin{figure}
    \centering
    \includegraphics[width=\columnwidth]{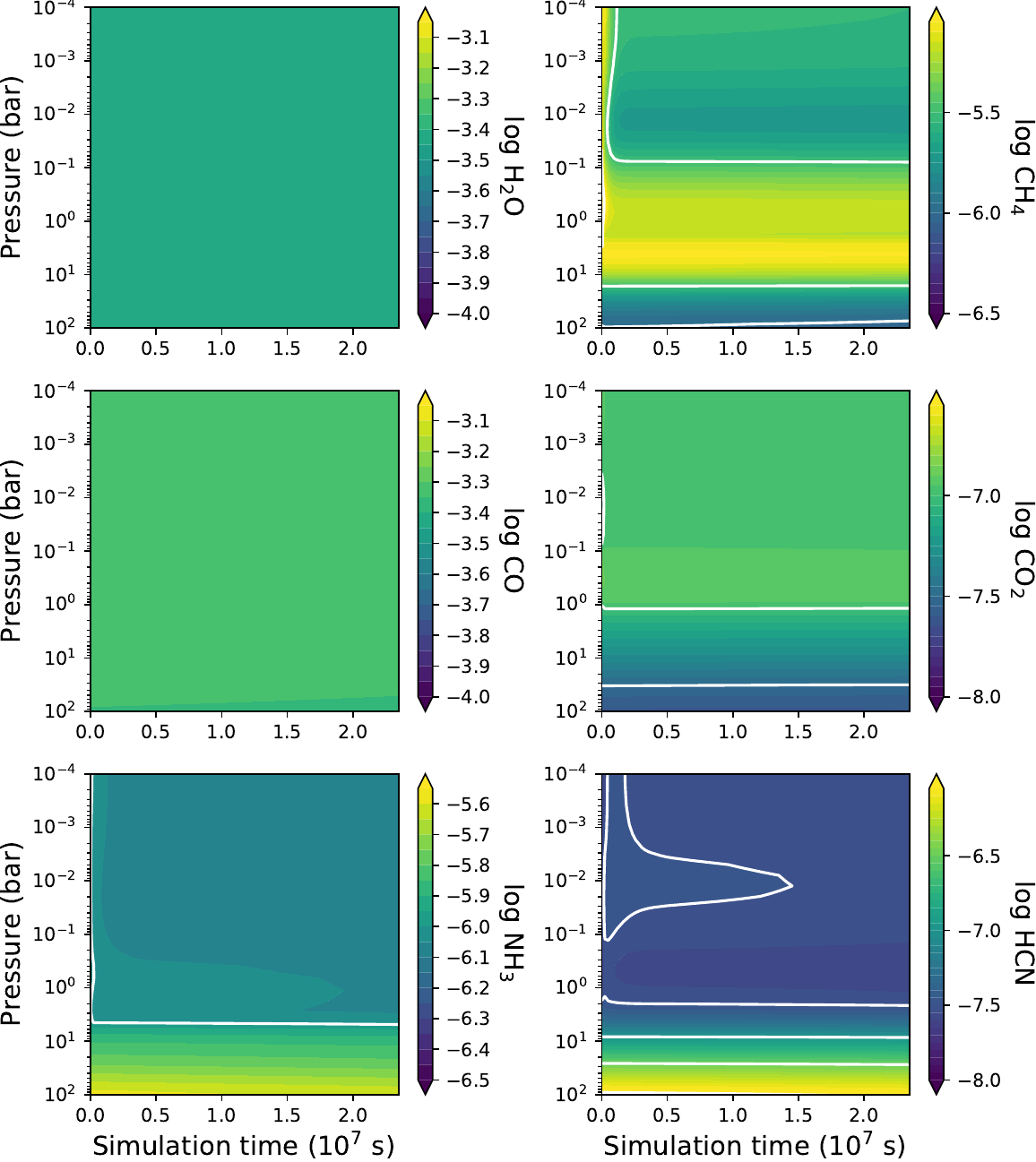}
    \caption{The number fractions at the night side of this exoplanet atmosphere model ($\Teff=1400$~K, $g=10$~m/s$^2$, G5) show convergence to a steady-state situation. Different species show different evolution speeds with respect to simulation time of the pseudo-2D chemistry code. Note that the colour maps for each molecule are different. White contours are spaced at intervals of 0.5 in log-space.}
    \label{fig_chem_convergence}
\end{figure}

To empirically check for convergence after 100 rotations, we plotted the number fractions at the morning terminator ($-90\degrees$) for some dominant molecules as a function of the total simulation time after 100 rotations (Fig.~\ref{fig_chem_convergence}). The model corresponds to a planet with $\Teff=1400$~K and g=$10$~m/s$^2$. We purposely plot a model with moderate temperature, since it would take the longest time to converge to a periodic steady-state. The required convergence times for cold and hot models would be lower, because of the small longitudinal temperature differences in the former, and short chemical time-scales in the latter case. 

The different molecules shown in the evolution plot (Fig.~\ref{fig_chem_convergence}) do not all evolve to a steady-state equally fast. Species such as \chem{H_2O} and \chem{CO} do not show deviations from their vertically quenched initial state. Other species, such as \chem{CH_4} and \chem{NH_3} do show evolution, because their initial concentration is further from the steady-state and they are zonally quenched. The nitrogen-bearing molecules \chem{NH_3} and \chem{HCN} exhibit the slowest evolution. After about $10^7$~s, most of the species have attained a final steady-state.

We conclude that an integration time of 100 rotations, as we have used in this work for each chemical model, is sufficient to reach a steady-state when starting from a converged 1D thermochemical kinetics model. 

\section{Common Adiabat}
\label{sec_appendix_adiabat}

\begin{figure}
    \centering
    \includegraphics[width=\columnwidth]{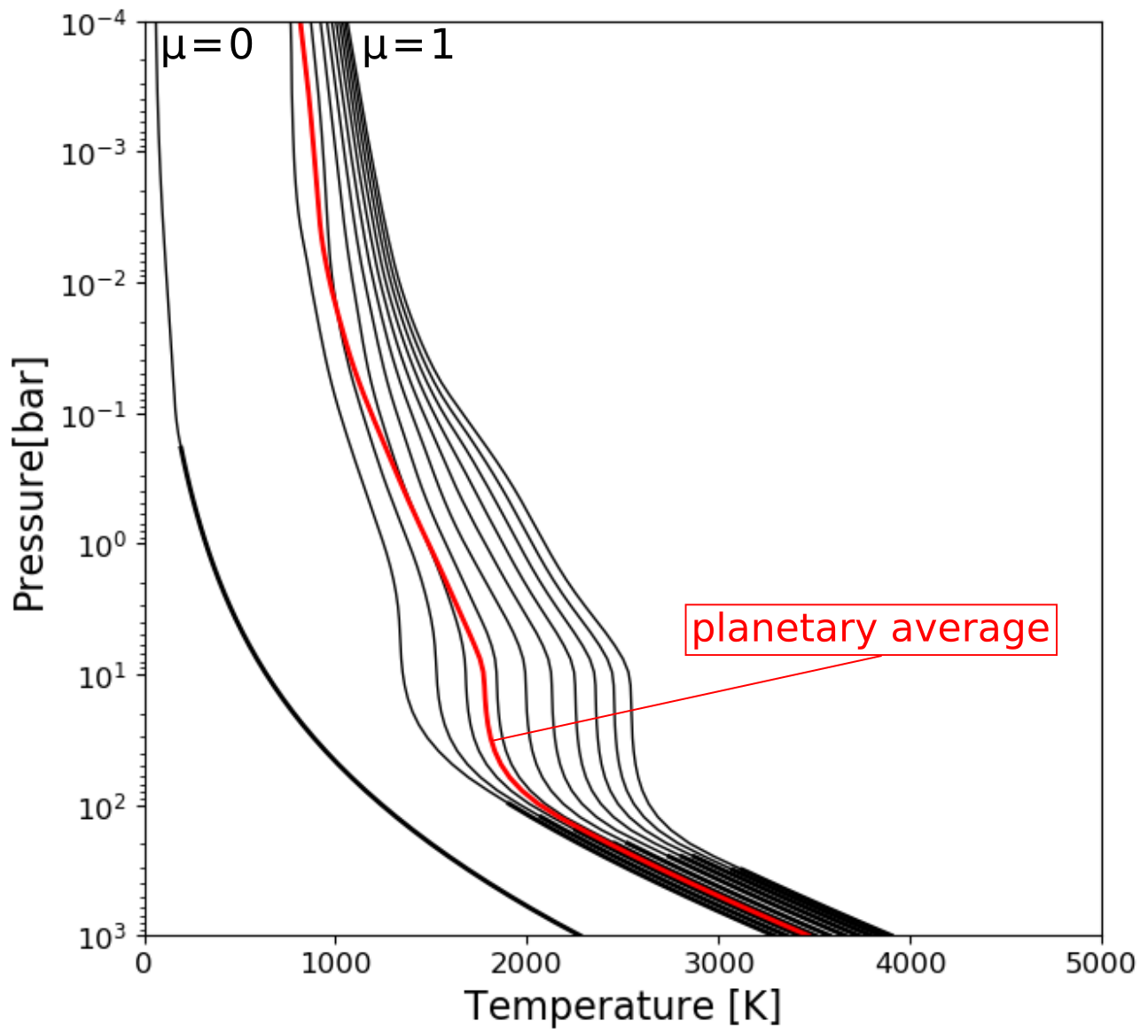}
    \caption{The pressure-temperature profiles corresponding to locations associated with different irradiation angles $\mu$, but constant intrinsic temperatures $T_{\textrm{int}} = 200$~K, end up on different adiabats in the deep convective part of the atmosphere. The profiles are calculated for a planet with $T_{\textrm{eff}}=1400$~K, $g=10$~m s$^{-2}$ and solar metallicity around a G5 star. The night-side profile ($mu$=0) is not irradiated, so the temperature-pressure profile is purely determined by the intrinsic temperature $T_{\textrm{int}}$. The bold part of the profiles signifies a convective regime. The red line shows the \textit{petitCODE} model associated with the planetary averaged distribution of the incident flux.}
    \label{fig_different_adiabats}
\end{figure}

The intrinsic temperature $T_{\textrm{int}}$ used in the computation of our \textit{petitCODE} models is associated with the intrinsic flux coming from the planet's interior. When pressure-temperature profiles for different irradiation angles $\mu$ are computed with a constant $T_{\textrm{int}}$, these profiles will end up at different adiabats in the deep, convective part of the atmosphere (see Fig.~\ref{fig_different_adiabats}). However, for the construction of our radiative-equilibrium profiles, we assumed that each profile corresponding to a certain irradiation angle converges to the same common adiabatic temperature gradient in the deep part of the atmosphere. To achieve this convergence, a different intrinsic temperature $T_{\textrm{int}}$ was imposed for every irradiation angle. This approach was previously used in 1D models of irradiated giant planets \citep{Barman2005} and in Newtonian-relaxed 3D GCMs for the hot Jupiters \hdtwenty{} and \hdeighteen{} \citep{Showman2008}. 

The assumption of a common adiabatic temperature lapse rate for all incidence angles is physically motivated. As these atmospheric layers are expected to be convective and lie well below the photosphere\footnote{The photosphere, the pressure layer at which the atmosphere becomes opaque to radiation, varies strongly with wavelength for planetary atmospheres, due to molecular opacities rather than a continuum opacity source \citep{Fortney2018}. In general, radiation originates from pressures between 0.01 and 1~bar.}, dynamical time-scales will be shorter than radiative time-scales. Therefore, these layers are likely to be well-mixed and have a common interior entropy \citep{Barman2004}. For fully convective models, it is a physical requirement that the irradiated and non-irradiated hemispheres of the same object reach a constant interior entropy at depth, as was argued for irradiated, convective stars in binary systems by \cite{Vaz1985, Brett1993}. Furthermore, 3D radiative transfer computations of irradiated planets also indicate a common radiative equilibrium in the deep layers of the atmosphere \citep{Hauschildt2008}.

Mixing towards a constant entropy value in the deep interior of convective atmospheres involves horizontal transport of energy from the irradiated side towards the non-irradiated side. This is, however, impossible to model rigorously in the 1D framework of \textit{petitCODE}. The total emerging flux of a converged \textit{petitCODE} model is equal to the sum of the intrinsic and irradiation fluxes of the planet \citep[see][]{Molliere2015}: 
\begin{equation}
    F_{\textrm{emerging}} = \sigma \left[ T_{\textrm{int}}^4 + \mu \frac{T_\ast^4}{4} \left( \frac{R_\ast}{d} \right)^2  \right],
\end{equation} where $R_\ast$ and $T_\ast$ are the stellar radius and temperature, $d$ is the distance from the star, and $\sigma$ is the Stefan-Boltzmann constant. Increasing the angle of incidence $\mu$ is equivalent to increasing the irradiation on the planet. This additional flux can reach the top layers of the convective zone and, since the temperature gradient is set by the adiabatic gradient, the extra heating causes an increase in temperature and entropy throughout the entire deep convective region. To counteract this and obtain a constant interior entropy, the intrinsic temperature can be lowered for larger angles of incidence and raised for smaller angles of incidence, thereby mimicking the horizontal transport of energy in these deep convective layers \citep{Barman2004}.

Thus, the intrinsic temperatures associated with each irradiation angle are reduced or increased to match a common adiabat. In this study, the common adiabatic temperature gradient is taken to be the one associated with the model of globally-averaged energy redistribution (indicated as the red line in Fig.~\ref{fig_different_adiabats}). We adopted a constant intrinsic temperature $T_{\textrm{int}}=200$~K for all planetary-averaged models, 
to facilitate both the intercomparison between the models in our grid, and the comparison with earlier works using \textit{petitCODE} \citep{Molliere2015, Molaverdikhani2019}. However, it is important to note that hot Jupiters may exhibit intrinsic temperatures that are much higher, ranging from 100~K to 700~K, depending on their irradiation and radius inflation \citep{Thorngren2019}. Furthermore, the intrinsic temperature and the extent of radius inflation have been shown to impact the atmospheric circulation of hot Jupiters \citep{Carone2020}. A dedicated study of the impact of the intrinsic temperature, however, is outside the scope of this study. 
The pressure-temperature profiles of a locus corresponding to a certain irradiation angle then have their intrinsic temperature adjusted, so that they end up on this planetary-averaged adiabat. 

\section{The Assumption of Synchronous Rotation}\label{sec_appendix_synchron}

Synchronous rotation is attained by tidal interaction of the planet with its host star. The efficiency of this interaction depends on the the physical properties of the star and the planet, and the distance between them. Since the semi-major axis in our grid is calculated based on the effective temperature of the planet and the star, it might be that for some combination of model parameters the tidal interaction efficiency is very low, and synchronous rotation may not be a valid assumption. 

In order to investigate the validity of synchronous rotation, we computed order-of-magnitude estimations of the synchronization times for each of our grid models, using equation~(1) in \cite{Showman2002}:
\begin{equation}\label{eq_tau_synch}
    \tau_{\rm syn} \approx Q \left( \frac{R_p^3}{GM_p} \right) \left| \omega - \omega_{\rm syn} \right| \left( \frac{M_p}{M_\ast} \right)^2 \left( \frac{a}{R_p} \right)^6.
\end{equation} Here, $M_p$, $R_p$ are the planetary mass and radius, $M_\ast$ is the stellar mass, $G$ is the gravitational constant, $a$ is the semi-major axis, $\omega$ is the starting rotational angular velocity and $\omega_{\rm syn}$ is the synchronized angular velocity, namely equal to the orbital angular velocity. Finally, $Q$ is the tidal dissipation factor. We followed the approach of \cite{Showman2002} in adopting $Q=10^5$ and choosing $\omega = 1.76 \cdot 10^{-4}$~s$^{-1}$, which is the current rotation rate of Jupiter. Likewise, Jupiter values $M_p=1$~$M_{\rm Jup}$ and $R_p = 1$~$R_{\rm Jup}$ were adopted for each model.
Other parameters have been calculated per grid model based on Section~\ref{sec_parameter_dependence}. In computing the synchronization times, we have assumed that increasing and decreasing the rotation rate of the planet due to tidal interaction is equally efficient \citep{Murray1999}.

\begin{figure}
    \centering
    \includegraphics[width=0.7\columnwidth]{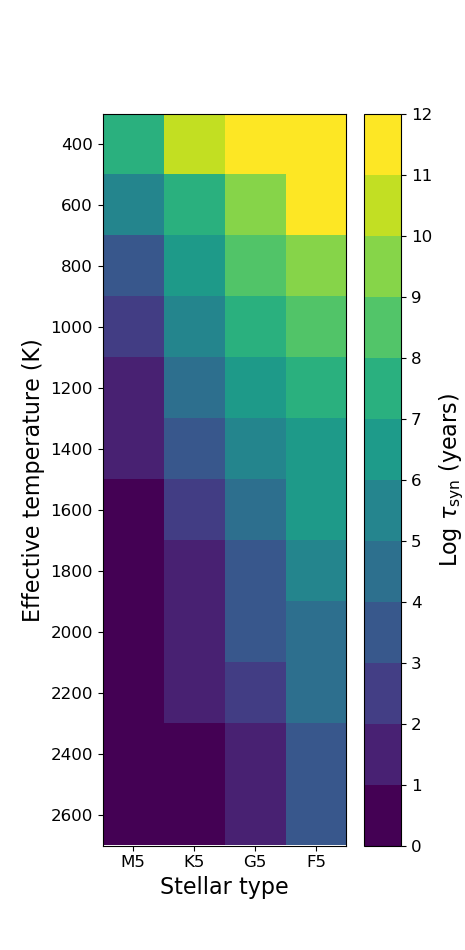}
    \caption{The tidal synchronization time-scales for each model in the grid are predominantly below 100~Myr, indicating that synchronous rotation is a good assumption for most models. However, for the colder planets orbiting hotter stars (\textit{upper right corner}), the time-scales are comparable to or longer than the typical evolution time-scale, and the assumption breaks down. The synchronization time-scales have been computed using equation~\eqref{eq_tau_synch} based on the physical parameters of Jupiter.}
    \label{fig_tau_synch}
\end{figure}

In Fig.~\ref{fig_tau_synch}, the synchronization time-scales based on our order-of-magnitude calculation are shown for all grid model configurations. We find that for all hot models ($\Teff > 1000$~K) and almost all models orbiting cool (M5, K5) stars, the synchronization time-scales are shorter than 100~Myr, which is at least one order of magnitude below the typical age of exoplanets \citep[e.g.][]{Silva_Aguirre2015, Maxted2015}. Hence, for these models synchronous rotation is a reasonable assumption. On the other hand, the coldest models orbiting G5- and F5-stars have synchronization time-scales higher than 10~Gyr and semi-major axes above 0.1~AU (see Table~\ref{tab:grid}). This makes it quite unlikely for these planets to be found in a 1:1 spin-orbit resonance, and for this subset of our grid synchronous rotation is not a valid assumption.
However, since this work is mainly a parameter study, we opt to still include the more improbable model configurations as well, since they can facilitate the derivation of general trends in the parameter space. For detailed studies of the atmospheric circulation on asynchronously rotating planets, we refer to \cite{Penn2018} (rocky planets) and \cite{Rauscher2014_andKempton, Auclair-Desrotour2018, Kazumasa2019} (giant planets).

\section{Sample of Radiative-Convective Equilibrium Models}
\label{sec_appendix_petitcode}

\begin{figure*}
    \centering
    \includegraphics[width=0.99\textwidth]{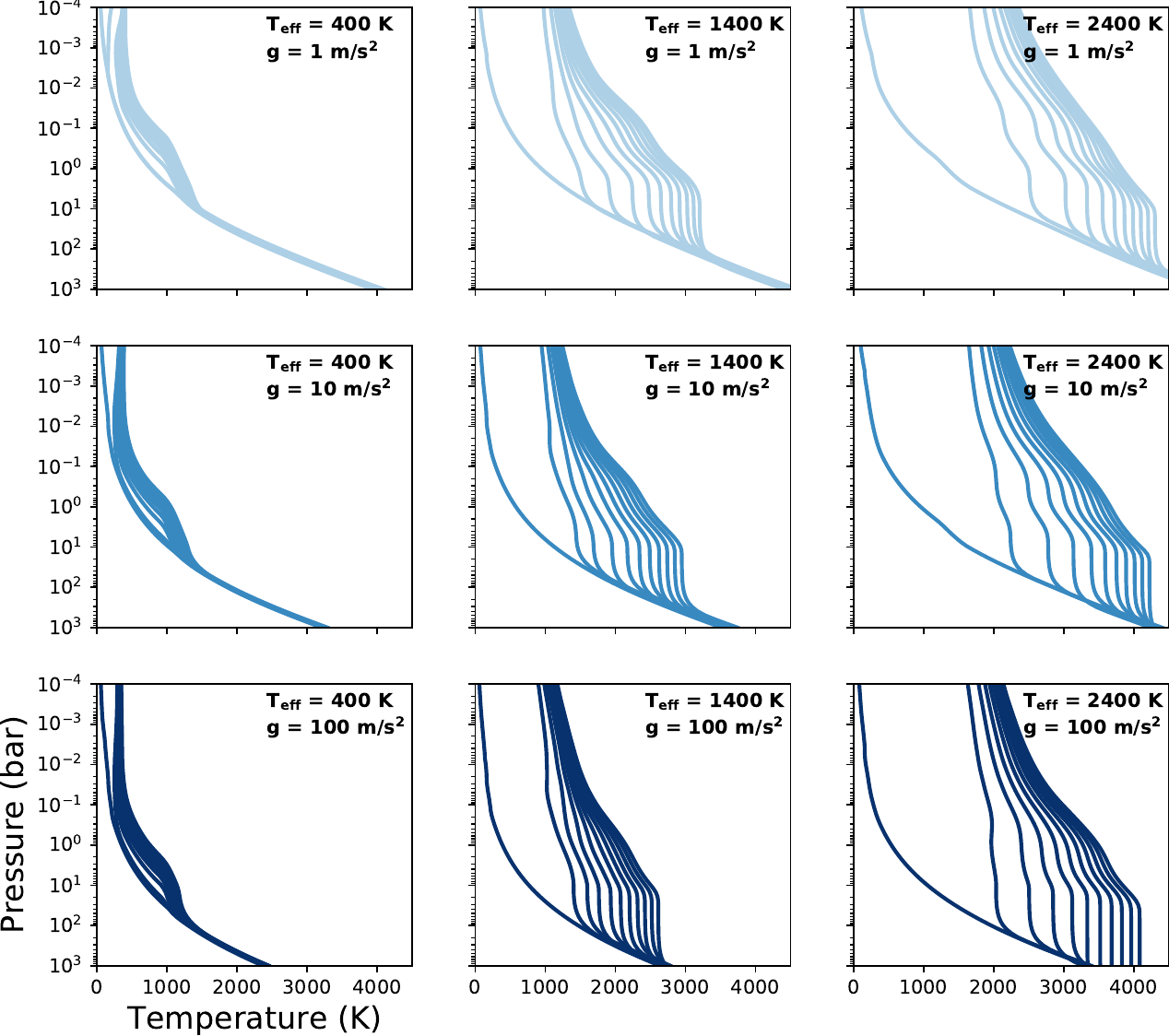}
    \caption{The radiative-equilibrium temperature for different angles of incidence, plotted per model as separate lines, increases with effective temperature, except for the coldest profile, which corresponds to the planet night side. Gravity acts to shift the temperature structure vertically to higher/lower pressures. This selection of \textit{petitCODE}\citep{Molliere2015} models corresponds to models with effective temperatures of 400~K, 1400~K, and 2400~K, and surface gravities of 1~m/s$^2$, 10~m/s$^2$, and 100~m/s$^2$.}
    \label{fig_petitcode_selection}
\end{figure*}

We present a selection of \textit{petitCODE} radiative-convective equilibrium temperatures as a function of pressure and the angle of incidence. These models have been constructed using the methodology laid out in Appendix~\ref{sec_appendix_adiabat}. Increasing the effective temperature of the exoplanet atmosphere leads to an increase in radiative-equilibrium temperature for all angles of incidence, except that of the night side (Fig.~\ref{fig_petitcode_selection}).

\section{Additional figures for Low and High Surface Gravities}
\label{sec_appendix_figures}

\begin{figure*}
    \centering
    \includegraphics[width=0.99\textwidth]{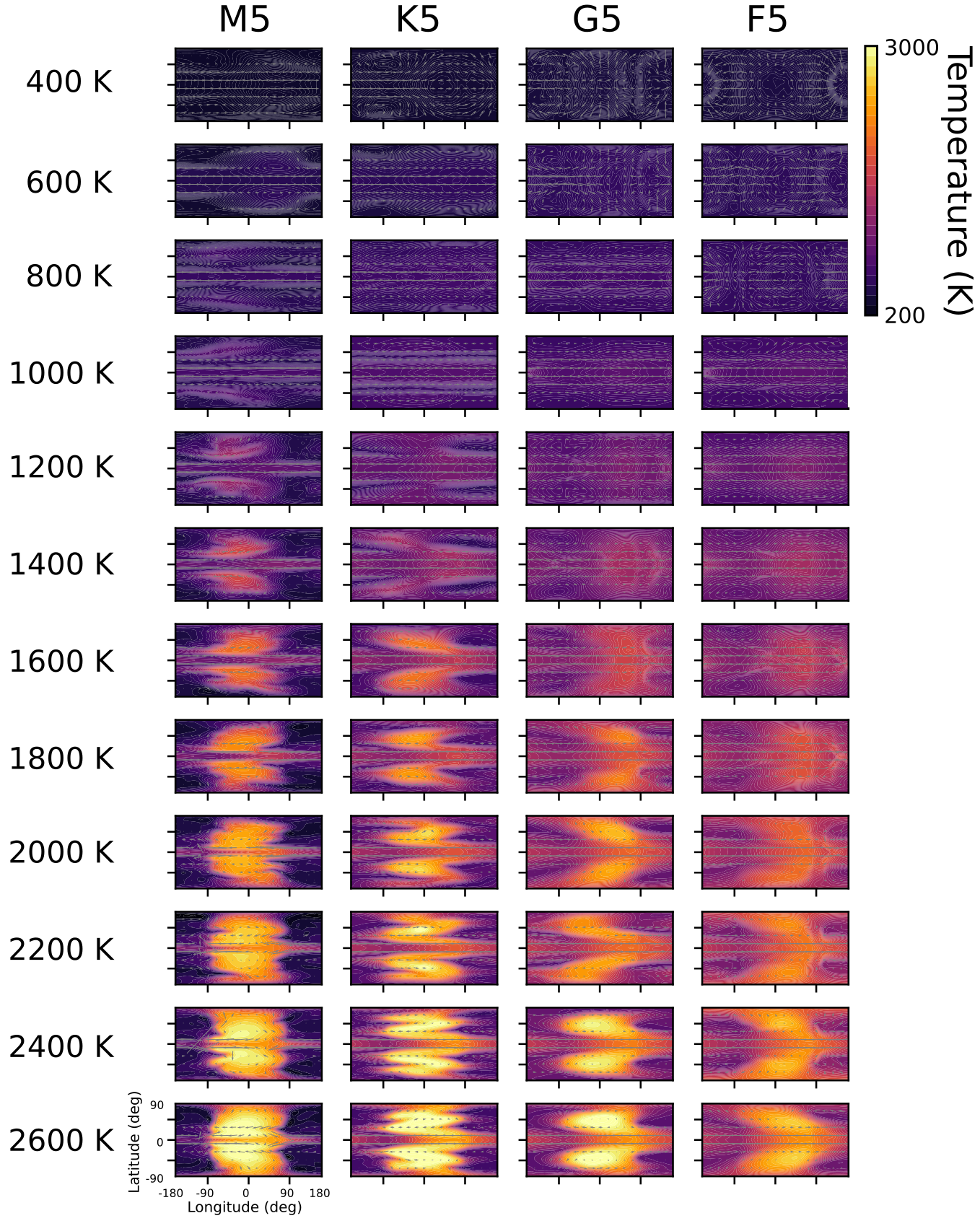}
    \caption{Temperature maps for the $g = 1$~m/s$^2$ models in our grid, plotted as isobaric slices at a pressure of 70~mbar. Grey arrows show the direction of the local horizontal winds, and the arrow length is scaled by the wind speed. The plots are ordered by host star type (horizontally) and planetary effective temperature (vertically). The substellar point is located at ($0\degrees, 0\degrees)$.}
    \label{fig_temperature_maps_appendix_g1}
\end{figure*}

\begin{figure*}
    \centering
    \includegraphics[width=0.99\textwidth]{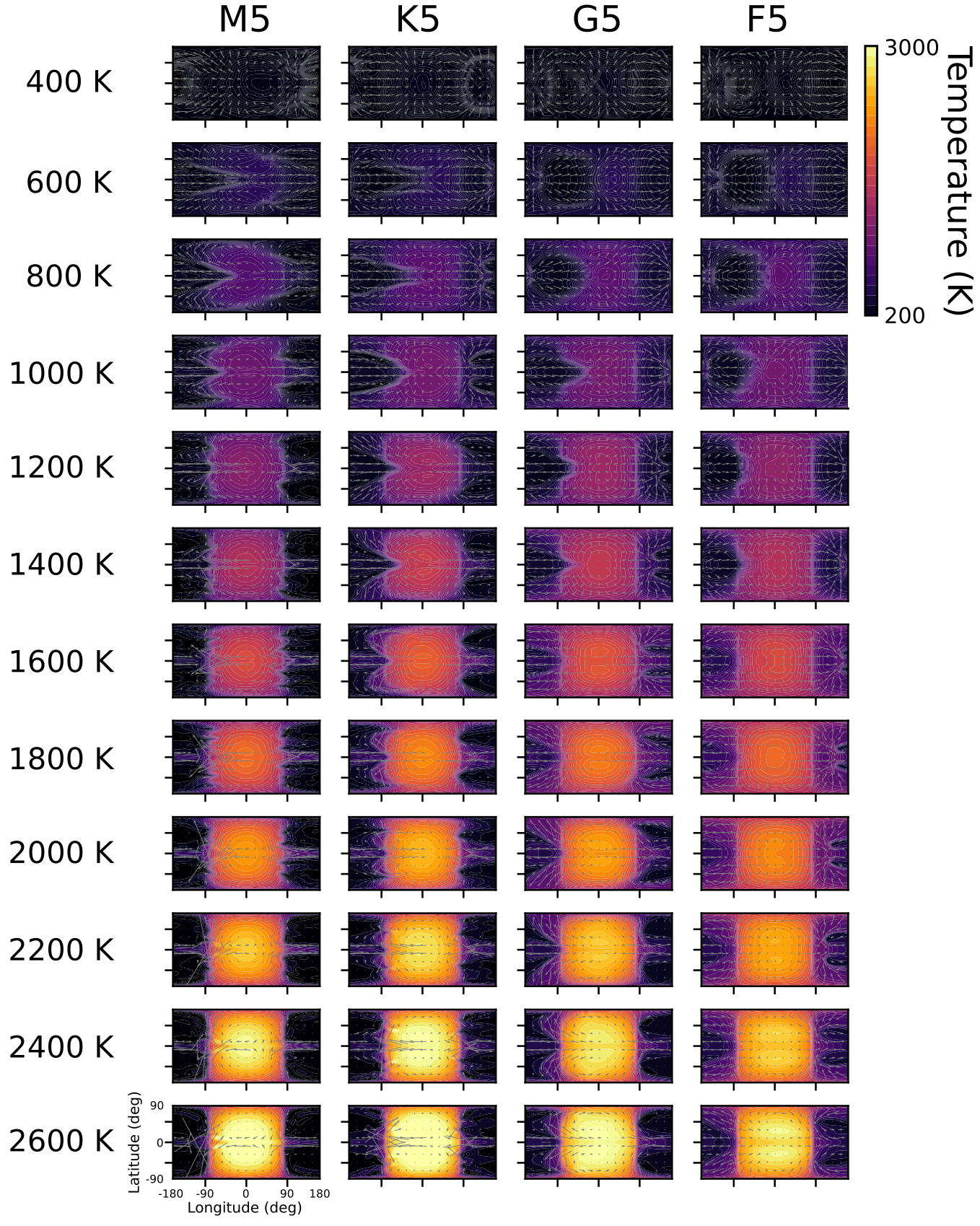}
    \caption{Temperature maps for the $g = 100$~m/s$^2$ models in our grid, plotted as isobaric slices at a pressure of 70~mbar. Grey arrows show the direction of the local horizontal winds, and the arrow length is scaled by the wind speed. The plots are ordered by host star type (horizontally) and planetary effective temperature (vertically). The substellar point is located at ($0\degrees, 0\degrees)$.}
    \label{fig_temperature_maps_appendix_g100}
\end{figure*}

\begin{figure*}
    \centering
    \includegraphics[width=0.99\textwidth]{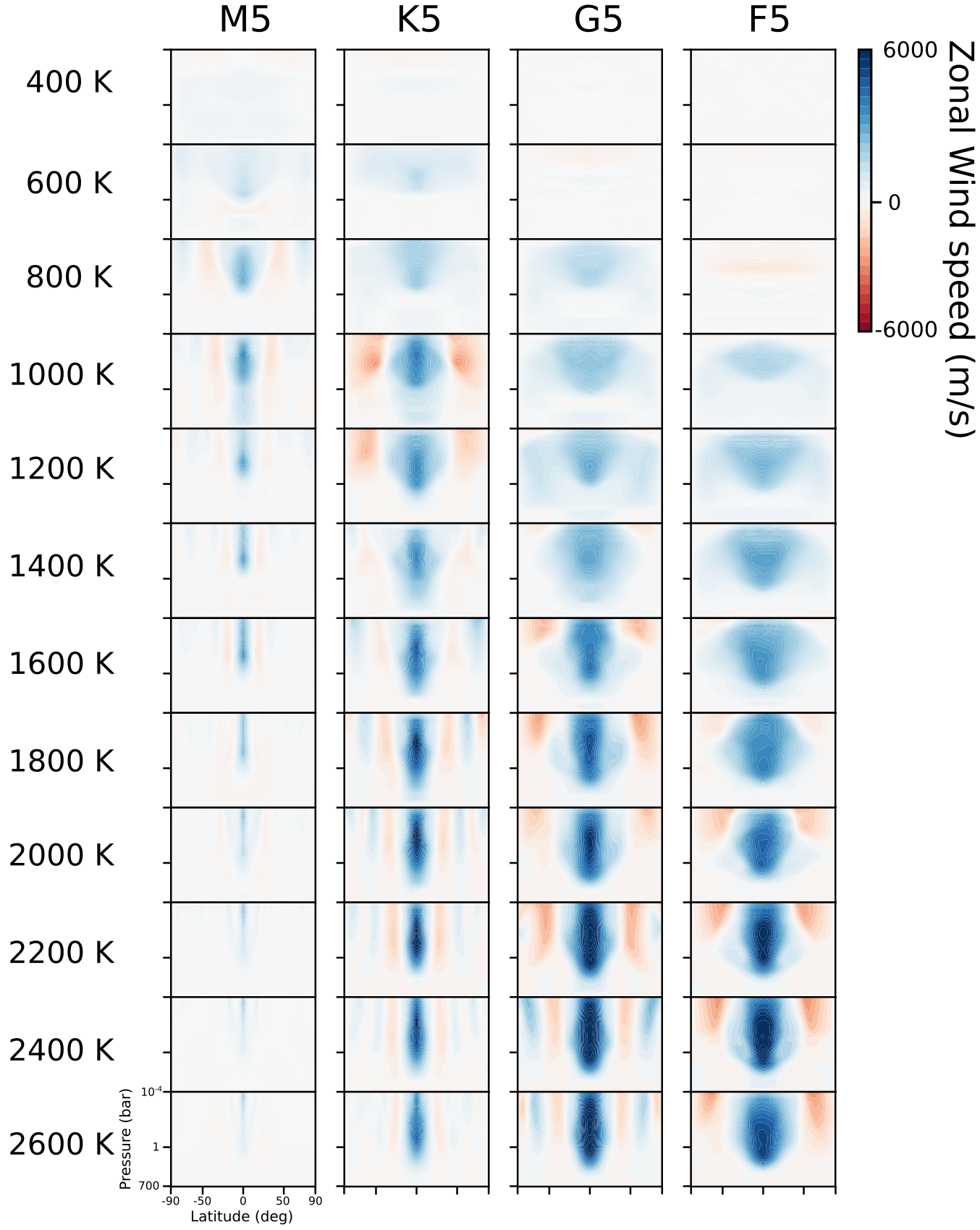}
    \caption{Zonally averaged zonal wind plots for the $g = 1$~m/s$^2$ models in our grid. Positive values of the wind speed (blue) correspond to prograde wind flow. The plots are ordered by host star type (horizontally) and planetary effective temperature (vertically).}
    \label{fig_zonalwind_plots_appendix_g1}
\end{figure*}

\begin{figure*}
    \centering
    \includegraphics[width=0.99\textwidth]{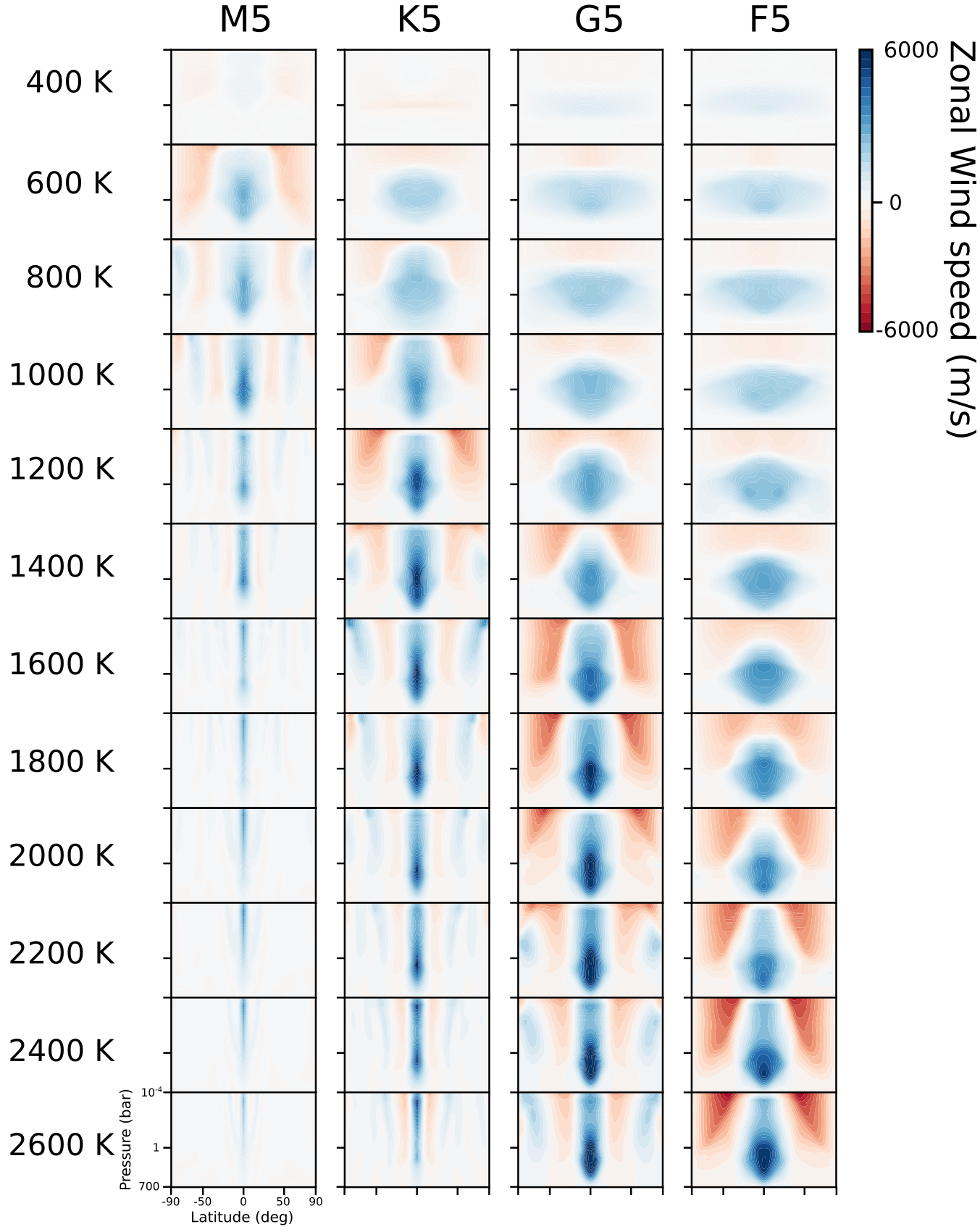}
    \caption{Zonally averaged zonal wind plots for the $g = 100$~m/s$^2$ models in our grid. Positive values of the wind speed (blue) correspond to prograde wind flow. The plots are ordered by host star type (horizontally) and planetary effective temperature (vertically).}
    \label{fig_zonalwind_plots_appendix_g100}
\end{figure*}

\begin{figure*}
    \centering
    \includegraphics[width=0.76\textwidth]{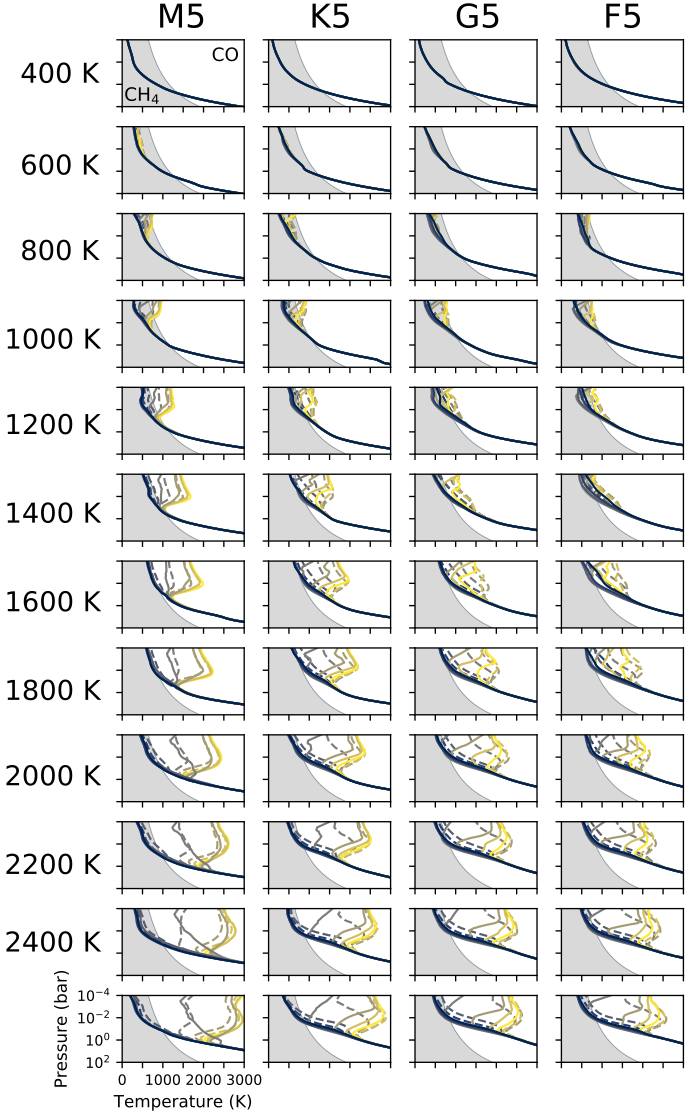}
    \caption{Temperature-pressure profiles from the $g = 1$~m/s$^2$ GCM models in our grid. Temperatures are meridionally averaged between $\pm20\degrees$, shown for longitudes spaced $30\degrees$ apart ($-180\degrees$, $-150\degrees$,\ldots , $150\degrees$), and colour coded with a gradual change from the anti-stellar (blue) to substellar point (yellow). Full and dashed lines mark locations west and east of the substellar point. The shaded background colour indicates the parameter space where methane is more abundant than CO in chemical equilibrium \citep{Visscher2012}. The plots are ordered by host star type (horizontally) and planetary effective temperature (vertically).}
    \label{fig_LPT_profiles_appendix_g1}
\end{figure*}

\begin{figure*}
    \centering
    \includegraphics[width=0.76\textwidth]{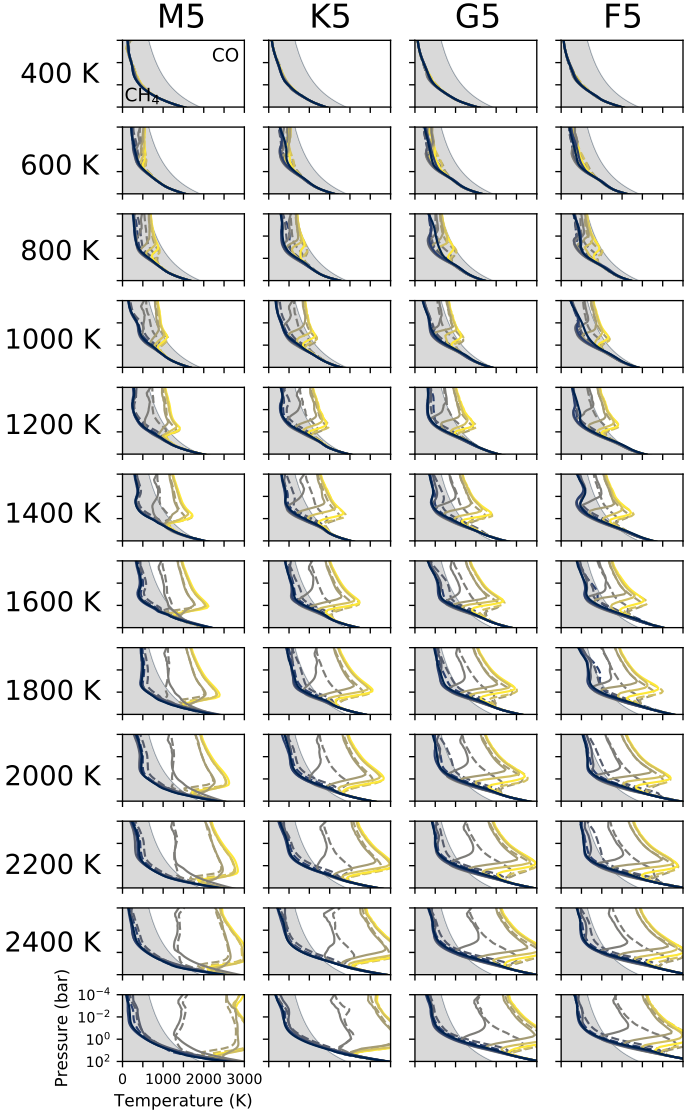}
    \caption{Temperature-pressure profiles from the $g = 100$~m/s$^2$ GCM models in our grid. Temperatures are meridionally averaged between $\pm20\degrees$, shown for longitudes spaced $30\degrees$ apart ($-180\degrees$, $-150\degrees$,\ldots , $150\degrees$), and colour coded with a gradual change from the anti-stellar (blue) to substellar point (yellow). Full and dashed lines mark locations west and east of the substellar point. The shaded background colour indicates the parameter space where methane is more abundant than CO in chemical equilibrium \citep{Visscher2012}. The plots are ordered by host star type (horizontally) and planetary effective temperature (vertically).}
    \label{fig_LPT_profiles_appendix_g100}
\end{figure*}

\begin{figure*}
    \centering
    \includegraphics[width=0.99\textwidth]{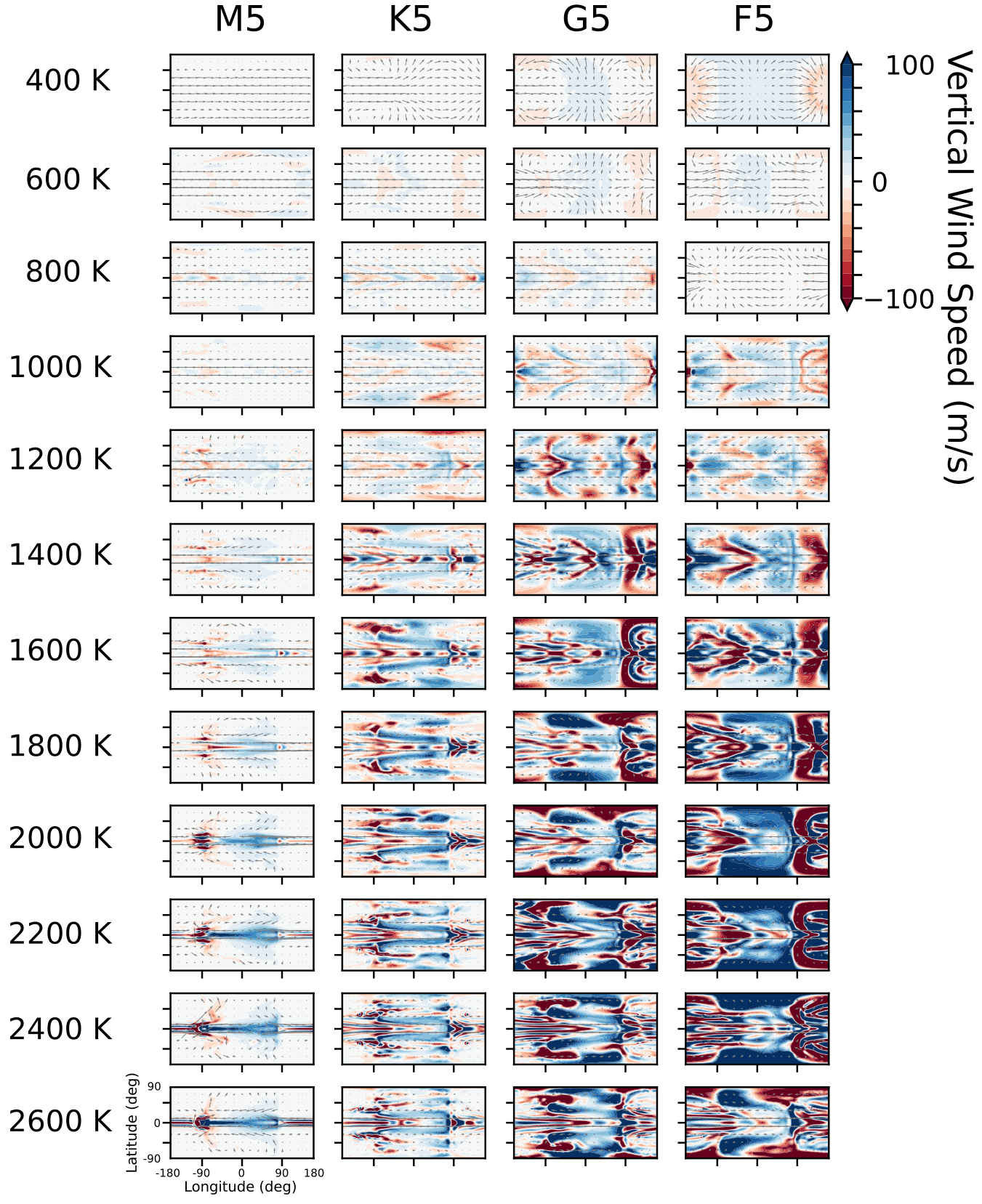}
    \caption{Maps of the vertical wind speed on a 70~mbar isobar for the $g = 1$~m/s$^2$ models in our grid. Upwelling motions (blue) are mostly present on the day side (longitudes between $-90\degrees$ and $+90\degrees$), whereas downwelling flow (red) is mostly visible on the night side, often as very localized fronts. The plots are ordered by host star type (horizontally) and planetary effective temperature (vertically). Thus, the rotation rate decreases along rows. For the sake of comparison, the colour scaling is kept consistent with Fig.~\ref{fig_verticalwind_plots} and \ref{fig_verticalwind_plots_appendix_g100}.}
    \label{fig_verticalwind_plots_appendix_g1}
\end{figure*}

\begin{figure*}
    \centering
    \includegraphics[width=0.99\textwidth]{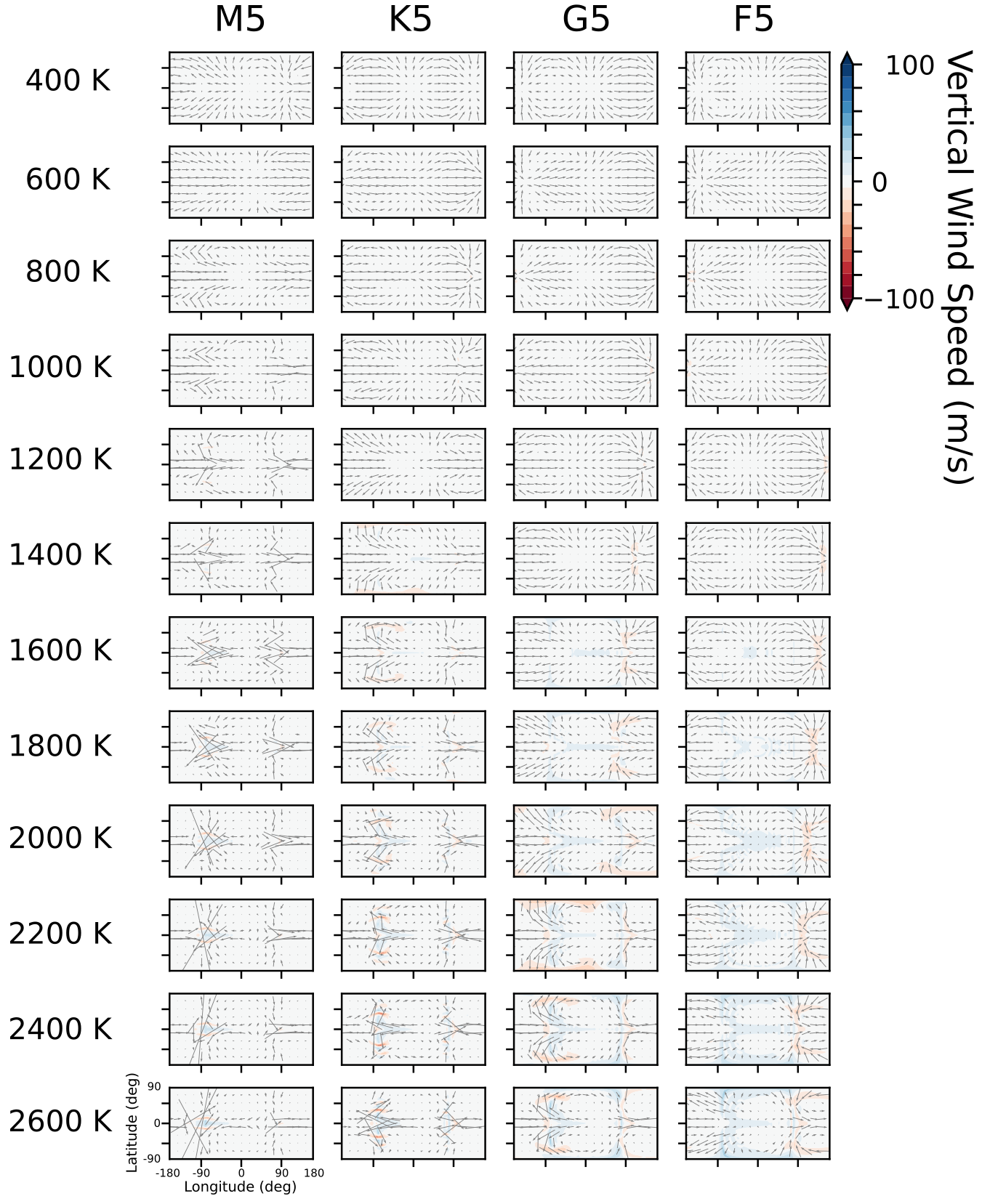}
    \caption{Maps of the vertical wind speed on a 70~mbar isobar for the $g = 100$~m/s$^2$ models in our grid. Upwelling motions (blue) are mostly present on the day side (longitudes between $-90\degrees$ and $+90\degrees$), whereas downwelling flow (red) is mostly visible on the night side, often as very localized fronts. The plots are ordered by host star type (horizontally) and planetary effective temperature (vertically). Thus, the rotation rate decreases along rows. For the sake of comparison, the colour scaling is kept consistent with Fig.~\ref{fig_verticalwind_plots} and \ref{fig_verticalwind_plots_appendix_g1}.}
    \label{fig_verticalwind_plots_appendix_g100}
\end{figure*}

\begin{figure*}
    \centering
    \includegraphics[width=\textwidth]{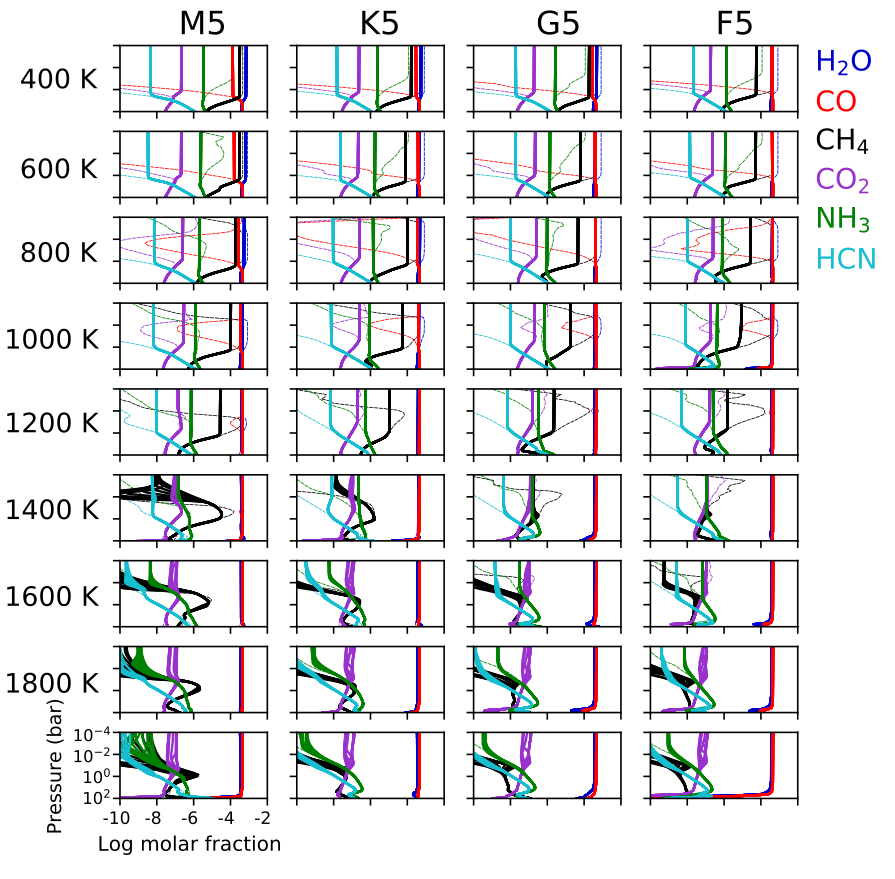}
    \caption{The abundances for the main molecular species, computed with the pseudo-2D chemistry code for $g = 1$~m/s$^2$ are shown as 12 vertical profiles plotted at different longitudes ($-180\degrees$, $-150\degrees$,\ldots , $150\degrees$). The dashed lines denote the chemical equilibrium composition at the substellar point. Note that models with effective temperatures of 2200~K, 2400~K and 2600~K are omitted.}
    \label{fig_2dchem_appendix_g1}
\end{figure*}

\begin{figure*}
    \centering
    \includegraphics[width=\textwidth]{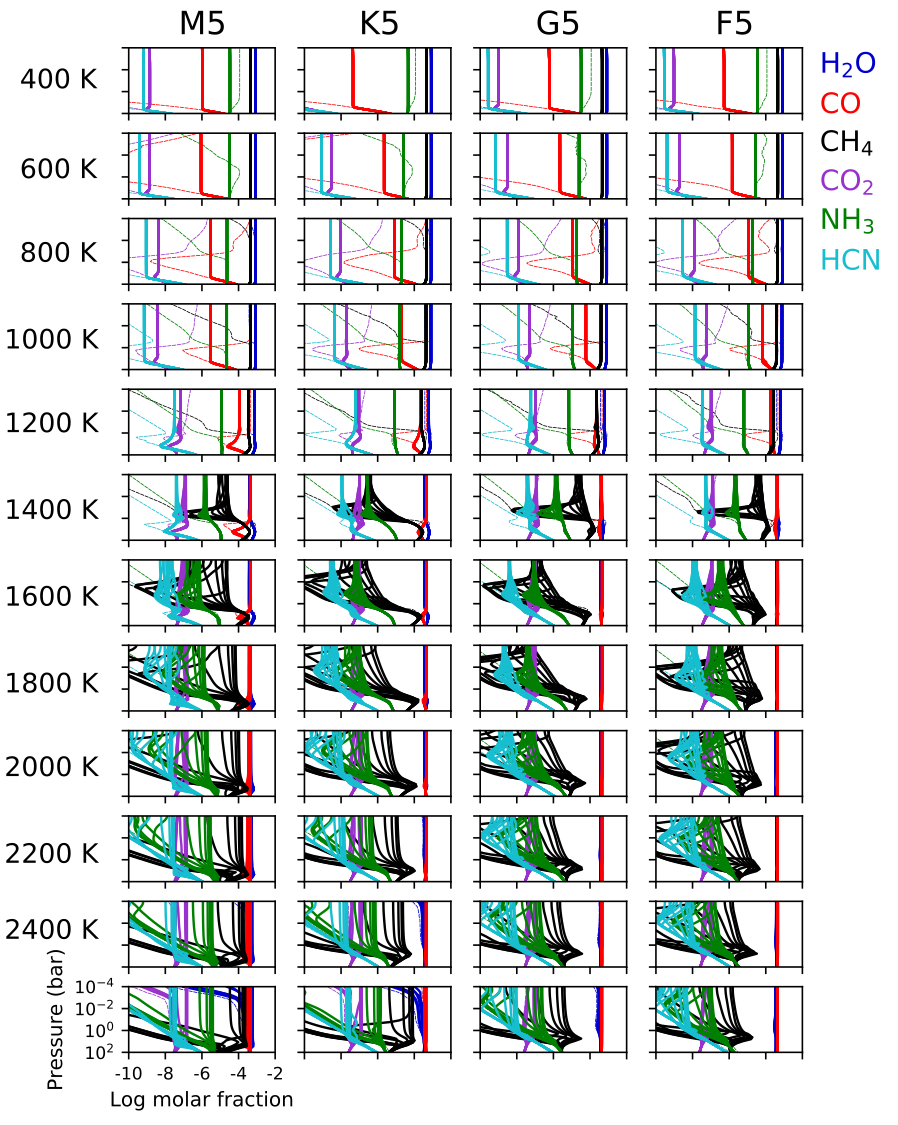}
    \caption{The abundances for the main molecular species, computed with the pseudo-2D chemistry code for $g = 100$~m/s$^2$ are shown as 12 vertical profiles plotted at different longitudes ($-180\degrees$, $-150\degrees$,\ldots , $150\degrees$). The dashed lines denote the chemical equilibrium composition at the substellar point.}
    \label{fig_2dchem_appendix_g100}
\end{figure*}

Here, we present additional figures of the low- ($g=1$~m/s$^2$) and high- ($g=100$~m/s$^2$) gravity models in our grid, which are analogous to the medium gravity models shown in Section~\ref{sec_results}. These consist of isobaric temperature maps (Fig.~\ref{fig_temperature_maps_appendix_g1} and \ref{fig_temperature_maps_appendix_g100}), zonal-mean zonal wind plots (Fig.~\ref{fig_zonalwind_plots_appendix_g1} and \ref{fig_zonalwind_plots_appendix_g100}), pressure-temperature profiles (Fig.~\ref{fig_LPT_profiles_appendix_g1} and \ref{fig_LPT_profiles_appendix_g100}), vertical wind maps (Fig.~\ref{fig_verticalwind_plots_appendix_g1} and \ref{fig_verticalwind_plots_appendix_g100}), and pseudo-2D chemistry plots (Fig.~\ref{fig_2dchem_appendix_g1} and \ref{fig_2dchem_appendix_g100}). For the low-gravity chemistry models, models with effective temperatures above 2000~K are omitted. For these cases, computations with the pseudo-2D chemical kinetics code did not result in physically acceptable solutions. The reason for the numerical problems has been identified to be linked to the atmospheric column height, which becomes disproportionately high in comparison with the planet radius, resulting in unphysically low column densities. Furthermore, planets with a low gravity of $g = 1$~m/s$^2$
and temperatures above 2000 K are not expected to retain their atmosphere, due to thermal evaporation. Therefore, we choose to omit these ultra-hot models for the low-gravity case.

\bsp	
\label{lastpage}
\end{document}